# Microstructural constitutive model for polycrystal viscoplasticity in cold and warm regimes based on continuum dislocation dynamics

S. Amir H. Motaman*, Ulrich Prahl

*Department of Ferrous Metallurgy, RWTH Aachen University, Intzestr. 1, D-52072 Aachen, Germany*

**A R T I C L E   I N F O**

**A B S T R A C T**

Viscoplastic flow of polycrystalline metallic materials is the result of motion and interaction of dislocations, line defects of the crystalline structure. In the microstructural (physics-based) constitutive model presented in this paper, the main underlying microstructural processes influencing viscoplastic deformation and mechanical properties of metals in cold and warm regimes are statistically described by the introduced sets of postulates/axioms for continuum dislocation dynamics (CDD). Three microstructural (internal) state variables (MSVs) are used for statistical quantifications of different types/species of dislocations by the notion of dislocation density. Considering the mobility property of dislocations, they are categorized to mobile and (relatively) immobile dislocations. Mobile dislocations carry the plastic strain (rate), while immobile dislocations contribute to plastic hardening. Moreover, with respect to their arrangement, dislocations are classified to cell and wall dislocations. Cell dislocations are those that exist inside cells/subgrains, and wall dislocations are packed in (and consequently formed) the subgrain walls/boundaries. Therefore, the MSVs incorporated in this model are cell mobile, cell immobile and wall immobile dislocation densities. The evolution of these internal variables is calculated by means of adequate equations that characterize the dislocation processes dominating material behavior during cold and warm monotonic viscoplastic deformation. The constitutive equations are then numerically integrated; and the constitutive parameters are determined/fitted for a widely used ferritic-pearlitic steel (20MnCr5).

## Contents

* Corresponding author. Tel: +49 241 80 90133; Fax: +49 241 80 92253
  Email: seyedamirhossein.motaman@iehk.rwth-aachen.de

| Nomenclature | | |
|---|---|---|
| *Symbol* | *Unit* | *Description* |
| $b$ | m | Burgers length (magnitude of Burgers vector) |
| $c$ | - | Material coefficient associated with probability amplitude of a dislocation process |
| $d$ | m | Critical distance for dislocation processes |
| $e$ | % | Relative error, residual/objective/fitness function |
| $f$ | - | Volume fraction |
| $G$ | Pa | Shear modulus |
| $H$ | Pa | Viscoplastic tangent modulus |
| $l$ | m | Length of a dislocation segment |
| $m$ | - | Strain rate sensitivity parameter |
| $n$ | - | Number of active slip/glide systems |
| $M$ | - | Taylor factor |
| $p$ | - | Probability amplitude (or frequency) of a dynamic dislocation process |
| $q$ | J.m$^{-3}$ | Volumetric heat generation |
| $r$ | - | Temperature sensitivity coefficient |
| $R$ | m | Radius |
| $s$ | - | Temperature sensitivity exponent |
| $\boldsymbol{S}$ | | Nonlocal microstructural state (a set containing all MSVs) |
| $t$ | s | Time |
| $T$ | K | Temperature |
| $u$ | J.m$^{-3}$ | Volumetric stored energy |
| $\boldsymbol{v}$ | m.s$^{-1}$ | Velocity vector |
| $w$ | J.m$^{-3}$ | Volumetric work |
| | | |
| $\alpha$ | - | Dislocation interaction strength/coefficient |
| $\beta$ | - | Dissipation factor, efficiency of plastic dissipation, or Taylor–Quinney coefficient |
| $\gamma$ | - | Shear strain in slip system, mean shear strain |
| $\varepsilon$ | - | Mean/nonlocal (normal) strain |
| $\theta$ | Pa | Plastic/strain hardening |
| $\varphi$ | Pa.s | Viscous/strain-rate hardening |
| $\kappa$ | - | Material constant associated with dissipation factor |
| $\Lambda$ | m | Dislocations spacing |
| $\rho$ | m$^{-2}$ | Dislocation density |
| $\sigma$ | Pa | Mean/nonlocal (normal) stress |
| $\tau$ | Pa | Resolved shear stress |
| | | |
| *Index* | | *Description* |
| ac | | Accumulation |
| an | | Annihilation |

| | |
|---|---|
| $d$ | Dynamic |
| gn | Generation |
| GN | Geometrically necessary |
| $i$ | Immobile |
| loc | Local |
| $m$ | Mobile, melt |
| $(n)$ | Time step index, previous time increment |
| $(n+1)$ | Current time increment |
| nc | Nucleation |
| $c$ | Cell |
| $p$ | Plastic |
| pn | Pinning |
| rm | Remobilization |
| $s$ | Static |
| sat | Saturated |
| SS | Statistically stored |
| $t$ | Total (subscript), time (superscript) |
| tr | Trapping |
| $v$ | Viscous |
| $w$ | Wall |
| $x$ | Cell, wall, or total ($x = c, w, t$) |
| $y$ | Mobile, immobile, or total ($y = m, i, t$), yield/flow |
| $z$ | Dislocation process ($z = \mathrm{gn, an, ac, tr, nc, rm,} spn, srm$) |
| | |
| $0$ | Reference, initial/undeformed state |
| $+$ | Increase/gain |
| $-$ | Decrease/Loss |
| $\hat{\ }$ | Normalized/dimensionless ($\hat{x} = \frac{x}{x_0}$) |
| $\smile$ | Function |
| $\sim$ | Statistical mean/average |
| $-$ | Equivalent |

| **Abbreviation** | **Description** |
|---|---|
| CB | Cell block |
| CDD | Continuum dislocation dynamics |
| CMD | Continuum microstructure dynamics |
| DDD | Discrete dislocation dynamics |
| DDW | Dense dislocation wall |
| DSA | Dynamic strain aging |
| DTH | Dynamic thermal hardening |
| DTS | Dynamic thermal softening |
| EBSD | Electron backscatter diffraction |
| EVP | Elasto-viscoplastic |
| FE | Finite element |
| FEMU | Finite element model updating |
| GB | Grain boundary |
| GNB | Geometrically necessary boundary |
| GND | Geometrically necessary dislocation |
| IDB | Incidental dislocation boundary |
| MB | Micro-band |

| MD | Molecular dynamics |
|---|---|
| MSV | Microstructural (internal) state variable |
| NMS | Nonlocal microstructural state |
| RMV | Representative material volume |
| RVE | Representative volume element |
| SC | Sub-cell |
| SFE | Stacking fault energy |
| TMM | Thermo-micro-mechanical, or thermo-mechanical-microstructural |
| TWIP | Twinning-induced plasticity |

## 1. Introduction

Nowadays, finite element (FE) simulation of manufacturing processes such as metal forming is an important part of process and product design and development in the industry. Correct and accurate description of material behavior and properties is always the biggest challenge in simulation of industrial manufacturing processes that are based on viscoplastic deformation. Dislocation-density-dependent physics-based constitutive models of metal plasticity while are computationally efficient and history-dependent, can accurately account for varying process parameters such as strain rate and temperature. Since these models are founded on essential phenomena dominating the deformation, they have a wide range of usability and validity. Moreover, they are suitable for manufacturing chain simulations as they can efficiently compute the cumulative effect of the various manufacturing processes by following the microstructure state through the entire manufacturing chain including interpass periods and give a realistic prediction of material behavior and final product properties. The constitutive models are mainly divided into the following main categories (Lin and Chen, 2011; Rusinek et al., 2010):

### 1.1. Empirical constitutive models

Empirical constitutive models provide description of the yield/flow stress based on empirical observations, and consist of some mathematical functions that lack the physical background. In these models, yield stress is usually an explicit function of accumulated plastic strain, strain rate and temperature, which makes empirical models not history dependent. Moreover, the problem with the accumulated plastic strain is that it is a non-measurable virtual variable. Furthermore, empirical constitutive models are normally characterized by reduced number of material constants and easy calibration. However, due to their empirical characteristics, they are usually covering limited range of applicability and flexibility and offer low accuracy. Empirical models are determined by fitting parameters of model equations to experimental data without considering the physical processes causing the observed material behavior. Empirical or phenomenological models are also named engineering models as they are more common in engineering applications than the physics-based material models. A number of common empirical constitutive models of metals viscoplasticity are usually incorporated in commercial FE programs (Follansbee and Kocks, 1988; Hockett and Sherby, 1975; Johnson and Cook, 1983; Khan and Liang, 1999; Rusinek and Klepaczko, 2001; Sung et al., 2010).

### 1.2. Microstructural constitutive models

Microstructural or physics-based constitutive models account for microstructural (physical) aspects of the material behavior. These are the models where knowledge about the underlying microstructural processes including dislocation processes, is applied to formulate the thermo-micro-mechanical (TMM) or thermo-mechanical-microstructural constitutive equations. In addition, since microstructural material models simulate the main microstructural phenomena influencing the overall mechanical response of the material to plastic deformation, they can be used in wide range of deformation parameters (strain rate and temperature) and loading/deformation modes (tension, compression, creep, and relaxation). Additionally, since in industrial metal forming processes, material usually undergoes variety of loading types and parameters, history-dependent microstructural constitutive models are much more suitable and robust for comprehensive simulations of complex industrial metal forming processes. Physics-based models may follow different approaches to describe microstructure evolution/kinetics in polycrystalline metallic materials under plastic deformation:

- _Discrete dislocation dynamics (DDD)_ in which slip/glide/motion and interaction of individual dislocations are considered; and thereby, the stress-strain response of the material is a result of direct simulation of a huge assemble of dislocations in a very small representative volume element (RVE). Some good examples can be found in works of van der Giessen and Needleman (1995), Zbib et al. (1998), Devincre et al. (2001), Zbib and La Diaz de Rubia (2002), Arsenlis et al. (2007), Groh et al. (2009), Zhou et al. (2010), Huang et al. (2012) and Chandra et al. (2018). The algorithms based on DDD are extremely costly in terms of computation time, they do not account for size effect, and cannot be readily implemented in standard FE software for industrial application. Nevertheless, DDD simulations are more efficient than those of molecular dynamics (MD) because the RVEs in DDD are much bigger in size than those used in MD which run in atomistic level. DDD simulations provide insights into larger scale behavior (mesoscale). Therefore, models based on MD and DDD are very helpful for studying of dislocation processes and construction of statistical continuum models based on dislocation density (Kubin, 2013; Li et al., 2014; Monavari et al., 2016).

Furthermore, DDD and MD have proven to be very useful tools for stochastic modeling of microplasticity experiments such as micro-pillar/column compression, micro-bending and nano-indentation, that together with (in-situ) electron microscopy provide a deep understanding into collective behaviors of dislocations such as dislocation sources, arrangements, configurations and interactions (Csikor et al., 2007; Cui et al., 2017, 2016a, 2016b; Derlet and Maaß, 2013; El-Awady et al., 2009; Greer et al., 2008; Lee et al., 2009; Miller et al., 2004; Motz et al., 2008; Ng and Ngan, 2008; Oh et al., 2009; Papanikolaou et al., 2018; Parthasarathy et al., 2007; Po et al., 2014; Shiari et al., 2005; Yamakov et al., 2002; Zaiser, 2013; Zhang et al., 2015; Zhu, 2004).

- _Continuum dislocation dynamics (CDD)_ describe the microstructure indirectly, so that the effects of the micro level processes are accounted for, in an average way on the macro level. Such type of approach is the subject of this study using the notion of dislocation density which unlike (accumulated) plastic strain, is measurable to some extent by electron microscopy and X-ray techniques. Due to their physical nature, besides plastic/strain/work hardening, constitutive models based on different types of dislocation densities have the potential of predicting many other important processes such as creep, relaxation, dynamic strain aging, static aging, and bake hardening.

  Opposite to the DDD approach, constitutive models based on CDD are formulated at the macro level, i.e. the microstructural (internal) state variables (MSVs) are calculated for a mesoscale representative material volume (RMV). In macroscale simulation of plastic deformation, material/integration points are considered to be RMVs. Additionally, an ideal test sample under homogenous uniaxial normal load (tension or compression) which is used for obtaining flow curve is assumed to be a RMV.

  With indirect approach of CDD, dislocation density-dependent constitutive models provide a bridge between the micro-level phenomena and macro-level continuum quantities, such as stress and strain rate. Furthermore, simulations performed using these constitutive models are much less costly (in the same range of common empirical constitutive models) and less complicated compared to the algorithms based on discrete dislocation dynamics. Hence, they can be easily implemented in standard FE software and are suitable for industrial applications. This study is limited to the isotropic case meaning that Bauschinger, asymmetrical and anisotropic effects are not included in the presented TMM constitutive model. However, the constitutive relations developed in this context can be applied in the crystal plasticity framework in order to account for anisotropic effects caused by nonuniform dislocation density evolution on each slip system which is negligible in steel alloys.

Viscoplastic deformation of crystalline materials with respect to temperature may occur in one of the following regimes/domains:

- cold regime: the maximum temperature in cold regime is normally characterized by temperatures above which diffusion controlled dislocation mechanisms such as dislocation climb and pinning become dominant (approximately $T < 0.3\,T_m$, where $T$ is the absolute temperature; and $T_m$ is the melting absolute temperature) (Galindo-Nava and Rae, 2016);
- warm regime: warm viscoplastic deformation of crystalline materials occur above cold but below hot temperature regime (approximately $0.3\,T_m < T < 0.5\,T_m$) (Doherty et al., 1997; Sherby and Burke, 1968); and
- hot regime: hot viscoplastic deformations are carried out above warm temperature regime. Hot metal forming processes are characterized by at least one of the hot/extreme microstructural processes such as recrystallization, phase transformation, notable precipitate processes, etc. (roughly $0.5\,T_m < T < T_m$).

Strain rate has different regimes as well, however, independent from the material (Field et al. (2004)):

- creep or static: $\dot{\varepsilon} < 10^{-4}\ s^{-1}$ (where $\dot{\varepsilon}$ is the strain rate);
- quasi-static: $10^{-4}\ s^{-1} \leq \dot{\varepsilon} < 10^{-2}\ s^{-1}$;
- intermediate-rates: $10^{-2}\ s^{-1} \leq \dot{\varepsilon} \leq 10\ s^{-1}$;
- dynamic: $10\ s^{-1} < \dot{\varepsilon} \leq 10^{3}\ s^{-1}$; and
- shock/highly-dynamic: $\dot{\varepsilon} > 10^{3}\ s^{-1}$.

In the present paper, the focus is on monotonic viscoplastic deformation of polycrystalline metallic materials in cold and warm temperature regimes with quasi-static to intermediate strain rate levels. Nevertheless, further studies must be conducted for generalization to hot regime probably in the framework of continuum micro-dynamics (CMD) which its scope encompasses CDD. Microstructural constitutive models based on CDD and CMD can be coupled with finite element method as microstructural solvers (in addition to the regular thermal and mechanical FE solvers) to simulate not only metal forming processes but also the entire material processing chain including casting, heat treatment, interpass periods, etc., one after the other. However, the microstructural constitutive model based on CMD are characterized with application of extra MSVs in addition to different types of dislocation density. These additional statistical MSVs can be phase volume fractions, void volume fraction, recrystallized volume fraction, twinned volume fraction, transformed volume fraction, precipitate concentrations, etc. In this paper, postulates of CDD are listed. Based on these postulates, a microstructural constitutive model for polycrystal isotropic viscoplasticity in cold and warm regimes is derived. Model's kinetics differential equations are then numerically integrated and subsequently its parameters are determined for a ferritic-pearlitic steel alloy 20MnCr5 which is widely used in industrial forging of automotive components such as bevel gears.

## 2. Background

The foundation of CDD was formed in 1930's when the pioneers of the theory, Orowan (1934) and Taylor (1934) introduced the concept of dislocation density and its relationship with plastic strain rate and yield stress. They considered the mean effect rather than individual aspects of dislocations motion and interactions in an attempt to describe macroscopic plastic flow. Johnston and Gilman (1959) were the first to propose an evolutionary equation for dislocation density, which was simply the superposition of a multiplication term and a recovery term. Webster (1966) applied an analogous methodology to creep by assuming that the time rate of change of dislocation density is due to multiplication, immobilization and annihilations processes. Later, widespread adoption of this approach was established by the works of Bergström (1970), Kocks (1976), Mecking and Kocks (1981).

Later, many physics-based constitutive models were proposed with more than one type of dislocation density (multi-MSV models). Ananthakrishna and Sahoo (1981), Bammann and Aifantis (1982), Estrin and Kubin (1986) and Hähner and Zaiser (1999) classified dislocations with respect to their mobility feature. They predicted the flow curve by constitutive models based on two MSVs, mobile/glissile/glide and immobile/sessile dislocation densities. 2-MSV models of mobile and immobile dislocation densities are still being developed and applied in different frameworks (Austin and McDowell, 2011; Hansen et al., 2013; Li et al., 2014). Mughrabi (1983), Nix et al. (1985) and Estrin et al. (1998) differentiated dislocations with regard to their arrangement in the dislocation network; and introduced models with two MSVs, cell and wall dislocation densities. Estrin et al. (1996) and Roters et al. (2000) proposed models with three MSVs, cell mobile, cell immobile, and wall immobile dislocation densities. These models appreciate different dislocations based on their mobility property and their arrangement. Likewise, Blum et al. (2002) approached the creep problem in metals by using a 2-MSV constitutive model that decomposed dislocations with respect to their singularity property, namely singular and dipolar dislocations. There exist models that account for dislocation character (edge and screw dislocations) (Cheong and Busso, 2004) and dislocation polarity (right-hand and left-hand) (Arsenlis and Parks, 2002; Roters, 2011). Ma and Roters (2004) classified dislocations further by allocating extra MSVs to their model for parallel and forest dislocation densities. Moreover, Estrin and Mecking (1992) incorporated effective grain size to the classical single variable Kocks-Mecking model Mecking and Kocks (1981).

Sandström and Lagneborg (1975), Busso (1998), Mukherjee et al. (2010) and Babu and Lindgren (2013) all developed multi-MSV microstructural constitutive models to characterize metals behavior under hot deformation by introduction of an additional statistical MSV, recrystallized fraction, with its corresponding kinetics equation. Fan and Yang (2011) and Bok et al. (2014) went further by allocating additional MSVs for each phase fraction to build a microstructural constitutive model for hot sheet metal forming.

Many researchers including Fleck et al. (1994), Fleck and Hutchinson (1997), Nix and Gao (1998), Gao et al. (1999), Gao (1999), Qiu et al. (2001), Gao and Huang (2001), Bhushan and Nosonovsky (2003), Huang et al. (2004), Abu Al-Rub and Voyiadjis (2004), Voyiadjis and Al-Rub (2005), Voyiadjis and Abed (2005), Brinckmann et al. (2006), Bardella (2006), Ardeljan et al. (2014), Lyu et al. (2015) and Nguyen et al. (2017b) distinguished between geometrically necessary dislocations (GNDs) and statistically stored dislocations (SSDs) to formulate strain (rate) gradient plasticity models. Furthermore, Busso (2000), Arsenlis and Parks (2002), Arsenlis (2004), Evers et al. (2004), Clayton et al. (2006), Ma et al. (2006), Beyerlein and Tomé (2008), Lim et al. (2011), Askari et al. (2013),

Li et al. (2014), Hochrainer et al. (2014), Sandfeld et al. (2015), and Askari et al. (2015) among many other authors applied the multi-MSV dislocation density-based approach to crystal plasticity framework.

Recently, some microstructural constitutive models have been developed for special purposes. Viatkina et al. (2007), Kitayama et al. (2013), Pham et al. (2013), Knezevic et al. (2013) and Zecevic and Knezevic (2015) proposed multi-MSV models for strain path dependent evolution of dislocation structures during cyclic plastic deformation to account for kinematic hardening and Bauschinger effect. Austin and McDowell (2011), Lloyd et al. (2014), Luscher et al. (2017) and Nguyen et al. (2017a) utilized dislocation density-based constitutive modeling for viscoplastic deformation of metals at dynamic and shock regimes. Patra and McDowell (2012) developed a physical–based constitutive model for inelastic deformation of irradiated bcc ferritic-martensitic steels by introduction of an extra statistical MSV, namely number of interstitial loops that are formed due to irradiation. Bouaziz and Guelton (2001), Allain et al. (2004) and Steinmetz et al. (2013) incorporated twinned volume fraction as an additional MSV to existing dislocation density-based models in order to reflect plastic hardening behavior of twinning-induced plasticity (TWIP) steels. These models were further developed and implemented in crystal plasticity framework by Wong et al. (2016) to account also for transformation-induced plasticity (TRIP) effect by incorporating one more MSV, namely transformed volume fraction. In addition, Kubin et al. (2002) and Ananthakrishna (2007) have published reviews on theoretical approaches for modeling of collective behavior of dislocations which readers are encouraged to read.

In the microstructural state variable approach, the mechanical state at a nonlocal/macroscale material point in a continuum body is characterized in terms of hidden or internal variables that statistically represent the nonlocal microstructural state (NMS) in addition to the observable statistical/nonlocal external variables such as temperature, strain rate and yield stress. To date, many statistical physics-based approaches have been applied for constitutive modeling of metals. Even though extensive work has been conducted in this area, there is not a universal agreement on the number and kind of MSVs to be used (Horstemeyer and Bammann, 2010) as well as the influencing microstructural processes and their associated kinetics equations that determine the values of MSVs. This study is an attempt for unification and completion of the previous works in the field of continuum dislocation dynamics, by gathering and modifying some of the most important postulates of the CDD theory. In the following section, it is argued that the statistical state of microstructure of pollycrystalline materials under monotonic and isotropic viscoplastic deformation in cold and warm regimes is fully defined by three MSVs that are cell mobile, cell immobile and wall immobile dislocation densities. For the first time, evolution of these MSVs with respect to time (or plastic strain) are described considering every statistically notable dislocation process/interaction affecting values of the aforementioned dislocation densities. Without comprehensive decomposition of dislocation types and processes as suggested in this paper, accurately capturing the mechanical response of complex metal alloys such as steels particularly in warm regime is not achievable.

## 3. Postulates and the constitutive model

To construct a constitutive model based on microstructural processes, first a set of axioms or postulates must be established as the basis for reasoning and subsequent derivation of constitutive relations. In the framework of CDD, the following postulates are introduced, although not all of them are independent. Additionally, for derivation of the constitutive equations, the consequences of each postulate in combination with the earlier ones (or the results of earlier postulates) are provided as well.

### 3.1. Fundamental postulates

1) <u>*Nonlocal microstructural state (NMS)*</u>: the mean/nonlocal yield/flow/critical shear stress resolved at slip systems ($\bar{\tau}_y$) as the nonlocal mechanical response of material is an implicit function of temperature, shear strain rate, and statistical state of microstructure, given by the following equation (Mecking and Kocks, 1981):

$$\bar{\tau}_y \equiv \check{\bar{\tau}}_y \left( T, \boldsymbol{S}, \dot{\bar{\gamma}}_p \right) ; \tag{1}$$

where $\check{\bar{\tau}}_y$ is the TMM constitutive function; $\dot{\bar{\gamma}}_p$ is mean plastic shear strain rate at slip systems; $T$ is temperature, and $\boldsymbol{S}$ is a set containing all MSVs and is referred to as nonlocal microstructural state since it represents the statistical state of microstructure:

$$\boldsymbol{S} \equiv \{ S_1, S_2, S_3, \dots \} ; \tag{2}$$

where $S_i$ is the i-th nonlocal MSV which can be various types of dislocation density, grain size, phase fractions, recrystallized fraction, precipitate concentration and size, etc. The evolution of each MSV which is often expressed as time rate of change of MSV, is a function of thermo-mechanical loading and $\boldsymbol{S}$ (Nadgornyi, 1988):

$$\dot{S}_k = \check{\tilde{S}}_k(T, \boldsymbol{S}, \dot{\tilde{\gamma}}_p) = \check{\tilde{S}}_k(T, S_1, S_2, S_3, \ldots, \dot{\tilde{\gamma}}_p) ; \tag{3}$$

where $\check{\tilde{S}}_k$ is the function that determines evolution of $S_k$ with respect to time. As shown in Fig. 1, $\boldsymbol{S}$ nonlocally represents the microstructural state of a material point on a macroscale continuum body. As deformation proceeds, the microstructure state $\boldsymbol{S}$ evolves towards a saturation/steady state $\boldsymbol{S}^{\text{sat}}$ (Estrin and Mecking, 1984).

Furthermore, it is emphasized that in this description, the NMS set and its components (MSVs) are treated based on the nonlocal principle/treatment (Eringen, 1983; Gao and Huang, 2001) as they are averaged over a mesoscale RMV. The mesoscale RMV must be a polycrystalline aggregate that represents the bulk material properly at the considered nonlocal continuum material point. Hence, it must consist of sufficient number of constituent single crystal grains in order to capture the size effect (Adams and Olson, 1998). This implies that inside the mesoscale RMV, locally the values of MSVs are not necessarily equal to their integral average over RMV due to their heterogeneous (local) distribution. For instance, some MSVs are highly concentrated at small regions while some are statistically distributed inside the grains of RMV (postulate (3)).

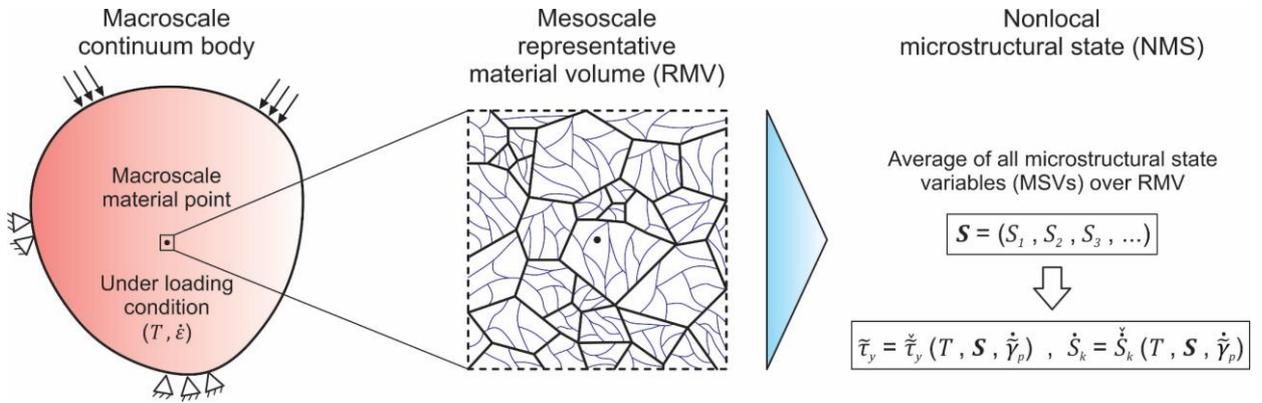

**Fig. 1.** Multiscale framework of CDD: schematic relation among macroscale continuum body under thermo-mechanical loading, mesoscale representative material volume, and nonlocal microstructural state.

It should be noted that accumulated plastic strain is not an MSV but a mechanical internal state variable and due to its virtual characteristic, it cannot be measured directly. Two identical material samples that are plastically (uniaxially) deformed to an equal amount of accumulated plastic strain but with different histories of temperature and strain rate (e.g. in cold or warm regime), if again deformed under an equal temperature and strain rate condition, do not necessarily yield the same stress response.

Furthermore, the mean/nonlocal normal yield/flow stress ($\sigma_y$) (or simply the yield stress) nonlocally applied on a polycrystalline aggregate is related to the mean yield shear stress resolved at its slip systems ($\bar{\tau}_y$) by Taylor factor (Kocks, 1970; Taylor, 1938):

$$M \equiv \frac{\sigma_y}{\bar{\tau}_y} = \frac{\dot{\tilde{\gamma}}_p}{\dot{\varepsilon}_p} = \frac{d\bar{\gamma}_p}{d\varepsilon_p} ; \tag{4}$$

where $\dot{\varepsilon}_p$ is the mean/nonlocal normal plastic strain rate (or simply the plastic strain rate). It is safe to assume constant $M \cong 3$ for a random orientation distribution of texture for bcc and fcc polycrystalline aggregates. Nevertheless, Taylor factor evolves as plastic strain accumulate; and it is also dependent on the deformation mode but for sufficiently random-textured polycrystals (weak texture or textureless) with random loading during deformation (isotropic case), these dependencies can be neglected (Kocks and Mecking, 2003).

2) *Dislocation mobility*: with respect to their mobility property, dislocations are divided into mobile and immobile dislocations. While mobile dislocations carry plastic strain, immobile dislocations contribute to plastic hardening (Estrin and Kubin, 1986; Hunter and Preston, 2015). Upon confronting obstacles, mobile dislocation segments may become fully or partially immobilized. Partial immobilization of a (prior) mobile dislocation segment and consequently its division to mobile and immobile dislocation segments is schematically shown in Fig. 2. As illustrated in Fig. 2, an individual dislocation might be consisted of several mobile and immobile dislocation segments. During movement, the length and mean bow-out radius of a bowed-out mobile segment bounded by adjacent immobile segments is reduced while the lengths of its bounding immobile dislocation segments are increased proportionally, as long as the immobile segments are not remobilized by remobilization mechanisms. As pointed out in postulate (1), according to the non-locality principle, mobile and immobile dislocation densities at each nonlocal material point are defined as follows:

$$\rho_{cm} \equiv \frac{1}{V} \sum_j l_{cm}^{(j)} \; ; \qquad l_{cm}^{(j)} \equiv \int_{l_{cm}^{(j)}} \mathrm{d}l_{cm}^{(j)} \; ; \tag{5}$$

$$\rho_{xi} \equiv \frac{1}{V} \sum_j l_{xi}^{(j)} \; ; \qquad l_{xi}^{(j)} \equiv \int_{l_{xi}^{(j)}} \mathrm{d}l_{xi}^{(j)} \; ; \qquad x = c, w \; ; \tag{6}$$

where $V$ is the considered volume which in nonlocal case, is the volume of mesoscale RMV; $l_{cm}^{(j)}$ and $l_{xi}^{(j)}$ are respectively lengths of j-th mobile dislocation segment and immobile dislocation segment of type $x$; $\mathrm{d}l_{cm}^{(j)}$ and $\mathrm{d}l_{xi}^{(j)}$ are infinitesimal elements of arc length along the dislocation segments $l_{cm}^{(j)}$ and $l_{xi}^{(j)}$, respectively; and $\rho_{cm}$ and $\rho_{xi}$ are (nonlocal) mobile dislocation density and immobile dislocation densities (of type $x$), respectively. $x$ can be $c$ or $w$ that respectively stand for cell and wall (see postulate (3)).

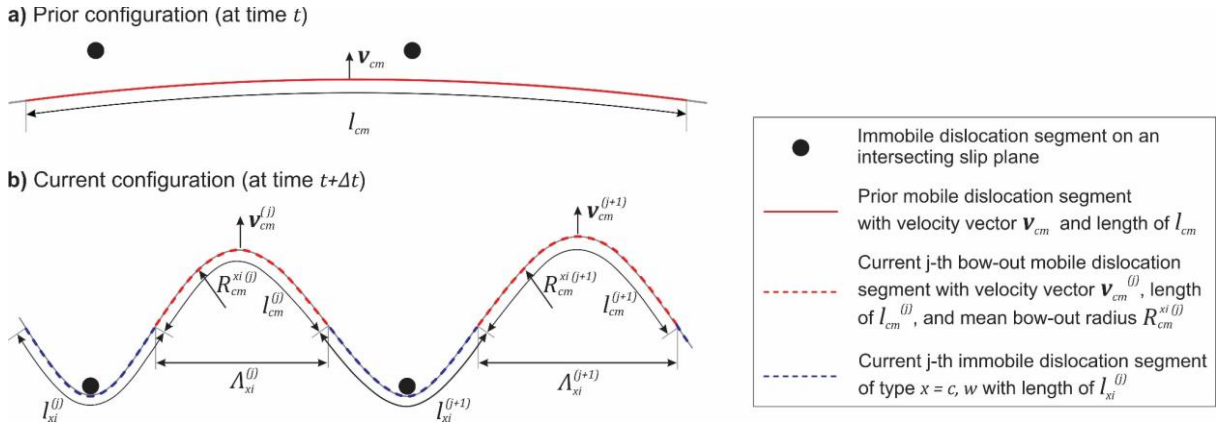

**Fig. 2.** Schematic of a mobile dislocation segment that is partially immobilized by immobile dislocations intersecting its slip plane (adopted from Hunter and Preston (2015)).

As shown in Fig. 2, it is postulated that motion of bowed out mobile dislocation segments moving on slip systems is always restricted at their both ends by immobile dislocation segments (either cell or wall immobile dislocations). Moving a bowed-out mobile dislocation segment requires a certain amount of shear stress acting on it which is inversely related to the mean radius of bowed-out segment (Hull and Bacon, 2011). In postulate (8), a statistical relation between bow-out radius and immobile dislocation density is derived. Hence, only the collection of immobile dislocations determines the required mean shear stress acting on bowed-out mobile dislocation segments to move them in order to accommodate plastic straining.

3) _Dislocation arrangement_: due to heterogeneous distribution of dislocations in crystal grains, dislocations with regard to their arrangement are categorized to cell dislocations that exist inside the cell/subgrain blocks/interiors and wall dislocations that form the cell walls/boundaries (Mughrabi, 1983). Wall dislocations are immobile and closely clustered (dense dislocation walls); and statistically, their Burgers vector is very similar to their surrounding wall dislocations. As such, they cause considerable lattice torsion/warp/twist/bending that is manifested by measurable misorientation angle across wall regions that can be detected and measured by electron backscatter diffraction (EBSD) (Gardner et al., 2010; Jiang et al., 2013) and X-ray methods. These dislocations are often referred to as geometrically necessary dislocations (GNDs) (Ashby, 1970; Nye, 1953) as they accommodate lattice curvature (incompatibility of plastic deformation) that arises by geometry change due to local gradient of plastic strain (Arsenlis and Parks, 1999; Gao and Huang, 2003) through formation of semi-planar geometrically necessary boundaries (GNBs) or dense dislocation walls (DDWs) (Kuhlmann-Wilsdorf and Hansen, 1991). Consequently, GNDs are in fact wall dislocations that contribute the most to plastic hardening due to the long range internal stress produced by them (Kassner et al., 2013; Mughrabi, 2006). As plastic straining proceeds, the degree of misorientation angle between adjacent cells or cell blocks (CBs) increases.

Physically, GNBs containing a high local dislocation density with a net Burgers vector are very different than spatially relatively random distributions of cell dislocations (Hughes et al., 2003). Cell dislocations that can be either mobile or immobile do not necessarily adopt any considerable particular semi-stable arrangement unless they become part of walls. Thus, cell dislocations are assumed to be statistically/homogenously distributed inside the CB structure (subgrain); and hence are known as statistically stored dislocations (SSD) (Ashby, 1970). However, stationary cell immobile dislocations form another type of semi-temporary accumulates/clusters/ pile-ups/bundles/tangles/nets named incidental dislocation boundaries (IDBs) (Kuhlmann-Wilsdorf and Hansen, 1991), or forest dislocations with relatively negligible misorientation angle. IDBs form secondary dislocation cells or sub-cells (SCs) inside subgrains. Moreover, the cell volumes bounded by GNBs may form tiny channel-shaped shear/deformation bands such as micro-bands (MBs) and lamellar bands (LBs) (Bay et al., 1992; Bay et al., 1989; Hughes, 1993; Hughes and Hansen, 1993). Fig. 3 schematically illustrates cell-wall substructure inside crystal grains. Therefore, there are three main independent types of dislocation density, cell mobile dislocation density ($\rho_{cm}$), cell immobile dislocation density ($\rho_{ci}$) and wall immobile dislocation density ($\rho_{wi}$):

$$\boldsymbol{S} \equiv \{\rho_{cm}, \rho_{ci}, \rho_{wi}\}; \tag{7}$$

$$\rho_{ct} \equiv \rho_{cm} + \rho_{ci} \equiv \rho_{SS}; \quad \rho_{wt} \equiv \rho_{wi} \equiv \rho_{GN}; \tag{8}$$

$$\rho_{ti} \equiv \rho_{ci} + \rho_{wi}; \quad \rho_{tm} \equiv \rho_{cm}; \tag{9}$$

$$\rho_t \equiv \rho_{tt} \equiv \rho_{ct} + \rho_{wt} = \rho_{tm} + \rho_{ti} = \rho_{cm} + \rho_{ci} + \rho_{wi}; \tag{10}$$

where $\rho$ denotes dislocation density; subscripts $t$, $i$, $m$, $c$, and $w$ respectively stand for total, immobile, mobile, cell and wall; and SS and GN represent statistically stored and geometrically necessary, respectively.

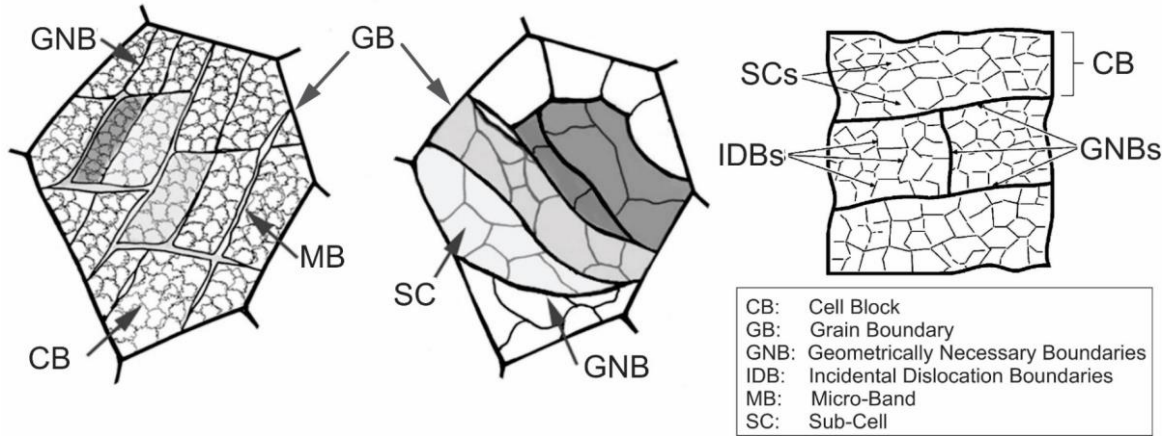

**Fig. 3.** Schematic of grain subdivision to cell-wall dislocation microstructure (Clayton et al., 2006; Hughes et al., 1998).

To sum up, wall immobile dislocations are locally (inside the RMV) highly concentrated/compacted at GNBs. Thus, local density of wall immobile dislocations at GNBs is much higher than density of cell immobile dislocations at IDBs or sub-cell interiors. Mobile dislocations are nearly homogeneously distributed in sub-cells. Consequently, local density of (cell) mobile dislocations is much lower than immobile dislocations.

Initial cell immobile dislocation density is determined by density of trapped and accumulated cell dislocations. Wall immobile dislocation density of an undeformed polycrystal is determined by the area of grain boundaries, phase and precipitate interfaces and other two-dimensional crystal defects. Even though grain boundaries and interfaces do not fit in classical definition of dislocations, as they are two-dimensional crystal defects like dislocations and because they interact with dislocations (acting as relatively impenetrable obstacles), technically, they can be included in undeformed wall immobile dislocation density. The equivalent dislocation density of these interfaces is relatively small enough that does not lead to a considerable error in the calculated initial yield stress, although they largely influence the initial plastic hardening. On the other hand, their equivalent density is not affected by plastic straining in cold and warm regime (constant during deformation). Since the interfaces are stable and strictly immobile in cold and warm regime, at the earliest stage of deformation they quickly transform to walls due to the misorientation (relative to undeformed configuration) they inflict, and hence are considered as sources of walls. Therefore, in the undeformed material state, (initial) wall immobile dislocation density depends on the effective grain size which includes the influence of all the interfaces (size effect). Furthermore, in some dual phase metals (e.g. ferritic-martensitic DP steels), due to phase transformation associated expansion during quenching, softer phase (ferrite) becomes plastically deformed by the harder phase (martensite) which results in formation of wall immobile dislocations (GNDs) in the softer phase. In such cases, initial wall immobile dislocation density has a relatively high value.

4) <u>*Viscoplastic decomposition*</u>: the mean yield shear stress at slip systems ($\tilde{\tau}_y$) has two major contributions that obeys the linear superposition rule (Kumar et al., 1968; Mecking and Kocks, 1981):

$$\tilde{\tau}_y = \tilde{\tau}_v + \tilde{\tau}_p \; ; \tag{11}$$

where $\tilde{\tau}_p$ is plastic/athermal/rate-independent/internal/back shear stress; and $\tilde{\tau}_v$ is referred to as viscous/rate-dependent/thermal/effective shear stress/drag, mean Peierls-Nabarro stress, or overstress which is the mean viscous lattice resistance to move mobile dislocations in a nearly obstacle/dislocation-free lattice (with relatively very low dislocation density) (Nabarro, 1997, 1952; Peierls, 1940). Viscous shear stress is affected by point defects such as vacancies, alloying elements and solute atoms (interstitial and substitutional). Additionally, plastic shear stress needs to be overcome to move the bowed-out mobile dislocations (Fig. 2). As mentioned in postulate (2), the lower the bow-out radius, the lower the required (extra) plastic shear stress to move it.

The viscoplastic decomposition can be depicted by the rheological model shown in Fig. 4, similar to Perzyna-type formulation (Perzyna, 1966), which consists of a parallel set of nonlinear dashpot/damper and nonlinear friction elements that are in series linkage with a linear spring element.

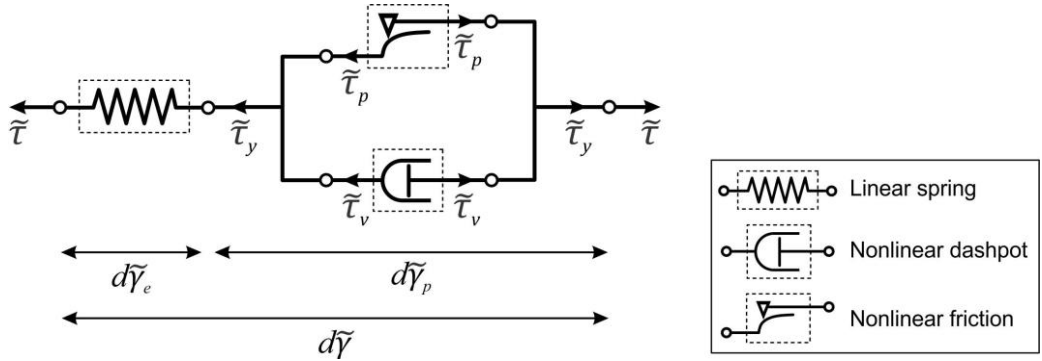

**Fig. 4.** Rheological model of polycrystal elasto-viscoplasticity; plastic flow occurs when the applied shear stress ($\tilde{\tau}$) becomes equal to the yield shear stress at slip systems ($\tilde{\tau} = \tilde{\tau}_y$).

Furthermore, given Eq. (4), Eq. (11) may take the following form:

$$\sigma_y = \sigma_v + \sigma_p \; ; \qquad \sigma_v \equiv \tilde{M}\tilde{\tau}_v \; ; \qquad \sigma_p \equiv \tilde{M}\tilde{\tau}_p \; . \tag{12}$$

5) <u>*Cell-wall decomposition*</u>: as emphasized in postulate (2), only immobile dislocations contribute to plastic hardening. The total surface density (total surface area per unit volume) of cells and walls are equal, as dislocation walls (GNBs) encompass the cell volumes (subgrains). In other words, the cell-wall dislocation substructure can be treated as a cellular composite. Hence, it is postulated that the two types of immobile dislocations contribute to plastic hardening in parallel (additive decomposition of plastic stress) in order to satisfy the compatibility for isochoric viscoplastic flow of material in which both cells and walls need to be deformed simultaneously at an equal rate. Thus, mean plastic shear stress is linearly decomposed to its constituent internal stresses corresponding to cell immobile dislocations (immobile SSDs) and wall immobile dislocations (immobile GNDs) (Columbus and Grujicic, 2002; Mughrabi, 1987; Voyiadjis and Al-Rub, 2005):

$$\tilde{\tau}_p = \tilde{\tau}_{pc} + \tilde{\tau}_{pw} \; ; \qquad (13)$$

where $\tilde{\tau}_{pc}$ and $\tilde{\tau}_{pw}$ are respectively contributions of cell and wall immobile dislocations to plastic shear stresses. $\tilde{\tau}_{pc}$ and $\tilde{\tau}_{pw}$ are axial and torsional plastic stress, respectively. In postulate (8), the relations of $\tilde{\tau}_{pc}$ and $\tilde{\tau}_{pw}$ with immobile dislocation densities are described. Considering Eq. (4), Eq. (13) is rewritten as follows:

$$\sigma_p = \sigma_{pc} + \sigma_{pw} \; ; \qquad \sigma_{pc} \equiv \tilde{M} \tilde{\tau}_{pc} \; ; \qquad \sigma_{pw} \equiv \tilde{M} \tilde{\tau}_{pw} \,. \qquad (14)$$

6) _Local dislocation density_: in saturation state, dislocation substructure and density is steady and statistically remain constant in a representative material volume. One can assume that the exact locations of walls (GNBs) inside RMV at saturation state are known prior to viscoplastic deformation. Hence, $V_w$ is assumed to be the summation of immediate (local) surrounding volumes of GNBs in the saturation state (including mobile dislocations constrained by GNBs constituent wall immobile dislocations). Cell immobile dislocations and the mobile dislocations bounded by them are envisaged to be uniformly distributed in the rest of RMV's volume ($V_c$). Therefore,

$$V = V_c + V_w \; ; \qquad V_c \cap V_w = \emptyset \,. \qquad (15)$$

Notice that $V$, $V_c$ and $V_w$ are virtual volumes. Moreover, $V$ (volume of RMV) is constant during deformation. In monotonic viscoplastic deformation, it is known that GNBs are relatively sharp boundaries containing locally dense wall immobile dislocations (postulate (3)). This indicates that $V_w \ll V_c$ and $V_w \ll V$. However, in materials under long and high-amplitude cyclic viscoplastic deformation, $V_w$ and $V_c$ become closer but still often $V_w < V_c$. Thus, in monotonic viscoplastic deformation which is the subject of this study, one can assume that cell and wall immobile dislocations during the entire deformation are homogenously locally distributed over constant volumes of $V_c$ and $V_w$, respectively. Therefore,

$$\rho_{xi}^{\text{loc}} \equiv \frac{1}{V_x} \sum_j l_{xi}^{(j)} \; ; \qquad x = c, w \,. \qquad (16)$$

According to Mughrabi (1983), constant cell volume fraction ($f_c$) and wall volume fraction ($f_w$) are defined as follows:

$$f_x \equiv \frac{V_x}{V} \; ; \qquad x = c, w \,. \qquad (17)$$

Combining Eqs (6), (16) and (17) results in (Mughrabi, 1983):

$$\rho_{xi}^{\text{loc}} = \frac{\rho_{xi}}{f_x} \; ; \qquad x = c, w \; ; \qquad (18)$$

$$f_c + f_w = 1 \,. \qquad (19)$$

From the view point of an observer who observes the macroscale material point shown in Fig. 1, viscoplastic deformation is homogenous. This leads to the assumption of nearly uniform local distribution of mobile dislocations which carry the plastic strain. Consequently, local and nonlocal mobile dislocation densities are almost equal ($\rho_{cm}^{\text{loc}} = \rho_{cm}$). Combining this with Eq. (18) and the fact that $f_x$ is constant gives:

$$\rho_{xy}^{\text{loc}} \propto \rho_{xy} \; ; \qquad \begin{cases} x = c, w, t \\ y = m, i, t \end{cases} . \qquad (20)$$

7) _Dislocations characteristic spacing_: dislocations intersect each other and form dislocation network with characteristic average spacing of $\tilde{\Lambda}$. It can be easily shown that the average spacing among dislocations (also known as mean free path of dislocations) of type $xy$ ($\tilde{\Lambda}_{xy}$) is inversely proportional to the square root of respective local dislocation density (Kocks, 1966; Nes, 1997; Seeger, 1955):

$$\widetilde{\Lambda}_{xy} \propto \frac{1}{\sqrt{\rho_{xy}^{\text{loc}}}} \propto \frac{1}{\sqrt{\rho_{xy}}} \; ; \quad \begin{cases} x = c, w, t \\ y = m, i, t \end{cases}. \tag{21}$$

It should be noted that $\widetilde{\Lambda}_{xy}$ has a local character. Additionally, it can be also readily shown that the volumetric number of junctions/intersection/nodes of dislocation type $xy$ or dislocation junction density of type $xy$ ($N_{xy}$) is proportional to dislocation density $\rho_{xy}^{\text{loc}}$ or $\rho_{xy}$ (Eq. (20)) while inversely proportional to the average spacing $\widetilde{\Lambda}_{xy}$ (Gottstein and Argon, 1987):

$$N_{xy} \propto \rho_{xy} \frac{1}{\overline{\Lambda}_{xy}^2} \propto \rho_{xy}^{3/2} \; ; \quad \begin{cases} x = c, w, t \\ y = m, i, t \end{cases}. \tag{22}$$

8) _Plastic stress_: as pointed out in postulate (2), the average bow-out radius of cell mobile dislocations that are constrained at both ends by immobile dislocations of type $x = c, w, t$ ($\bar{R}_{cm}^{xi}$) (Fig. 2) is proportional to average spacing of immobile dislocations of type $x$ (Nes, 1997):

$$\bar{R}_{cm}^{xi} \propto \widetilde{\Lambda}_{xi} \; ; \quad x = c, w, t. \tag{23}$$

On the other hand, the mean plastic resolved shear stress at slip systems is directly proportional to Burgers length (magnitude of Burgers vector) ($b$) and shear modulus ($G$) and inversely proportional to $\bar{R}_{cm}^{xi}$ (Gao et al., 1999; Hull and Bacon, 2011; Nabarro, 1952):

$$\tilde{\tau}_{px} \propto \frac{bG}{\bar{R}_{cm}^{xi}} \; ; \quad x = c, w. \tag{24}$$

where $\tilde{\tau}_{px}$ is the mean plastic resolved shear stress associated with immobile dislocations of type $x$. Given Eqs. (21) and (23), the Taylor relation (Bailey and Hirsch, 1960; Seeger et al., 1957; Taylor, 1934) is derived:

$$\tilde{\tau}_{px} = bG\tilde{\alpha}^{\text{loc}}\sqrt{\rho_{xi}^{\text{loc}}} \; ; \quad x = c, w. \tag{25}$$

where $\alpha^{\text{loc}}$ is a material constant known as local dislocation interaction strength/coefficient; and $\tilde{\alpha}^{\text{loc}}$ is statistical average of $\alpha^{\text{loc}}$ for different configurations of interacting mobile-immobile dislocations in various slip systems at the considered local point. Given Eq. (18):

$$\tilde{\tau}_{px} = bG\tilde{\alpha}_x\sqrt{\rho_{xi}} \; ; \quad \tilde{\alpha}_x = \frac{\tilde{\alpha}^{\text{loc}}}{\sqrt{f_x}} \; ; \quad x = c, w. \tag{26}$$

where $\tilde{\alpha}_x$ is the nonlocal interaction strength related to local density and geometrical arrangement of immobile dislocations of cell and wall species ($x = c, w$). Variation of $\tilde{\alpha}_x$ with plastic strain is assumed to be negligible (Kocks and Mecking, 2003). Since most of the plastic hardening is due to wall immobile dislocations, essentially $\tilde{\alpha}_w > \tilde{\alpha}_c$. Further, combining Eqs. (19) and (26) yields:

$$f_c = \frac{\tilde{\alpha}_w^2}{\tilde{\alpha}_c^2 + \tilde{\alpha}_w^2} \; ; \quad f_w = \frac{\tilde{\alpha}_c^2}{\tilde{\alpha}_c^2 + \tilde{\alpha}_w^2} \; ; \quad \tilde{\alpha}^{\text{loc}} = \frac{\tilde{\alpha}_c\tilde{\alpha}_w}{\sqrt{\tilde{\alpha}_c^2 + \tilde{\alpha}_w^2}}. \tag{27}$$

Finally, given Eq. (4), Eq. (26) may be rewritten as:

$$\sigma_{px} = MbG\tilde{\alpha}_x\sqrt{\rho_{xi}} \; ; \quad x = c, w. \tag{28}$$

9) _Kinetics superposition_: dynamic evolution of dislocation density is described as the rate of change of dislocation density with respect to time or plastic strain which is linear superposition of increase/production (with positive sign) and decrease/elimination (negative sign) terms (Johnston and Gilman, 1959; Kocks, 1976; Webster, 1966):

$$\dot{\rho}_{xy} = \dot{\rho}_{xy}^{+} - \dot{\rho}_{xy}^{-} \; ; \qquad \begin{cases} x = c, w, t \\ y = m, i, t \end{cases} . \tag{29}$$

Therefore,

$$\partial_{\bar{\gamma}_p} \rho_{xy} = \partial_{\bar{\gamma}_p} \rho_{xy}^{+} - \partial_{\bar{\gamma}_p} \rho_{xy}^{-} \; ; \qquad \partial_{\varepsilon_p} \rho_{xy} = \partial_{\varepsilon_p} \rho_{xy}^{+} - \partial_{\varepsilon_p} \rho_{xy}^{-} \; ; \qquad \begin{cases} x = c, w, t \\ y = m, i, t \end{cases} ; \tag{30}$$

where $\partial_{\bar{\gamma}_p} \equiv \frac{\partial}{\partial \bar{\gamma}_p}$ and $\partial_{\varepsilon_p} \equiv \frac{\partial}{\partial \varepsilon_p}$ are partial derivative operator with respect to mean shear plastic strain at slip systems and mean plastic strain, respectively. Static and overall (combination of static and dynamic) evolution of dislocation densities are represented by time derivative relations:

$$\left( \dot{\rho}_{xy} \right)_s = \left( \dot{\rho}_{xy}^{+} \right)_s - \left( \dot{\rho}_{xy}^{-} \right)_s \; ; \qquad \begin{cases} x = c, w, t \\ y = m, i, t \end{cases} ; \tag{31}$$

$$\left( \dot{\rho}_{xy} \right)_d = \dot{\bar{\gamma}}_p \, \partial_{\bar{\gamma}_p} \rho_{xy} = \dot{\varepsilon}_p \, \partial_{\varepsilon_p} \rho_{xy} \; ; \qquad \begin{cases} x = c, w, t \\ y = m, i, t \end{cases} ; \tag{32}$$

$$\dot{\rho}_{xy} = \left( \dot{\rho}_{xy} \right)_d + \left( \dot{\rho}_{xy} \right)_s = \dot{\varepsilon}_p \partial_{\varepsilon_p} \rho_{xy} + \left( \dot{\rho}_{xy} \right)_s \; ; \qquad \begin{cases} x = c, w, t \\ y = m, i, t \end{cases} ; \tag{33}$$

where subscripts $s$ and $d$ respectively represent static and dynamic states.

Furthermore, for convenience in calculations and dimensional balancing of equations, the normalized/dimensionless dislocation density of type $xy$ ($\hat{\rho}_{xy}$) is defined as follows:

$$\hat{\rho}_{xy} \equiv \frac{\rho_{xy}}{\rho_0} \; ; \qquad \begin{cases} x = c, w, t \\ y = m, i, t \end{cases} ; \tag{34}$$

where the hat-sign ($\hat{\;}$) indicates the normalization; and $\rho_0$ is constant reference dislocation density. Therefore, in all the previous relations that are presented so far in this paper (except Eq. (26) and (28)) $\rho_{xy}$ can be replaced by $\hat{\rho}_{xy}$. Thereby, $\hat{\bar{\Lambda}}_{xy}$ and $\hat{N}_{xy}$ are respectively normalized average spacing among dislocations of type $xy$ and normalized dislocation junction density of type $xy$.

10) _Dynamic dislocation processes_: there are six main classes of dynamic dislocation processes that are statistically considerable in constitutive modeling of metal viscoplasticity in cold and warm regimes:

- Generation/multiplication of mobile dislocations: Motion/displacement of mobile dislocations leads to their elongation.
- Annihilation of dislocations: dislocations of all three types, upon contact, can be annihilated by mobile dislocations with opposite Burgers vector slipping on neighboring parallel slip planes.
- Accumulation of immobile dislocations: mobile dislocations can be immobilized by immobile dislocations and produce accumulates of immobile dislocations.
- Trapping of mobile dislocations: mobile dislocations interacting each other can be immobilized by trapping process. Trapping process has two underlying mechanisms/reactions that result in immobilization of infected parts of mobile dislocations: locking of mobile dislocations; and pinning of mobile dislocations by interstitial solute/impurity atoms. Dynamic strain aging (DSA) effect in part is associated with pinning of mobile dislocations. At macroscopic level, pinning process reveals itself in the existence of enhanced upper initial yield stress which is followed by yield point elongation associated with Lüders bands (Hahn, 1962; Hall, 1970) in stress-strain curves. Pinning phenomena and their macroscopic effects are classified in Fig. 5.
- Nucleation of wall dislocations: cell immobile dislocations can become wall dislocations if the accumulates they belong to, while are being immobilized sufficiently strong, reach a critical size and local density.
- Remobilization/mobilization of immobile dislocations: immobile dislocations can be mobilized and contribute to plastic straining.

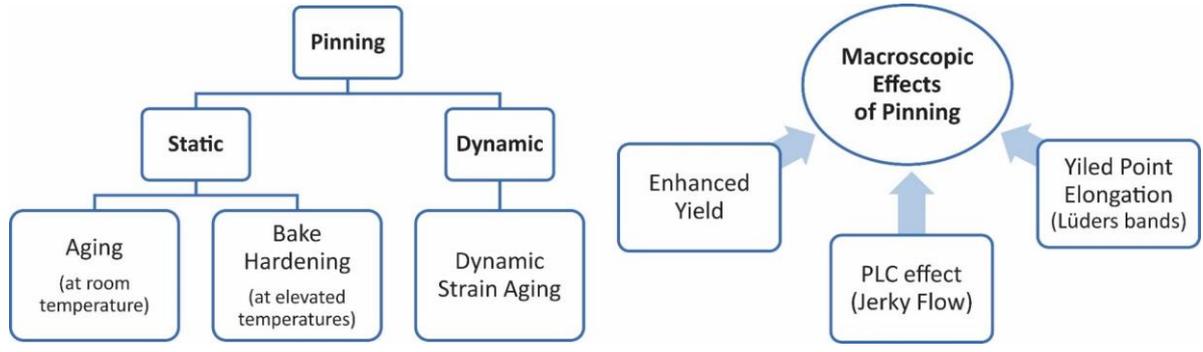

**Fig. 5.** Pinning phenomena and their macroscopic effects.

11) <u>*Static dislocation processes*</u>: there are two major static (at nonlocal $\dot{\bar{\gamma}}_p \approx 0$) dislocation processes during viscoplastic deformation of metals in cold and warm regimes (Pham et al., 2015):

   - Static pinning (Fig. 5) of cell dislocations by diffusion of interstitial solute atoms such as Carbon and Nitrogen to surroundings of dislocation cores and junctions (Cottrell cloud) which are statistically distributed (SSDs). This is due to higher stored elastic energy density of dislocation surroundings compared to defect-free regions of lattice (Cottrell and Bilby, 1949).
   - Static remobilization of immobile dislocations by means of thermal mechanisms such as junction dissociation and dislocation climb (vacancy diffusion).

   It should be noted that the aforementioned dynamic and static dislocation processes often consist of multiple individual underlying mechanisms. In the following sections, evolution of different types of dislocations by the above-listed dislocation processes is quantitatively and statistically described through several postulates.

### 3.2. Dynamic dislocation processes

12) <u>*Dynamic generation*</u>: motion of bowing mobile dislocation segments that are held at both ends by immobile dislocations generates more mobile dislocations by different mechanisms (Stricker et al., 2018) such as Frank-Read mechanism (Frank and Read, 1950), multiple cross-slip mechanism (Koehler, 1952), or mobile junctions mechanism (Stricker and Weygand, 2015; Weygand, 2014). Thus, the rate of dynamic generation rate of mobile dislocations in plastic shear increment ($\partial_{\bar{\gamma}_p} \rho_{cm}^{gn}$) is proportional to cell mobile dislocation density. The generation rate is higher where average bow-out radius of mobile dislocation segments bounded by immobile dislocations ($\bar{R}_{cm}^{ti}$) and the average length of mobile dislocation segments ($\bar{\Lambda}_{ti} \propto \bar{R}_{cm}^{ti}$) is larger (Nadgornyi, 1988; Steif and Clifton, 1979):

$$\partial_{\bar{\gamma}_p} \rho_{cm}^{gn} \propto \rho_{cm} \, \bar{R}_{cm}^{ti} \, . \tag{35}$$

Given Eqs. (4), (21), (23) and (34):

$$\partial_{\bar{\gamma}_p} \rho_{cm}^{gn} = p_{cm}^{gn} \, \rho_{cm} \, ; \qquad p_{cm}^{gn} = c_{cm}^{gn} \, \frac{1}{\sqrt{\hat{\rho}_{ti}}} \, ; \tag{36}$$

$$\partial_{\varepsilon_p} \hat{\rho}_{cm}^{gn} = M \, c_{cm}^{gn} \, \frac{\hat{\rho}_{cm}}{\sqrt{\hat{\rho}_{ti}}} \, ; \tag{37}$$

where $\partial_{\varepsilon_p} \hat{\rho}_{cm}^{gn}$ is normalized dynamic generation rate of cell mobile dislocation density with respect to plastic strain; $p_{cm}^{gn}$ is probability amplitude of dynamic generation of cell mobile dislocations; and $c_{cm}^{gn}$ is material coefficient associated with probability amplitude of dynamic generation of cell mobile dislocations. Frequency of occurrence of a dynamic dislocation process with probability amplitude of $p$ is equal to $\nu = p \, \dot{\bar{\gamma}}_p$. In addition, dynamic generation of mobile dislocations is an athermal dislocation process meaning that it is not directly dependent on the temperature.

13) <u>*Dynamic annihilation:*</u> the annihilation process takes place between a mobile dislocation and another dislocation that can be of any type. Thus, there are three sorts of dislocation annihilation: mutual annihilation of cell mobile dislocations, mutual annihilation of cell mobile and cell immobile dislocations, and mutual annihilation of cell mobile and wall immobile dislocations. Hence, the magnitude of dynamic annihilation rate of mobile dislocations and dislocations of type $xy$ with respect to plastic shear strain increment ($\partial_{\bar{\gamma}_p}\rho_{xy}^{an}$) is proportional to cell mobile dislocation density and the dislocation density of type $xy$ (Ananthakrishna and Sahoo, 1981):

$$\partial_{\bar{\gamma}_p}\rho_{xy}^{an} \propto \rho_{cm}\,\rho_{xy}\,; \quad xy = cm, ci, wi\,. \tag{38}$$

Given Eqs. (4) and (34):

$$\partial_{\bar{\gamma}_p}\rho_{xy}^{an} = p_{xy}^{an}\,\rho_{xy}\,; \quad p_{xy}^{an} = c_{xy}^{an}\,\hat{\rho}_{cm}\,; \quad xy = cm, ci, wi\,; \tag{39}$$

$$\partial_{\varepsilon_p}\hat{\rho}_{xy}^{an} = M\,c_{xy}^{an}\,\hat{\rho}_{cm}\,\hat{\rho}_{xy}\,; \quad xy = cm, ci, wi\,; \tag{40}$$

where $\partial_{\varepsilon_p}\hat{\rho}_{xy}^{an}$ is normalized dynamic annihilation rate of dislocation density of type $xy$ with respect to plastic strain; $p_{xy}^{an}$ is probability amplitude of dynamic annihilation of dislocation density of type $xy$; And $c_{xy}^{an}$ are material coefficients related to frequency of dynamic annihilation of dislocation density of type $xy$. Annihilation events take place between a mobile dislocation and another dislocation with opposite/antiparallel Burgers vector on parallel planes when two dislocations are within a critical distance from each other. Two near-screw dislocations of opposite sign, slipping on two neighboring parallel slip planes can annihilate by cross slip of one of them (Brown, 2002; Essmann and Mughrabi, 1979; Nix et al., 1985; Oren et al., 2017; Pauš et al., 2013; Püschl, 2002; Seeger et al., 1957). Two near-edge dislocations may annihilate as well by spontaneous formation and disintegration of a very narrow unstable dislocation dipole which can be assisted by dislocation climb. (Eisenlohr and Blum, 2005; Monavari and Zaiser, 2018; Vegge and Jacobsen, 2002). Thus, thermally activated dislocation climb and cross-slip mechanisms facilitate dislocation annihilation. Therefore, dynamic annihilation of dislocations are thermal dislocation processes meaning that probability amplitudes associated with dislocation annihilation processes have temperature dependence (increase with increasing temperature). On the other hand, since at least dislocation climb as a contributing mechanism for dynamic annihilation processes is controlled by time-dependent diffusion of vacancies, annihilation processes have strain rate dependence with negative strain rate sensitivity. Besides, dislocation cross-slip can be as well considered a rate-controlled mechanism (Nes, 1997).

Thus, one might suggest that when a mobile dislocation and another dislocation (of type $xy$) are on the same slip system within the critical distance (some authors call it capture radius) for annihilation of dislocations of type $xy$ ($d_{xy}^{an}$), they will likely annihilate each other, in case of having opposite Burgers vectors (Essmann and Mughrabi, 1979). In light of this, for instance, the dimensionless material parameter associated with frequency of dynamic annihilation among cell mobile dislocations ($c_{cm}^{an}$) can be decomposed as $c_{cm}^{an} = \frac{d_{cm}^{an}}{\bar{n}b}$, where $\bar{n}$ is the average number of active slip systems ($\bar{n} \geq 3$). Statistically, one can assume equal density of dislocations on all active slip systems (in isotropic case) which gives rise to normalization factor $\frac{1}{\bar{n}}$ (Roters et al., 2000).

14) *Dynamic accumulation:* immobilization of mobile dislocations by immobile dislocations produces or increases the size of immobile dislocation accumulates. This dislocation process is known as dislocation accumulation; and its magnitude with respect to plastic shear strain increment ($\partial_{\bar{\gamma}_p}\rho_{xi}^{ac}$) is proportional to cell mobile dislocation density. It is also inversely proportional to average spacing among immobile dislocations of type $x = c, w$ ($\bar{\Lambda}_{xi}$) (Kocks, 1976) as well as the average radius of mobile dislocation segments bounded by immobile dislocations of type $x = c, w$ ($\bar{R}_{cm}^{xi} \propto \bar{\Lambda}_{xi}$):

$$\partial_{\bar{\gamma}_p}\rho_{xi}^{ac} \propto \rho_{cm}\,\frac{1}{\bar{\Lambda}_{xi}}\,; \quad x = c, w\,. \tag{41}$$

Having Eqs. (4), (21) and (34):

$$\partial_{\bar{\gamma}_p}\rho_{xi}^{ac} = p_{xi}^{ac}\,\rho_{cm}\,; \quad p_{xi}^{ac} = c_{xi}^{ac}\,\sqrt{\hat{\rho}_{xi}}\,; \quad x = c, w\,; \tag{42}$$

$$\partial_{\varepsilon_p}\hat{\rho}_{xi}^{ac} = M\,c_{xi}^{ac}\,\sqrt{\hat{\rho}_{xi}}\,\hat{\rho}_{cm}\,; \quad x = c, w\,; \tag{43}$$

where $\partial_{\varepsilon_p}\hat{\rho}_{xi}^{\mathrm{ac}}$ is normalized dynamic accumulation rate of immobile dislocations of type $x$ with respect to plastic strain; $p_{xi}^{\mathrm{ac}}$ is probability amplitude of dynamic accumulation of immobile dislocations of type $x$; and $c_{xi}^{\mathrm{ac}}$ are material parameters associated with probability of dynamic accumulation of immobile dislocations of type $x$. Dynamic accumulation of immobile dislocations is an athermal process.

15) _Dynamic trapping:_ interaction between mobile dislocations moving in different intersecting slip systems results in the formation of dislocation junctions/intersections/nodes. These dislocation junctions can be lock junctions such as Lomer-Cottrell lock junctions (Cottrell, 1953; Lomer, 1951), Hirth lock junctions (Hirth, 1961), collinear (Madec et al., 2003) and coplanar lock junctions (Thompson, 1953). Lock junctions restrict movement of involved mobile dislocations by immobilizing parts of them near the formed lock junction, leading to latent hardening (Franciosi, 1985). In addition, some junctions formed due to interaction of mobile dislocation pairs are not lock junctions (depending on the configuration of respective slip systems and mobile dislocations). They can as well be mobile/glissile or temporal/temporary junctions which do not lead to immediate immobilization directly.

Essentially, dislocation junctions due to their relative high energy density are attractive regions for diffusion and accumulation of interstitial solute atoms. Therefore, some of the mobile and temporal junctions become pinned/arrested/anchored due to diffusion and accumulation of interstitial solutes at their vicinity; which leads to subsequent immobilization of respective mobile dislocations or infected parts of them (Cottrell and Bilby, 1949; Mulford and Kocks, 1979; van den Beukel and Kocks, 1982). Density of potential junctions of interacting mobile dislocations, as mentioned in postulate (7), is proportional to density of mobile dislocations ($\rho_{cm}$) and inversely proportional to the average spacing among mobile dislocations ($\bar{\Lambda}_{cm}$). Thus, frequency of occurrence of mobile dislocation trapping events comprised of two major underlying mechanisms (locking and pinning), must be proportional to junction density of cell mobile dislocations ($N_{cm}$). Henceforth,

$$\partial_{\bar{\gamma}_p}\rho_{cm}^{\mathrm{tr}} \propto N_{cm} \,. \tag{44}$$

Given Eqs. (4), (22) and (34):

$$\partial_{\bar{\gamma}_p}\rho_{cm}^{\mathrm{tr}} = p_{cm}^{\mathrm{tr}}\,\rho_{cm} \,; \qquad p_{cm}^{\mathrm{tr}} = c_{cm}^{\mathrm{tr}}\,\sqrt{\hat{\rho}_{cm}} \,; \tag{45}$$

$$\partial_{\varepsilon_p}\hat{\rho}_{cm}^{\mathrm{tr}} = M\,c_{cm}^{\mathrm{tr}}\,\hat{\rho}_{cm}^{3/2} \,; \tag{46}$$

where superscript tr denotes the trapping process; $\partial_{\varepsilon_p}\hat{\rho}_{cm}^{\mathrm{tr}}$ is normalized dynamic trapping rate of cell mobile dislocations with respect to plastic strain; $p_{cm}^{\mathrm{tr}}$ is probability amplitude of dynamic trapping of cell mobile dislocations; and $c_{cm}^{\mathrm{tr}}$ is a material coefficient associated with probability amplitude of dynamic trapping of cell mobile dislocations. Since the pinning mechanism of trapping process is a diffusion controlled reaction, dynamic trapping of cell mobile dislocations is a thermal dislocation process that its magnitude increases with increasing temperature. In other words, by increasing temperature, particularly in metals with notable interstitial concentration, the rate of immobilization of dislocation junctions (trapping) becomes higher. Additionally, since pinning occurs due to diffusion of interstitial solute atoms to dislocation cores and junction, $c_{cm}^{\mathrm{tr}}$ is a function of concentration of interstitial solutes atoms such as Carbon and Nitrogen.

Moreover, one might argue that $c_{cm}^{\mathrm{tr}}$ has a negative strain rate sensitivity because the pinning process is associated with time-dependent diffusion of interstitial solute atoms. The DSA effect is observed when the interstitial atoms reorientation time is shorter than the waiting time of temporal and mobile junctions at the locations nearby interstitial solute complex (Kubin et al., 1988; Kubin and Estrin, 1990). Hence, there exist a critical strain rate beyond which the intensity of mobile dislocation pinning abruptly drops. That strain rate belongs to the dynamic regime.

16) _Dynamic wall nucleation_: accumulates of cell immobile dislocations that are immobile sufficiently strong (meaning that their remobilization requires higher stress than mean yield resolved shear stress) will grow and become denser by immobilizing more mobile dislocations at their vicinity (cell immobile accumulation process). These cell immobile pile-ups that reach a critical size and local density, subsequently become adequately stable to produce sufficient local stress concentration which results in necessary lattice curvature in their vicinity to make them part of walls. This process is called wall nucleation. From this point of view, wall nucleation process converts some cell immobile to wall immobile dislocations. As such, its rate is proportional to cell immobile dislocation density. On the other hand, the accumulation of cell immobile dislocations increases the probability of wall nucleation process. Hence, statistical magnitude of dynamic nucleation rate of wall immobile dislocations with respect to plastic shear strain increment ($\partial_{\bar{\gamma}_p}\rho_{wi}^{nc}$) is proportional to cell immobile dislocation density ($\rho_{ci}$) and the rate of cell immobile dislocation accumulation ($\partial_{\bar{\gamma}_p}\rho_{ci}^{ac}$):

$$\partial_{\bar{\gamma}_p}\rho_{wi}^{nc} \propto \rho_{ci}\,\partial_{\bar{\gamma}_p}\rho_{ci}^{ac}. \tag{47}$$

Given Eqs. (4), (21), (34) and (41):

$$\partial_{\bar{\gamma}_p}\rho_{wi}^{nc} = p_{wi}^{nc}\,\rho_{ci}\,; \qquad p_{wi}^{nc} = c_{wi}^{nc}\,\sqrt{\hat{\rho}_{ci}}\,\hat{\rho}_{cm}\,; \tag{48}$$

$$\partial_{\varepsilon_p}\hat{\rho}_{wi}^{nc} = M\,c_{wi}^{nc}\,\hat{\rho}_{ci}^{3/2}\,\hat{\rho}_{cm}\,; \tag{49}$$

where $\partial_{\varepsilon_p}\hat{\rho}_{wi}^{nc}$ is normalized dynamic nucleation rate of wall immobile dislocations with respect to plastic strain; $p_{wi}^{nc}$ is probability amplitude of dynamic nucleation of wall immobile dislocations; and $c_{wi}^{nc}$ is a material coefficient associated with probability amplitude of dynamic nucleation of wall immobile dislocations. Pinning of cell immobile dislocations at their junctions by solute atoms contributes to stronger immobilization of cell immobile accumulates and consequently enhances the stability of the respective IDBs which are about to transform to GNBs. This is also suggested by existence of the factor $\hat{\rho}_{ci}^{3/2}$ in Eq. (49) which is proportional to junction density of cell immobile dislocations ($N_{ci} \propto \rho_{ci}^{3/2}$), considering that the pinning process occurs mainly at dislocation junctions. From another perspective, given Eq. (26), presence of the factor $\sqrt{\hat{\rho}_{ci}}$ in Eq. (48) is proportional to the average plastic shear stress concentration at IDBs.

With this approach, twin formation can be considered as a wall nucleation mechanism, in case of slip-dominated metals. Regular dislocation walls are formed by high local stress concentration that causes local elastic bending of crystal in metals with relatively high stacking fault energy (SFE). Analogously, twins are also boundaries of lattice misorientation (GNBs) that plastically nucleate due to local stress concentration at sufficiently stable dislocation pile-ups (IDBs) (Christian and Mahajan, 1995; Kibey et al., 2007; Venables, 1964) in metals with adequately low SFE. Each stacking fault that nucleates a twin is bounded by partial dislocations such as Shockley and Frank partial dislocations (Hirth and Lothe, 1982; Hull and Bacon, 2011) that are geometrically necessary (Mahajan and Chin, 1973) and hence are treated as wall dislocations. Therefore, $c_{wi}^{nc}$ also reflects frequency of twin nucleation which in turn depends on the mean SFE at the nonlocal material point under consideration. Twin forming dislocations must overcome the intrinsic SFE barrier to nucleate twin's associated stacking fault. Further, most of the energy required for twinning is spent on creating its associated stacking fault. Thus, especially in relatively low SFE metals such as high manganese TWIP steels, $c_{wi}^{nc}$ has a relatively high value. It is emphasized that pinning of cell immobile dislocations facilitates twin nucleation by increasing stress concentration in the cells to reach the critical shear stress for dislocation separation/dissociation/splitting mechanism (Byun, 2003; Koyama et al., 2015). In slip dominated plastic deformation in which twinning occurs, the share of TWIP effect in total accumulated plastic strain is negligible compared to that of slip. However, as already highlighted, twinning has a remarkable influence on plastic hardening by facilitating the formation of dislocation walls.

As mentioned before, pinning is a diffusion-controlled mechanism that its magnitude increases by increasing temperature in metals with adequate interstitial solute content. Moreover, the rate of stacking fault formation due to dislocation separation mechanism depends on SFE which in turn is generally an increasing function of temperature. In view of this, dynamic nucleation of wall immobile dislocations is a thermal dislocation process that its statistical magnitude by increasing temperature might increase or decrease depending on the metal structure and composition.

Further, since in nucleation of wall immobile dislocations, time-dependent diffusion-controlled pinning mechanism is partly involved, one might argue that $c_{wi}^{nc}$ has negative strain rate sensitivity. On the other hand, wall and twin nucleation can be amplified by increasing strain rate provoking positive strain rate sensitivity of $c_{wi}^{nc}$. Therefore, overall strain rate sensitivity of $c_{wi}^{nc}$ can be negative, zero or positive (Yang et al., 2017).

17) *Dynamic remobilization*: probability of recurrence of remobilization of cell or wall immobile dislocations depends on their density (Bergström, 1970):

$$\partial_{\bar{\gamma}_p}\rho_{xi}^{rm} \propto \rho_{xi}\,; \qquad x = c, w\,; \tag{50}$$

where $\partial_{\bar{\gamma}_p}\rho_{xi}^{\mathrm{rm}}$ is the remobilization rate of immobile dislocations of type $x$ with respect to plastic shear strain increment. Given Eqs. (4) and (34):

$$\partial_{\bar{\gamma}_p}\rho_{xi}^{\mathrm{rm}} = p_{xi}^{\mathrm{rm}}\,\rho_{xi}\;;\qquad p_{xi}^{\mathrm{rm}} = c_{xi}^{\mathrm{rm}}\;;\qquad x = c, w\;; \tag{51}$$

$$\partial_{\varepsilon_p}\hat{\rho}_{xi}^{\mathrm{rm}} = M\,c_{xi}^{\mathrm{rm}}\,\hat{\rho}_{xi}\;;\qquad x = c, w\,. \tag{52}$$

where $\partial_{\varepsilon_p}\hat{\rho}_{xi}^{\mathrm{rm}}$ is normalized dynamic remobilization rate of immobile dislocations of type $x$ with respect to plastic strain; $p_{wi}^{\mathrm{rm}}$ is probability amplitude of dynamic remobilization of immobile dislocations of type $x$; and $c_{xi}^{\mathrm{rm}}$ are material parameters associated with probability amplitude of dynamic remobilization of immobile dislocations of type $x$. Dislocation remobilization process is performed by different underlying mechanisms such as climb, cross-slip, bow-out, junction dissociation/unzipping, etc. (Hunter and Preston, 2015). Owing to the involvement of thermal mechanisms in the remobilization processes, such as dislocation climb and cross-slip, they are treated as thermal dislocation processes. At low and medium temperatures, dynamic dislocation remobilization and annihilation processes (together known as dynamic recovery) are mainly governed by the cross-slip mechanism while at high temperatures dislocation climb is the dominant mechanism of dynamic recovery processes (Essmann and Mughrabi, 1979; Galindo-Nava et al., 2012; Kubin et al., 1992; Nix et al., 1985; Püschl, 2002; Rivera-Díaz-del-Castillo and Huang, 2012). Essentially, by definition inherently $\partial_{\varepsilon_p}\hat{\rho}_{wi}^{\mathrm{rm}}$ should be much lower than $\partial_{\varepsilon_p}\hat{\rho}_{ci}^{\mathrm{rm}}$ ($\partial_{\varepsilon_p}\hat{\rho}_{wi}^{\mathrm{rm}} \ll \partial_{\varepsilon_p}\hat{\rho}_{ci}^{\mathrm{rm}}$).

Finally, since dislocation climb as an underlying mechanism for dynamic remobilization processes is controlled by time-dependent diffusion of vacancies, remobilization processes might have strain rate dependence with negative strain rate sensitivity. On the other hand, by increasing strain rates, the viscous stress is also increased which assists the remobilization process. This provokes positive strain rate sensitivity of remobilization parameters. Therefore, $c_{xi}^{\mathrm{rm}}$ might have negative, zero or positive strain rate sensitivities.

By approaches analogous to the one adopted in postulate (13) (dynamic annihilation), one may also physically interpret the constitutive parameters associated with frequency of other dislocation processes ($c_{xy}^z$) by defining various mean critical distances, local densities, shear stresses, etc. Nevertheless, the mean critical physical parameters have virtual characters (like probability amplitude), as they are extremely difficult and mostly even impossible to be determined accurately enough by means of independent experimental measurements. Now that every main dynamic dislocation process is statistically and quantitatively characterized, the approach first introduced by Ananthakrishna and Sahoo (1981), and Estrin and Kubin (1986) is applied to construct the overall kinetics equations for dynamic evolution of each dislocation type.

18) _Kinetics of wall immobile dislocations_: dynamic nucleation and accumulation of wall immobile dislocations, contribute to dynamic increase of wall immobile dislocation density while dynamic decrease of wall immobile dislocations occurs as the consequence of annihilation and remobilization of wall immobile dislocations:

$$\partial_{\varepsilon_p}\hat{\rho}_{wi}^+ = \partial_{\varepsilon_p}\hat{\rho}_{wi}^{\mathrm{nc}} + \partial_{\varepsilon_p}\hat{\rho}_{wi}^{\mathrm{ac}}\;; \tag{53}$$

$$\partial_{\varepsilon_p}\hat{\rho}_{wi}^- = \partial_{\varepsilon_p}\hat{\rho}_{wi}^{\mathrm{an}} + \partial_{\varepsilon_p}\hat{\rho}_{wi}^{\mathrm{rm}}\,. \tag{54}$$

Combining Eqs. (30), (53), and (54) yields:

$$\partial_{\varepsilon_p}\hat{\rho}_{wi} = \partial_{\varepsilon_p}\hat{\rho}_{wi}^{\mathrm{nc}} + \partial_{\varepsilon_p}\hat{\rho}_{wi}^{\mathrm{ac}} - \left(\partial_{\varepsilon_p}\hat{\rho}_{wi}^{\mathrm{an}} + \partial_{\varepsilon_p}\hat{\rho}_{wi}^{\mathrm{rm}}\right)\,. \tag{55}$$

Eq. (55) is the dynamic evolutionary/kinetics equation of wall immobile dislocation density.

19) _Kinetics of cell immobile dislocations_: dynamic increase of cell immobile dislocation density is the result of dynamic trapping of cell mobile dislocations and accumulation of cell immobile dislocations, whereas dynamic decrease of cell immobile dislocations takes place by means of annihilation and remobilization of cell immobile dislocations and also nucleation of wall immobile dislocations:

$$\partial_{\varepsilon_p}\hat{\rho}_{ci}^+ = \partial_{\varepsilon_p}\hat{\rho}_{cm}^{\mathrm{tr}} + \partial_{\varepsilon_p}\hat{\rho}_{ci}^{\mathrm{ac}}\;; \tag{56}$$

$$\partial_{\varepsilon_p}\hat{\rho}_{ci}^- = \partial_{\varepsilon_p}\hat{\rho}_{ci}^{\mathrm{an}} + \partial_{\varepsilon_p}\hat{\rho}_{ci}^{\mathrm{rm}} + \partial_{\varepsilon_p}\hat{\rho}_{wi}^{\mathrm{nc}}. \tag{57}$$

Combining Eqs. (30), (56), and (57) reads:

$$\partial_{\varepsilon_p}\hat{\rho}_{ci} = \partial_{\varepsilon_p}\hat{\rho}_{cm}^{\mathrm{tr}} + \partial_{\varepsilon_p}\hat{\rho}_{ci}^{\mathrm{ac}} - \left(\partial_{\varepsilon_p}\hat{\rho}_{ci}^{\mathrm{an}} + \partial_{\varepsilon_p}\hat{\rho}_{ci}^{\mathrm{rm}} + \partial_{\varepsilon_p}\hat{\rho}_{wi}^{\mathrm{nc}}\right). \tag{58}$$

Eq. (58) is the dynamic evolutionary equation of cell immobile dislocation density.

20) _Kinetics of cell mobile dislocations_: dynamic generation of cell mobile dislocations and remobilization of cell and wall immobile dislocations contribute to dynamic increase of cell mobile dislocations, while dynamic decrease of cell mobile dislocations occurs through dynamic annihilation of cell mobile, cell immobile, and wall immobile dislocations, accumulation of cell and wall immobile dislocations, and pinning of mobile dislocations:

$$\partial_{\varepsilon_p}\hat{\rho}_{cm}^+ = \partial_{\varepsilon_p}\hat{\rho}_{ci}^{\mathrm{gn}} + \partial_{\varepsilon_p}\hat{\rho}_{ci}^{\mathrm{rm}} + \partial_{\varepsilon_p}\hat{\rho}_{wi}^{\mathrm{rm}}; \tag{59}$$

$$\partial_{\varepsilon_p}\hat{\rho}_{cm}^- = 2\,\partial_{\varepsilon_p}\hat{\rho}_{cm}^{\mathrm{an}} + \partial_{\varepsilon_p}\hat{\rho}_{ci}^{\mathrm{an}} + \partial_{\varepsilon_p}\hat{\rho}_{wi}^{\mathrm{an}} + \partial_{\varepsilon_p}\hat{\rho}_{ci}^{\mathrm{ac}} + \partial_{\varepsilon_p}\hat{\rho}_{wi}^{\mathrm{ac}} + \partial_{\varepsilon_p}\hat{\rho}_{cm}^{\mathrm{tr}}. \tag{60}$$

The term $\partial_{\varepsilon_p}\hat{\rho}_{cm}^{\mathrm{an}}$ is considered twice in dynamic decrease of cell mobile dislocations because two mobile dislocations annihilate each other in the process of dynamic annihilation of cell mobile dislocations. Combining (30), (59), and (60) gives:

$$\partial_{\varepsilon_p}\hat{\rho}_{cm} = \partial_{\varepsilon_p}\hat{\rho}_{cm}^{\mathrm{gn}} + \partial_{\varepsilon_p}\hat{\rho}_{ci}^{\mathrm{rm}} + \partial_{\varepsilon_p}\hat{\rho}_{wi}^{\mathrm{rm}}$$
$$- \left(2\,\partial_{\varepsilon_p}\hat{\rho}_{cm}^{\mathrm{an}} + \partial_{\varepsilon_p}\hat{\rho}_{ci}^{\mathrm{an}} + \partial_{\varepsilon_p}\hat{\rho}_{wi}^{\mathrm{an}} + \partial_{\varepsilon_p}\hat{\rho}_{ci}^{\mathrm{ac}} + \partial_{\varepsilon_p}\hat{\rho}_{wi}^{\mathrm{ac}} + \partial_{\varepsilon_p}\hat{\rho}_{cm}^{\mathrm{tr}}\right). \tag{61}$$

Eq. (61) is the dynamic evolutionary equation of cell mobile dislocation density. Considering Eqs. (10) and (30), dynamic evolution of total dislocation density reads:

$$\partial_{\varepsilon_p}\hat{\rho}_t^+ = \partial_{\varepsilon_p}\hat{\rho}_{cm}^{\mathrm{gn}}; \tag{62}$$

$$\partial_{\varepsilon_p}\hat{\rho}_t^- = 2\left(\partial_{\varepsilon_p}\hat{\rho}_{cm}^{\mathrm{an}} + \partial_{\varepsilon_p}\hat{\rho}_{ci}^{\mathrm{an}} + \partial_{\varepsilon_p}\hat{\rho}_{wi}^{\mathrm{an}}\right); \tag{63}$$

$$\partial_{\varepsilon_p}\hat{\rho}_t = \partial_{\varepsilon_p}\hat{\rho}_{cm}^{\mathrm{gn}} - 2\left(\partial_{\varepsilon_p}\hat{\rho}_{cm}^{\mathrm{an}} + \partial_{\varepsilon_p}\hat{\rho}_{ci}^{\mathrm{an}} + \partial_{\varepsilon_p}\hat{\rho}_{wi}^{\mathrm{an}}\right). \tag{64}$$

Eq. (64) is the dynamic evolutionary equation of total dislocation density.

### 3.3. Static dislocation processes

21) _Static pinning_: nonlocal static pinning of cell (mobile and immobile) dislocations is very similar to dynamic pinning (postulate (15) and (16)) with the difference that it occurs at very low local strain rates in relatively long durations. Time rate of static pinning of cell dislocations ($\dot{\rho}_{cy}^{\mathrm{spn}}$) by diffusion of solute atoms at their junctions is proportional to their junction density ($N_{cy}$):

$$\dot{\rho}_{cy}^{\mathrm{spn}} \propto N_{cy}; \quad y = m, i. \tag{65}$$

Given Eqs. (22) and (34):

$$\dot{\rho}_{cy}^{\mathrm{spn}} = c_{cy}^{\mathrm{spn}}\rho_{cy}^{3/2}; \quad y = m, i; \tag{66}$$

where $c_{cy}^{\mathrm{spn}}$ is a material coefficient associated with frequency of static pinning of cell dislocations of type $y = m, i$. Since pinning is a diffusion-controlled process, static pinning of cell dislocations is a thermal dislocation process that its magnitude increases with increasing temperature. As mentioned in postulate (15), since pinning process occurs due to diffusion of interstitial solute atoms to dislocation surrounding, $c_{cy}^{spn}$ is a function of interstitial solute content.

22) _Static remobilization_: nonlocal static remobilization process consists of the same contributing mechanisms of its dynamic counterpart. Time rate of static remobilization of immobile dislocations of type $x = c, w$ ($\dot{\rho}_{xi}^{srm}$) is proportional to density of them:

$$\dot{\rho}_{xi}^{srm} \propto \rho_{xi} ; \quad x = c, w . \tag{67}$$

Given Eqs. (22) and (34):

$$\dot{\hat{\rho}}_{xi}^{srm} = c_{xi}^{srm} \hat{\rho}_{xi} ; \quad x = c, w ; \tag{68}$$

where $c_{xi}^{srm}$ is a material parameter associated with frequency of static remobilization of immobile dislocations of type $x = c, w$. Moreover, static remobilization is a thermal dislocation process due to thermal character of its underlying mechanisms such as dislocation climb and junction dissociation.

23) _Static kinetics_: static pinning of cell mobile dislocations reduces cell mobile and increases cell immobile dislocation density, while static remobilization processes of immobile dislocations increase cell mobile and decrease immobile dislocation density. Thus, given Eq. (31), static evolution of dislocation densities reads:

$$\left(\dot{\hat{\rho}}_{cm}\right)_s = \dot{\hat{\rho}}_{ci}^{srm} + \dot{\hat{\rho}}_{wi}^{srm} - \dot{\hat{\rho}}_{cm}^{spn} ; \tag{69}$$

$$\left(\dot{\hat{\rho}}_{ci}\right)_s = \dot{\hat{\rho}}_{cm}^{spn} - \dot{\hat{\rho}}_{ci}^{srm} - \dot{\hat{\rho}}_{ci}^{spn} ; \tag{70}$$

$$\left(\dot{\hat{\rho}}_{wi}\right)_s = \dot{\hat{\rho}}_{ci}^{spn} - \dot{\hat{\rho}}_{wi}^{srm} . \tag{71}$$

Notice that pinning of already immobilized cell dislocations (cell immobile dislocations), strengthens their immobilization (resistance to remobilization). Consequently, upon application of external stress, just prior to local yielding, some of those pinned cell immobile dislocations will convert to GND due to elastic bending of their surrounding lattice to a sufficient mean misorientation angle.

### 3.4. Temperature and strain rate dependencies

24) _Temperature dependence of dislocation processes_: as pointed out earlier, each thermal dislocation process comprised of at least one underlying thermally-activated mechanism. Temperature is a statistical variable (representative of mean amplitude of atomic fluctuations) as well, which directly affects probability amplitude of different thermal dislocation processes. It is postulated that the change in probability amplitude of a thermal dislocation process with respect to temperature in constant strain rate is proportional to a power of temperature change:

$$\Delta p_{xy}^z \propto (\Delta T)^{s_{xy}^z} ; \quad xy = cm, ci, wi . \tag{72}$$

Therefore,

$$\frac{\Delta c_{xy}^z}{c_{xy0}^z} = r_{xy}^z \left(\frac{\Delta T}{T_0}\right)^{s_{xy}^z} ; \quad \Delta T = T - T_0 ; \quad \Delta c_{xy}^z = c_{xy}^z - c_{xy0}^z ; \quad xy = cm, ci, wi ; \tag{73}$$

where $T$ is absolute temperature; $T_0$ is reference absolute temperature; $p_{xy}^z$ is probability amplitude associated with thermal dislocation process $z$ that involves dislocations of type $xy$; $c_{xy}^z$ and $c_{xy0}^z$ are respectively, current (at current temperature) and reference (at reference temperature and strain rate) material parameters associated with probability amplitude of dislocation process $z$ of dislocations of type $xy$; $r_{xy}^z$ and $s_{xy}^z$ are respectively temperature sensitivity coefficient and exponent associated with dislocation process $z$ of dislocations of type $xy$. Reference temperature is assumed to be the minimum temperature in the temperature regime under consideration. Thus, in case of cold and warm regimes, reference temperature is the room temperature. Likewise, the reference strain rate is assumed to be the lowest strain rate in the investigated strain rate regime. The reference strain rate must be lower than the maximum/critical strain rate beyond which the viscoplastic deformation cannot be considered isothermal anymore because of adiabatic heat generation.

Referring to the postulates of dynamic and static dislocation processes (sections 3.2 and 3.3), thermal dislocation processes are dynamic annihilation of dislocations, dynamic trapping of cell mobile dislocation, dynamic nucleation of wall immobile dislocations, dynamic remobilization of immobile dislocation, static pinning of cell mobile dislocations, and static remobilization of immobile dislocations:

$$r_{xy}^z \begin{cases} > 0: & z = \text{an}, \text{tr}, \text{rm}, spn, srm \\ \gtreqless 0: & z = \text{nc} \\ = 0: & z = \text{gn}, \text{ac} \end{cases} \quad ; \quad s_{xy}^z \begin{cases} > 0: & z = \text{an}, \text{tr}, \text{nc}, \text{rm}, spn, srm \\ = 0: & z = \text{gn}, \text{ac} \end{cases} . \quad (74)$$

Temperature dependencies of frequency of different thermal dislocation processes characterized via Eq. (73) are the general monotonically increasing or decreasing functions of temperature ($\partial c/\partial T \geq 0$ or $\partial c/\partial T \leq 0$) with unchanging concavity ($\partial^2 c/\partial T^2 \geq 0$ or $\partial^2 c/\partial T^2 \leq 0$) throughout the entire temperature domain (in cold and warm regimes). The temperature dependence of dislocation mobility (stress dependence of velocity of mobile dislocations) and underlying mechanisms of thermal dislocation processes, e.g. dislocation climb (speed), are often much more complex than that can be described purely by Arhenius relation and the respective activation energy barriers (Amodeo and Ghoniem, 1990; Argon and Moffatt, 1981; Blum et al., 2002; Eisenlohr and Blum, 2005; Gu et al., 2015; Hirth and Lothe, 1982; Yuan et al., 2018). In addition, the temperature dependence of some of the involved thermal mechanisms has not been properly understood yet. Therefore, it is reasonable to assume a phenomenological power-law description for temperature dependence of constitutive parameters associated with probability amplitude of different thermal dislocation processes.

25) *Strain-rate dependence of dislocation processes*: the strain rate dependence of probability amplitude of rate-dependent dynamic dislocation processes is suggested to be described by power-law relation as:

$$p_{xy}^z \propto \hat{\dot{\gamma}}_p^{m_{xy}^z} ; \quad xy = cm, ci, wi ; \quad (75)$$

where $m_{xy}^z$ is strain rate sensitivity associated with dislocation process $z$ of dislocations of type $xy$. Therefore, at reference temperature:

$$\hat{c}_{xy}^z = \hat{\dot{\gamma}}_p^{m_{xy}^z}; \quad \hat{\dot{\gamma}}_p \equiv \frac{\dot{\gamma}_p}{\dot{\gamma}_0} ; \quad \hat{c}_{xy}^z \equiv \frac{c_{xy}^z}{c_{xy0}^z} ; \quad xy = cm, ci, wi ; \quad (76)$$

where $\dot{\gamma}_0$ is reference shear strain rate; and $\hat{\dot{\gamma}}_p$ is normalized mean plastic shear strain rate. Combination of Eqs. (73) and (76) produce:

$$\hat{c}_{xy}^z = \left[ 1 + r_{xy}^z \left( \hat{T} - 1 \right)^{s_{xy}^z} \right] \hat{\dot{\gamma}}_p^{m_{xy}^z} ; \quad \hat{T} \equiv \frac{T}{T_0} ; \quad xy = cm, ci, wi ; \quad (77)$$

where $\hat{T}$ is normalized absolute temperature; $\hat{c}_{xy}^z$ is normalized material coefficient associated with frequency of dislocation process $z$ that involves dislocations of type $xy$. Moreover, given Eqs. (4) and (76), and owing to Taylor factor ($M$) being constant in the isotropic case:

$$\hat{\dot{\varepsilon}}_p \equiv \frac{\dot{\varepsilon}_p}{\dot{\varepsilon}_0} = \frac{\dot{\gamma}_p}{\dot{\gamma}_0} \equiv \hat{\dot{\gamma}}_p ; \quad \dot{\varepsilon}_0 \equiv \frac{\dot{\gamma}_0}{M} ; \quad (78)$$

where $\dot{\varepsilon}_0$ is reference strain rate. Furthermore, as pointed out, material coefficients associated with probability amplitudes of dynamic thermal dislocation processes depend on strain rate, although their rate sensitivity in quasi-static and intermediate regimes is often negligible. Therefore,

$$m_{xy}^z \begin{cases} < 0: & z = \text{an}, \text{tr} \\ \gtreqless 0: & z = \text{nc}, \text{rm} \\ = 0: & z = \text{gn}, \text{ac}, spn, srm \end{cases} \quad ; \quad xy = cm, ci, wi . \quad (79)$$

26) <u>*Temperature dependence of plastic stress*</u>: strain rate sensitivity of mean dislocation interaction strengths ($\tilde{\alpha}_x$) are reportedly negligible compared to strain rate sensitivity of viscous stress (Mecking and Kocks, 1981). Nonetheless, mean interaction strengths decrease with increasing temperature for each material in a characteristic way (Kassner, 2015; Kocks and Mecking, 2003). Shear modulus ($G$) has a mild temperature dependence (decreasing with increasing temperature) as well which can be expressed by power-law relation (Argon, 2012; Galindo-Nava and Rae, 2016; Ghosh and Olson, 2002). Variation of temperature-dependent factors in Eqs. (26) and (28), $\tilde{\alpha}_x$ and $G$, with respect to temperature can be described simultaneously similar to temperature dependence relation of frequency of thermal dislocation processes:

$$G\widehat{\tilde{\alpha}}_x = 1 + r_{\alpha x}^G \left(\hat{T} - 1\right)^{s_{\alpha x}^G}; \quad G\widehat{\tilde{\alpha}}_x \equiv \frac{G\tilde{\alpha}_x}{G_0\tilde{\alpha}_{x0}}; \quad r_{\alpha x}^G < 0; \quad s_{\alpha x}^G > 0; \quad x = c, w; \tag{80}$$

where $G_0$ and $\tilde{\alpha}_{x0}$ are respectively reference (at reference temperature) shear modulus and mean dislocation interaction strength associated with immobile dislocations of type $x$; $G\widehat{\tilde{\alpha}}_x$ is normalized $G\tilde{\alpha}_x$; and $r_{\alpha x}^G$ and $s_{\alpha x}^G$ are temperature sensitivity coefficient and exponent associated with $G\tilde{\alpha}_x$.

27) <u>*Viscous stress and its temperature dependence*</u>: as shown in Fig. 4, the viscous contribution of yield shear stress at slip systems is represented by a nonlinear dashpot in rheological representation of the elasto-viscoplastic (EVP) constitutive model. Stress response of the nonlinear dashpot is:

$$\bar{\tau}_v = \bar{\tau}_{v0} \hat{\dot{\gamma}}_p^{m_v}; \quad m_v > 0; \tag{81}$$

where $\bar{\tau}_{v0}$ is mean viscous shear stress at reference strain rate; and $m_v$ is strain rate sensitivity parameter of viscous stress. At constant reference strain rate, by the assumption of $\Delta\bar{\tau}_v \propto (\Delta T)^{s_v}$, temperature dependence of viscous resistance can be expressed as:

$$\bar{\tau}_{v0} = \bar{\tau}_{v00}\left[1 + r_v\left(\hat{T} - 1\right)^{s_v}\right]; \quad r_v < 0; \quad 0 < s_v \leq 1; \tag{82}$$

where $\bar{\tau}_{v00}$ is the reference viscous shear stress (at reference temperature and strain rate); and $r_v$ and $s_v$ are respectively viscosity's temperature sensitivity coefficient and exponent. Eq. (82) is similar to the relations proposed by Kocks et al. (1975) and Argon (2012) for temperature dependence of viscous stress. Inserting Eq. (82) to Eq. (81) leads to:

$$\hat{\bar{\tau}}_v = \left[1 + r_v\left(\hat{T} - 1\right)^{s_v}\right]\hat{\dot{\gamma}}_p^{m_v}; \quad \hat{\bar{\tau}}_v \equiv \frac{\bar{\tau}_v}{\bar{\tau}_{v00}}; \tag{83}$$

where $\hat{\bar{\tau}}_v$ is normalized mean viscous shear stress. Considering Eqs. (12) and (83), with constant Taylor factor (in isotropic case):

$$\hat{\sigma}_v \equiv \frac{\sigma_v}{\sigma_{v00}} = \frac{\bar{\tau}_v}{\tau_{v00}} \equiv \hat{\bar{\tau}}_v; \quad \sigma_{v00} \equiv M\tau_{v00}; \tag{84}$$

where $\sigma_{v00}$ is the reference viscous stress (at reference temperature and strain rate); and $\hat{\sigma}_v$ is normalized viscous stress. The reason for assuming an ambiguous phenomenological description for temperature dependence of viscous stress is that the temperature dependence of mean (mixed/curved) dislocations mobility function is extremely complex and has not been yet properly understood (Argon, 2012; Fleischer, 1962; Gilbert et al., 2011; Gilman, 1965; Kocks et al., 1975; Li, 1967; Tang and Marian, 2014).

28) <u>*Rate dependence of strain rate sensitivities*</u>: strain rate sensitivity of viscous drag increases with increasing temperature (Cereceda et al., 2016; Khan and Liu, 2012; Kocks, 1976; Rusinek and Rodríguez-Martínez, 2009) and is assumed to have a similar form as aforementioned temperature and strain rate dependence relations:

$$\hat{m}_v = \left[1 + r_v^m\left(\hat{T} - 1\right)^{s_v^m}\right]\hat{\dot{\gamma}}_p^{m_v^m}; \quad \hat{m}_v \equiv \frac{m_v}{m_{v0}}; \quad r_v^m, s_v^m \geq 0; \tag{85}$$

where $m_{v0}$ is reference strain rate sensitivity (at reference temperature and strain rate); $\hat{m}_v$ is normalized strain rate sensitivity of viscous stress; $r_v^m$ and $s_v^m$ are respectively temperature sensitivity coefficient and exponent associated with strain rate sensitivity of viscous stress; and $m_v^m$ is strain rate sensitivity parameter associated with strain rate sensitivity of viscous stress. Moreover, strain rate sensitivity coefficients associated with dislocation processes are assumed to have similar form of temperature and strain rate dependencies:

$$\hat{m}_{xy}^z = \left[1 + r_{z_{xy}}^m \left(\hat{T} - 1\right)^{s_{z_{xy}}^m}\right] \hat{\hat{\gamma}}_p^{\,m_{z_{xy}}^m} \;; \qquad \hat{m}_{xy}^z \equiv \frac{m_{xy}^z}{m_{xy0}^z} \;; \quad xy = cm, ci, wi \;; \tag{86}$$

where $m_{z_{xy}}^m$, $m_{z_{xy}0}^m$ and $\hat{m}_{xy}^z$ are respectively, current, reference (at reference temperature) and normalized strain rate sensitivities associated with dislocation process $z$ of dislocations of type $xy$; and $r_{z_{xy}}^m$ and $s_{z_{xy}}^m$ are temperature sensitivity coefficient and exponent associated with strain sensitivity of dislocation process $z$ of dislocations of type $xy$, respectively. Further, $m_v^m$ and $m_{z_{xy}}^m$ are very small compared to $m_v$ and $m_{xy}^z$. Thus, only often shock regimes they have considerable impact.

29) *Plastic dissipation/adiabatic heating*: volumetric adiabatic heat generation rate due to plastic work ($\dot{q}_p$) which is a fraction of volumetric plastic power ($\dot{w}_p$) (Taylor and Quinney, 1934), is obtained as follows:

$$\dot{q}_p = \beta \, \dot{w}_p = \dot{w}_p - \dot{u} \;; \qquad \dot{w}_p = \dot{\hat{\gamma}}_p \tilde{\tau}_c = \sigma_y \, \dot{\varepsilon}_p \;; \qquad \beta \equiv 1 - \frac{\dot{u}}{\dot{w}_p} \;; \tag{87}$$

where $\dot{u}$ is volumetric stored elastic power which is a fraction of volumetric plastic power that is stored in material by dislocations; and $\beta$ is referred to as dissipation/conversion factor, inelastic heat fraction, efficiency of plastic dissipation, or the Taylor–Quinney coefficient. In other words, $\beta$ is the fraction of plastic power that is not stored elastically and consequently is converted to heat. Taylor and Quinney (1934) emphasized that the fraction $\beta$ increases as plastic deformation progresses until saturation state where $\beta = 1$. In other words, in saturation state, the entire input volumetric plastic power is converted to heat ($\dot{q}_p = \dot{w}_p$). Moreover, Rosakis et al. (2000) and Zehnder (1991) proposed models for variation of $\beta$ as a function of plastic strain and plastic hardening ($\theta \equiv \partial_{\varepsilon_p} \sigma_y$) where $\beta$ approaches one by decreasing $\theta$ as plastic strain increases. At the beginning of plastic deformation of a nearly undeformed/annealed polycrystalline metallic material, large portion of the input plastic energy is stored in the crystal structure by generation of dislocations and dislocation structures. However, as the deformation progresses the generation rate of dislocations diminishes whereas the annihilation rate increases (annihilation releases the previously stored energy of dislocations in form of heat) until the saturation state where these two rates are equivalent. Therefore, one can assume:

$$\beta = \left(\frac{\partial_{\varepsilon_p} \hat{\rho}_t^-}{\partial_{\varepsilon_p} \hat{\rho}_t^+}\right)^{\kappa} \;; \quad \kappa > 0 \;; \tag{88}$$

where $\kappa$ is a material constant associated with dissipation factor.

Furthermore, given Eqs. (12), (14), (28), (34), (78), (83) and (84), plastic/strain hardening ($\theta$) can be calculated as follows:

$$\theta \equiv \partial_{\varepsilon_p} \sigma_y = \partial_{\varepsilon_p} \sigma_p = \theta_c + \theta_w \;; \tag{89}$$

$$\theta_x \equiv \partial_{\varepsilon_p} \sigma_{px} = \frac{MbG\tilde{\alpha}_x}{2\sqrt{\rho_{xi}}} \, \partial_{\varepsilon_p} \rho_{xi} = \frac{\partial_{\varepsilon_p} \hat{\rho}_{xi}}{2\hat{\rho}_{xi}} \sigma_{px} \;; \qquad x = c, w \;; \tag{90}$$

where $\theta_x$ is plastic hardening associated with dislocations of type $x$. In addition, according to Eqs. (12), (14), (28), (34), (55), (58), (78), (83), (84) and (85) viscous/strain-rate hardening ($\varphi$) is obtained as follows:

$$\varphi \equiv \partial_{\dot{\varepsilon}_p} \sigma_y = \varphi_v + \varphi_p \;; \tag{91}$$

$$\varphi_v \equiv \partial_{\dot{\varepsilon}_p} \sigma_v = \frac{m_v}{\dot{\varepsilon}_p} \left[1 + m_v^m \, \breve{\ln}(\hat{\dot{\varepsilon}}_p)\right] \sigma_v \;; \tag{92}$$

$$\varphi_p \equiv \partial_{\dot{\varepsilon}_p} \sigma_p = \varphi_{pc} + \varphi_{pw} \;; \qquad \varphi_{px} \equiv \partial_{\dot{\varepsilon}_p} \sigma_{px} = \frac{\partial_{\dot{\varepsilon}_p} \hat{\rho}_{xi}}{2\hat{\rho}_{xi}} \sigma_{px} = \frac{\partial \varepsilon_p}{\partial \dot{\varepsilon}_p} \theta_x \;; \qquad x = c, w \;; \tag{93}$$

where $\partial_{\dot{\varepsilon}_p} \equiv \frac{\partial}{\partial \dot{\varepsilon}_p}$ is partial derivative operator with respect to plastic strain rate ($\dot{\varepsilon}_p$); $\varphi_v$ and $\varphi_p$ are viscous hardenings respectively associated with viscous and plastic stress; and $\varphi_{px}$ is viscous hardening associated with plastic stress of type $x$.

# 4. Numerical integration and parameter identification

## 4.1. Numerical integration

Since there is no analytical solution to overall evolution rate of dislocation densities, they must be solved numerically. Hence, the differential continuum equations expressed in previous sections must be numerically integrated with respect to time. Simulation time is discretized to small increments. Consider a (pseudo) time interval $\left[t^{(n)}, t^{(n+1)}\right]$, such that $\Delta t^{(n+1)} \equiv t^{(n+1)} - t^{(n)}$ is the time increment at $(n+1)$-th time step. Accordingly,

$$\Delta(\bullet)^{(n+1)} \equiv (\bullet)^{(n+1)} - (\bullet)^{(n)} ; \quad (\dot{\bullet})^{(n+1)} \equiv \frac{\Delta(\bullet)^{(n+1)}}{\Delta t^{(n+1)}} ; \tag{94}$$

where $(\bullet)$ can be any time-dependent variable; and superscripts $(n)$ and $(n + 1)$ respectively, represent the value of corresponding time-dependent variable at the beginning and the end of $(n + 1)$-th time step. Application of forward/explicit Euler method for numerical integration of dislocation densities gives:

$$\hat{\rho}_{xy}^{(n+1)} = \hat{\rho}_{xy}^{(n)} + \Delta\hat{\rho}_{xy}^{(n)} ; \quad \Delta\hat{\rho}_{xy}^{(n)} = \Delta\varepsilon_p^{n+1}\, \partial_{\varepsilon_p}\hat{\rho}_{xy}^{(n)} = \Delta t^{(n+1)}\, \dot{\hat{\rho}}_{xy}^{(n)} ; \quad \hat{\rho}_{xy}^{(n=0)} = \hat{\rho}_{xy0} ; \quad \begin{cases} x = c, w, t \\ y = m, i, t \end{cases} ; \tag{95}$$

where $\partial_{\varepsilon_p}\hat{\rho}_{xy}^{(n)}$ is computed by Eqs. (55), (58) and (61).  Given Eqs. (12), (14), (28), (34), (77), (78), (80), (83), (84), (85) and (86):

$$\sigma_y^{(n+1)} = \sigma_v^{(n+1)} + \sigma_{pc}^{(n+1)} + \sigma_{pw}^{(n+1)} ; \tag{96}$$

$$\sigma_{px}^{(n+1)} = Mb(G\tilde{\alpha}_x)^{(n)}\sqrt{\rho_0\,\hat{\rho}_{xi}^{(n+1)}} ; \quad x = c, w ; \tag{97}$$

$$(G\tilde{\alpha}_x)^{(n)} = G_0\tilde{\alpha}_{x0}\left[1 + r_{\alpha x}^G\left(\hat{T}^{(n)} - 1\right)^{s_{\tilde{\alpha}x}^G}\right] ; \quad \hat{T}^{(n)} \equiv \frac{T^{(n)}}{T_0} ; \quad x = c, w ; \tag{98}$$

$$\sigma_v^{(n+1)} = \sigma_{v00}\left[1 + r_v\left(\hat{T}^{(n)} - 1\right)^{s_v}\right]\left(\hat{\dot{\varepsilon}}_p^{(n+1)}\right)^{m_v^{(n)}} ; \quad \hat{\dot{\varepsilon}}_p^{(n+1)} \equiv \frac{\dot{\varepsilon}_p^{(n+1)}}{\dot{\varepsilon}_0} = \frac{\Delta\varepsilon_p^{(n+1)}}{\Delta t^{(n+1)}\,\dot{\varepsilon}_0} ; \tag{99}$$

$$m_v^{(n)} = m_{v0}\left[1 + r_v^m\left(\hat{T}^{(n)} - 1\right)^{s_v^m}\right]\left(\hat{\dot{\varepsilon}}_p^{(n+1)}\right)^{m_v^m} ; \tag{100}$$

$$c_{xy}^{z\,(n)} = c_{xy0}^z\left[1 + r_{xy}^z\left(\hat{T}^{(n)} - 1\right)^{s_{xy}^z}\right]\left(\hat{\dot{\varepsilon}}_p^{(n+1)}\right)^{m_{xy}^{z\,(n)}} ; \quad xy = cm, ci, wi ; \tag{101}$$

$$m_{xy}^{z\,(n)} = m_{xy0}^z\left[1 + r_{zxy}^m\left(\hat{T}^{(n)} - 1\right)^{s_{zxy}^m}\right]\left(\hat{\dot{\varepsilon}}_p^{(n+1)}\right)^{m_{zxy}^m} ; \quad xy = cm, ci, wi . \tag{102}$$

In addition, discretized forms of Eqs. (87) and (88) in combination with Eqs. (62) and (63) read:

$$\Delta q_p^{(n+1)} = \beta^{(n)}\frac{\sigma_y^{(n)} + \sigma_y^{(n+1)}}{2}\Delta\varepsilon_p^{(n+1)} ; \tag{103}$$

$$\beta^{(n)} = \left(\frac{2\left(\partial_{\dot{\varepsilon}_p}\hat{\rho}_{cm}^{\text{an}\,(n)} + \partial_{\dot{\varepsilon}_p}\hat{\rho}_{ci}^{\text{an}\,(n)} + \partial_{\dot{\varepsilon}_p}\hat{\rho}_{wi}^{\text{an}\,(n)}\right)}{\partial_{\dot{\varepsilon}_p}\hat{\rho}_{cm}^{\text{gn}\,(n)}}\right)^\kappa . \tag{104}$$

For FE implementation, in return mapping (using Newton-Raphson scheme) of EVP constitutive models as well as calculation of viscoplastic consistent tangent operator (implicit FE framework), the viscoplastic tangent modulus ($H$) is required. Given Eqs. (40), (46), (49), (52), (55), (58), (79), (95), (96), (97), (99), (100), (101) and (102):

$$H^{(n+1)} \equiv \frac{\partial \sigma_y^{(n+1)}}{\partial \Delta \varepsilon_p^{(n+1)}} = \frac{m_v^{(n)}\big[1 + m_{t_v}^m \widetilde{\ln}(\dot{\hat{\varepsilon}}_p^{(n+1)})\big]}{\Delta \varepsilon_p^{(n+1)}} \sigma_v^{(n+1)}$$ 

$$+ \frac{\partial_{\varepsilon_p} \hat{\rho}_{ci}^{(n)} + m_{cm}^{\mathrm{tr}\,(n)}\big[1 + m_{\mathrm{tr}_{cm}}^m \widetilde{\ln}(\dot{\hat{\varepsilon}}_p^{(n+1)})\big]\partial_{\varepsilon_p}\hat{\rho}_{cm}^{\mathrm{tr}\,(n)} - \sum_{z_{xy}=\substack{\mathrm{an}_{ci}\\\mathrm{rm}_{ci}\\\mathrm{nc}_{wi}}} m_{xy}^{z\,(n)}\big[1 + m_{z_{xy}}^m \widetilde{\ln}(\dot{\hat{\varepsilon}}_p^{(n+1)})\big]\partial_{\varepsilon_p}\hat{\rho}_{xy}^{z\,(n)}}{2\hat{\rho}_{ci}^{(n+1)}} \sigma_{pc}^{(n+1)}$$

$$+ \frac{\partial_{\varepsilon_p} \hat{\rho}_{wi}^{(n)} + m_{wi}^{\mathrm{nc}\,(n)}\big[1 + m_{\mathrm{nc}_{wi}}^m \widetilde{\ln}(\dot{\hat{\varepsilon}}_p^{(n+1)})\big]\partial_{\varepsilon_p}\hat{\rho}_{wi}^{\mathrm{nc}\,(n)} - \sum_{z_{xy}=\substack{\mathrm{an}_{wi}\\\mathrm{rm}_{wi}}} m_{xy}^{z\,(n)}\big[1 + m_{z_{xy}}^m \widetilde{\ln}(\dot{\hat{\varepsilon}}_p^{(n+1)})\big]\partial_{\varepsilon_p}\hat{\rho}_{xy}^{z\,(n)}}{2\hat{\rho}_{wi}^{(n+1)}} \sigma_{pw}^{(n+1)}.$$

(105)

Details of FE implementation of the above discretized constitutive equations in the framework of hypo-EVP finite deformation based on isotropic associative $J_2$ plasticity is planned to be published soon.

### 4.2. Identification of constitutive parameters

Generally, constitutive/material parameters/constants of constitutive models are directly derived from experimental flow curves (yield stress versus plastic strain) through parameter identification techniques. Here, as an important hypothesis, it is envisaged that having the flow curves of a material in various temperatures, strain rates, etc. in the considered regimes is sufficient for calibration of a properly devised microstructural constitutive model using a robust parameter identification procedure. After all, under viscoplastic deformation, the underlying microstructural processes statistically result in a specific material response that is macroscopically observable through flow curves. However, although due to their physical nature, some of the constitutive parameters can be obtained by independent characterization methods (other than flow curves), it is suggested that the most accurate and effective approach for finding them is the simultaneous parameter identification, given the flow curves as the reference of fitting. Identification of microstructural parameters is carried out in two steps:

1) In the first step, isothermal uniaxial compression/upsetting or tensile tests are conducted in reference strain rate and various temperatures. Then, based on the measured experimental flow curves, most of the constitutive parameters are determined by means of pointwise parameter optimization using RMV-based analysis (RMVA).
2) In the last step, nonisothermal uniaxial tests are performed with different strain rates (in intermediate strain rate regime) and various initial temperatures (in cold and warm regime). Subsequently, the remaining constants which are strain rate sensitivity parameters ($m$), temperature sensitivity coefficients ($r$) and exponents ($s$) associated with strain rate sensitivities, as well as the material parameter associated with dissipation factor ($\kappa$) are simultaneously calibrated by parameter optimization using finite element model updating (FEMU) method also known as finite element-based inverse strategy (Steenackers et al., 2007).

After conducting uniaxial tests, force-displacement data is processed to true stress-strain, and then to flow curves. Afterwards, noise of flow curve data is reduced (here, "smooth" function in MATLAB library is used for noise filtration) and interpolated for constant plastic strain intervals. For the first step of parameter identification by RMVA, the following mean relative error which is also called objective, fitness, or residual function, is constructed as follows to be optimized/minimized:

$$\bar{e}(\boldsymbol{C}) \equiv \frac{1}{p} \sum_{i=1}^{p} \frac{\omega_i}{q_i} \sum_{j=0}^{q_i} \varpi_j \frac{\big|\sigma_{ij}^{\mathrm{num}}(\boldsymbol{C}, T_i, \Delta\varepsilon_p, \boldsymbol{S}^j) - \sigma_{ij}^{\mathrm{exp}}(T_i, \varepsilon_p^j = j\Delta\varepsilon_p)\big|}{\sigma_{ij}^{\mathrm{exp}}(T_i, \varepsilon_p^j = j\Delta\varepsilon_p)} \; ;$$

(106)

where $\boldsymbol{C}$ is the set containing constitutive constants for fitting; $p$ is total number of experiments at different temperatures with constant reference strain rate $\dot{\varepsilon}_0$; $q_i$ is total number of data points of isothermal experimental flow curve at temperature $T_i$; $\Delta\varepsilon_p$ is the constant plastic strain interval between data points; $\sigma_{ij}^{\mathrm{num}}$ is yield stress calculated numerically by the constitutive model at temperature $T_i$ which is corresponding to $\varepsilon_p^j = j\Delta\varepsilon_p$; $\boldsymbol{S}^j$ is the set of MSVs of $j$-th step; $\sigma_{ij}^{\mathrm{exp}}$ is yield stress in isothermal experimental flow curve at temperature $T_i$ and plastic strain $\varepsilon_p^j = j\Delta\varepsilon_p$; $\omega_i$ and $\varpi_j$ are weighting factors respectively for isothermal flow curve at temperature $T_i$, and data point $j$ corresponding to $\varepsilon_p^j = j\Delta\varepsilon_p$. In default, $\omega_i = 1$ and $\varpi_j = 1$, unless some individual flow curve or ranges of plastic strain, have higher weight/importance ($\omega_i$ , $\varpi_j > 1$) or lower importance ($\omega_i$ , $\varpi_j < 1$) than the rest. For the example provided in the next section, $\Delta\varepsilon_p = 0.001$ and $\omega_i = \varpi_j = 1$.

For many cases, finding the global solution/minimum to this optimization problem is difficult and requires complex and robust mathematical optimization methods which follow the steepest descent with the crossover functionality. Many methods are available for searching for the global solution such as gradient-free minimization methods and evolutionary algorithms including genetic algorithm. These methods exist in literature and their corresponding computer codes are developed and embedded in many commercial mathematical software. For the example provided in the next section, minimization of Eq. (106) for calibration of material parameters is carried out by a script written in MATLAB software. The optimization script employs MATLAB's global search class "GlobalSearch" and genetic algorithm "ga" which are coupled with its local minimization solver for constrained nonlinear multivariate functions "fmincon". The local minimization solver optionally utilizes various robust optimization algorithms such as interior point and reflective trust region.

For the second step of fitting based on FEMU, firstly the constitutive model must be implemented and programmed as a material subroutine based on hypo-EVP finite deformation using the constitutive parameters derived in the first step of calibration and with an initial guess for the constitutive parameters yet to be determined in the second step. Then, depending on the type of conducted experimental uniaxial tests, an FE model of either uniaxial compression or tension test must be constructed and linked to the material subroutine. Next, optimization of remaining constitutive parameters is done by iterative running of FE simulation of uniaxial tests with consecutives correction and update of remaining parameters. Running of FE simulation of uniaxial tests with updated parameters can be done by an optimization script that controls the pre-processing and post-processing of FE simulation. However, with a good initial guess for the remaining material constants, the second step of optimization can be done manually as well. For the example provided in the next section, the presented constitutive model has been implemented as implicit and explicit user-defined material subroutines for hypo-EVP finite deformation in ABAQUS FE package based on isotropic associative $J_2$ plasticity.

## 5. Results, discussion and validation

In order to determine constitutive parameters, and also for validation of the constitutive model, uniaxial compression tests at various temperatures and strain rates using a deformation-type dilatometer (DIL-805A/D by TA Instruments) were conducted as dilatometer provides better precision than conventional compression test machines. Temperature measurement was done by thermocouple welded on the middle point in height direction at the lateral surface of the compression specimen. Temperature and strain rate were controlled by the dilatometer's integrated PID controller. Moreover, GND density measurements using high resolution EBSD for deformed compression specimens at different plastic strains are carried out. The material used in dilatometer compression is case-hardenable ferritic-pearlitic steel alloy 20MnCr5 (1.7147/1.7149) which is widely used in industrial forging of automotive components such as bevel gear. The chemical composition of 20MnCr5 used in experiments is presented in Table 1. The material constants required by the constitutive model that are independently measurable are also presented in Table 2.

**Table 1**
Chemical composition of the investigated steel 20MnCr5 [mass%].

| C | Si | Mn | P | S | Cr | Mo | Ni | Cu | Al | N |
|---|----|----|----|----|----|----|----|----|----|---|
| 0.210 | 0.191 | 1.350 | 0.014 | 0.025 | 1.270 | 0.074 | 0.076 | 0.149 | 0.040 | 0.010 |

**Table 2**
Reference parameters and independently measurable required material constants.

| $T_0$ [°C] | $\dot{\varepsilon}_0$ [s⁻¹] | $\rho_0$ [m⁻²] | $G_0$ [GPa] | $M$ [-] | $b$ [m] |
|---|---|---|---|---|---|
| 20 | 0.01 | $10^{12}$ | 82.5 | 3.0 | $2.55 \times 10^{-10}$ |

As already mentioned, reference temperature and strain rate are selected as the minimum possible temperature and strain rate in the considered domain (Table 2). Since for the steel studied here, the minimum temperature for onset of dynamic recrystallization of ferrite is slightly above 600 °C, the maximum temperature for constitutive modeling in warm regime is selected to be 600 °C. For the first step of parameter identification, experimental dilatometer compression tests at reference strain rate and 13 different temperatures (20, 50, 100, 150, ..., 600 °C) are performed. For each temperature at least two compression tests are carried out. Cylindrical dilatometer compression samples with diameter of 3.5mm and height of 5mm without lubrication pockets are used. However, lubricant is applied on contact surfaces. Subsequently, corresponding flow curves including yield point elongation are derived, smoothed, and interpolated in constant plastic strain intervals. Averages of the resultant flow curves that have the same nominal test parameters (temperature and strain rate) are calculated. These flow curves are then used for calculation of constitutive parameters using RMVA scheme. The constitutive constants that are determined in the first calibration step using RMVA method are presented in Table 3; and their corresponding temperature sensitivity coefficients and exponents are listed in Table 4.

**Table 3**
Constitutive parameters determined in the first step of parameter identification by RMVA scheme.

| $c_{cm}^{gn}$ [-] | $c_{cm0}^{an}$ [-] | $c_{ci0}^{an}$ [-] | $c_{ci0}^{an}$ [-] | $c_{ci}^{ac}$ [-] | $c_{wi}^{ac}$ [-] | $c_{cm0}^{tr}$ [-] | $c_{wi0}^{nc}$ [-] |
|---|---|---|---|---|---|---|---|
| $6.2970 \times 10^2$ | 0.1492 | 0.0133 | 0.0312 | 0.4989 | 0.1280 | 1.4184 | $1.5534 \times 10^{-3}$ |

| $c_{ci0}^{rm}$ [-] | $c_{wi0}^{rm}$ [-] | $\bar{\alpha}_{c0}$ [-] | $\bar{\alpha}_{w0}$ [-] | $\hat{\rho}_{cm0}$ [-] | $\hat{\rho}_{ci0}$ [-] | $\hat{\rho}_{wi0}$ [-] | $\sigma_{v00}$ [MPa] |
|---|---|---|---|---|---|---|---|
| 0.2261 | 0.0217 | 0.1001 | 0.4725 | $2.2573 \times 10^1$ | $2.6427 \times 10^1$ | 0.9234 | 318.84 |

**Table 4**
Temperature sensitivity coefficients and exponents determined in the first step of parameter identification by RMVA scheme.

| $r_{cm}^{an}$ [-] | $r_{ci}^{an}$ [-] | $r_{wi}^{an}$ [-] | $r_{cm}^{tr}$ [-] | $r_{wi}^{nc}$ [-] | $r_{ci}^{rm}$ [-] | $r_{wi}^{rm}$ [-] | $r_{ac}^{G}$ [-] | $r_{aw}^{G}$ [-] | $r_v$ [-] |
|---|---|---|---|---|---|---|---|---|---|
| 0.0547 | 2.0581 | 0.2045 | 3.9680 | 6.1587 | 5.0910 | 2.0631 | - 0.0835 | - 0.0288 | - 0.3376 |

| $s_{cm}^{an}$ [-] | $s_{ci}^{an}$ [-] | $s_{wi}^{an}$ [-] | $s_{cm}^{tr}$ [-] | $s_{wi}^{nc}$ [-] | $s_{ci}^{rm}$ [-] | $s_{wi}^{rm}$ [-] | $s_{ac}^{G}$ [-] | $s_{aw}^{G}$ [-] | $s_v$ [-] |
|---|---|---|---|---|---|---|---|---|---|
| 8.6725 | 0.9988 | 4.0282 | 1.5593 | 4.8075 | 5.5999 | 3.4306 | 2.8735 | 2.5451 | 0.5115 |

Mean relative error of the fitting is $\bar{e} = 0.39\%$ which is in the same range of experimental scatter/precision (around 0.43%). constitutive parameters identified in the second step of calibration procedure by FEMU technique using nonisothermal uniaxial compression tests at different strain rates and various initial temperatures are presented in Table 5.

**Table 5**
Constitutive parameters identified in the second step of parameter calibration by FEMU method.

| $m_{v0}$ [-] | $r_v^m$ [-] | $s_v^m$ [-] | $m_v^m$ [-] | $\kappa$ [-] |
|---|---|---|---|---|
| 0.027 | 0.0785 | 5.0 | 0.0 | 2.0 |

Temperature dependencies of material coefficients associated with frequencies of various thermal dynamic dislocation processes are shown in Fig. 6. For discussing the temperature dependence of different dynamic dislocation processes demonstrated in Fig. 6, it is useful to consider an intermediate temperature regime as the transition between cold and warm regimes which in case of the investigated material ranges from 200 °C to 400 °C. As shown in Fig. 6, probability amplitudes associated with dynamic annihilation of cell mobile and wall immobile dislocations, dynamic nucleation of wall immobile dislocations and dynamic remobilization of immobile dislocations are nearly constant or have almost linear temperature dependencies in cold (0 °C < $T$ < 200 °C) temperatures. However, they have appreciable nonlinear (exponential Arhenius-type) temperature dependencies in the intermediate (200 °C < $T$ < 400 °C) and warm (400 °C < $T$ < 600 °C) regimes. This is partly ascribed to different temperature dependencies of various thermal dislocation mechanisms in different temperature regimes that contribute unevenly to the aforementioned dynamic dislocation processes.

As Fig. 6 suggests, frequencies of annihilation of cell mobile and wall immobile dislocations as well as remobilization of immobile dislocations are intensified by transition from cold to intermediate and warm regimes. This can be in part explained by the well-accepted phenomenon that the principal recovery mechanism changes from cross-slip to climb by transition from cold to warm regime because climb is energetically more favorable in warm regime (Essmann and Mughrabi, 1979; Galindo-Nava et al., 2012; Kubin et al., 1992; Nix et al., 1985; Püschl, 2002; Rivera-Díaz-del-Castillo and Huang, 2012). As plotted in Fig. 6 (e), probability amplitude of dynamic remobilization of cell immobile dislocations has the strongest temperature dependence compared to the rest of thermal dynamic dislocation processes. At 600 °C, it is roughly 230 times higher than room temperature. However, dynamic annihilation of wall immobile dislocations at 600 °C is only about 10 times higher than that of room temperature.

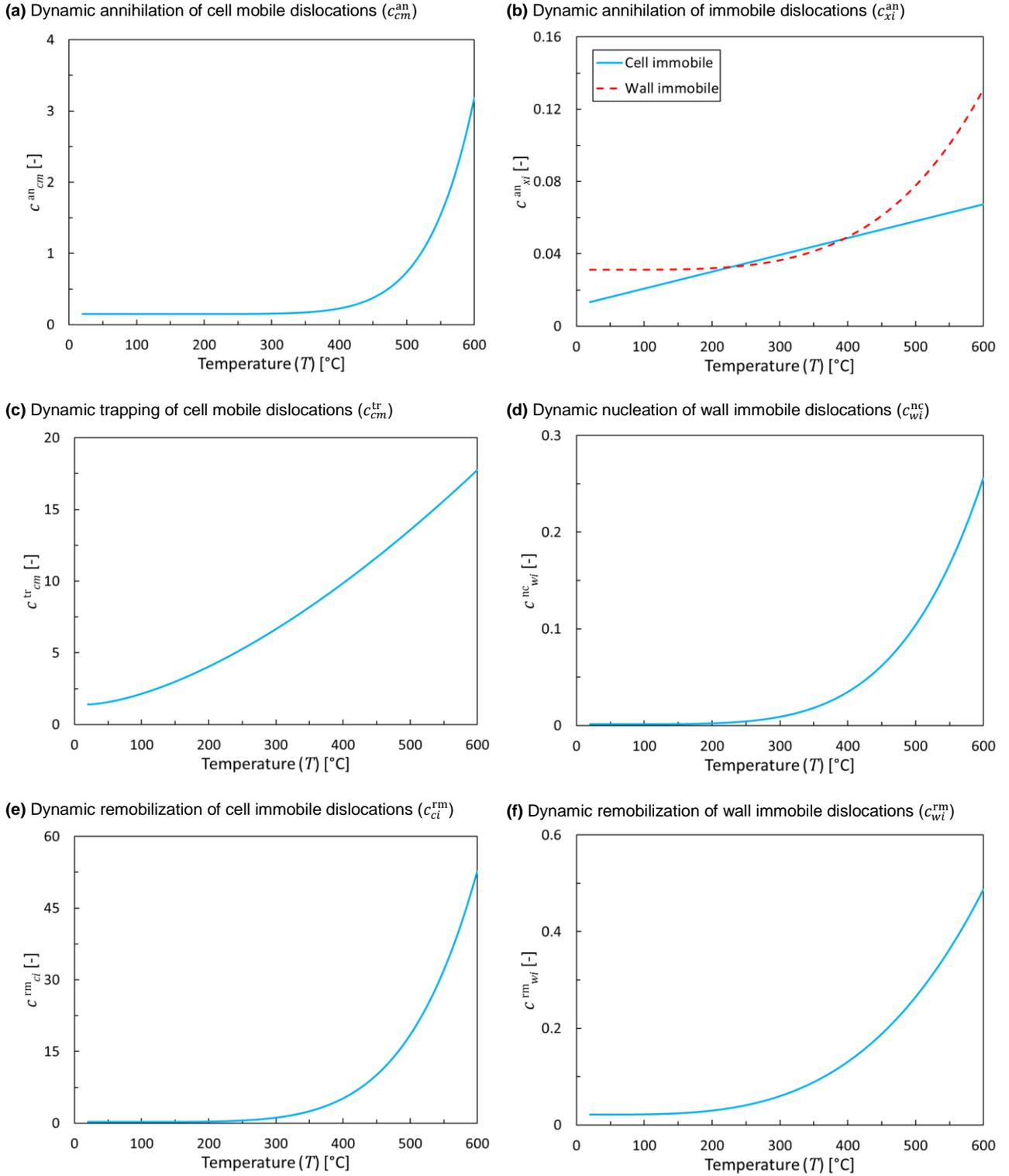

**Fig. 6.** Temperature dependencies of material coefficients associated with probability amplitudes of different thermal dynamic dislocation processes.

As mentioned in postulates (13) and (17), dynamic annihilation and remobilization (dynamic recovery) processes share some of the underlying thermal mechanisms. Moreover, as shown in Fig. 6 (b), probability amplitude of dynamic annihilation of cell immobile dislocations is almost linear in the entire temperature domain (cold and warm regimes) while the probability amplitude of the other dynamic annihilation processes have nonlinear temperature dependencies (Fig. 6 (a) and (b)). The reason is that the thermally-activated and diffusion-controlled dislocation recovery mechanisms such as dislocation climb operating on a cell immobile dislocation in warm regime, more likely results in its remobilization than annihilation compared to its wall immobile counterpart. However, after being remobilized, this former cell immobile dislocation that is now another cell mobile dislocation can become annihilated by another cell mobile dislocation (dynamic annihilation of cell mobile dislocations). This also can partly explain why the frequency of cell mobile dislocation annihilation is more sensitive to temperature raise. Hence, the climb mechanism has much stronger effect on dynamic annihilation of wall immobile dislocations than dynamic annihilation of cell immobile dislocations. Combining this with the fact that the climb mechanism is controlled by vacancy diffusion which has a high nonlinear temperature dependence in warm regime, will explain the difference between forms of temperature dependencies of probability amplitudes of dynamic annihilation of different types of immobile dislocations. As inferred from Fig. 6 (e), the climb mechanism substantially influences cell immobile dislocations because of their immobility and particular loose local arrangement compared to wall immobile dislocations. In cold regime, as shown in Fig. 6 (e), the climb mechanism is not strong enough in order to notably impact the remobilization rate of cell immobile dislocations. However, as shown in Fig. 6 (b), together with cross-slip, its intensity in cold regime is sufficient to linearly affect the annihilation rate of cell immobile dislocations.

Probability amplitude of remobilization of cell immobile dislocations for the investigated material, as mentioned, grows intensively in warm regime by increasing temperature, while it is almost constant in cold regime. Accordingly, cell immobile dislocations and their associated dislocation substructures (IDBs) have relatively high stability in cold regime. However, their stability diminishes abruptly by transition from cold to warm temperatures. Hence, cell immobile dislocation density must be much lower during plastic deformation at warm temperatures than that of cold temperatures. From this it can be inferred that the studied metal tends to formation of dislocation sub-cells (bounded by IDBs) in cold regime while in warm regime it leans toward formation of dislocation cells (bounded by dislocation walls or GNBs).

Furthermore, temperature dependencies of shear modulus, interaction strengths, viscous stress, and strain rate sensitivity are depicted in Fig. 7.

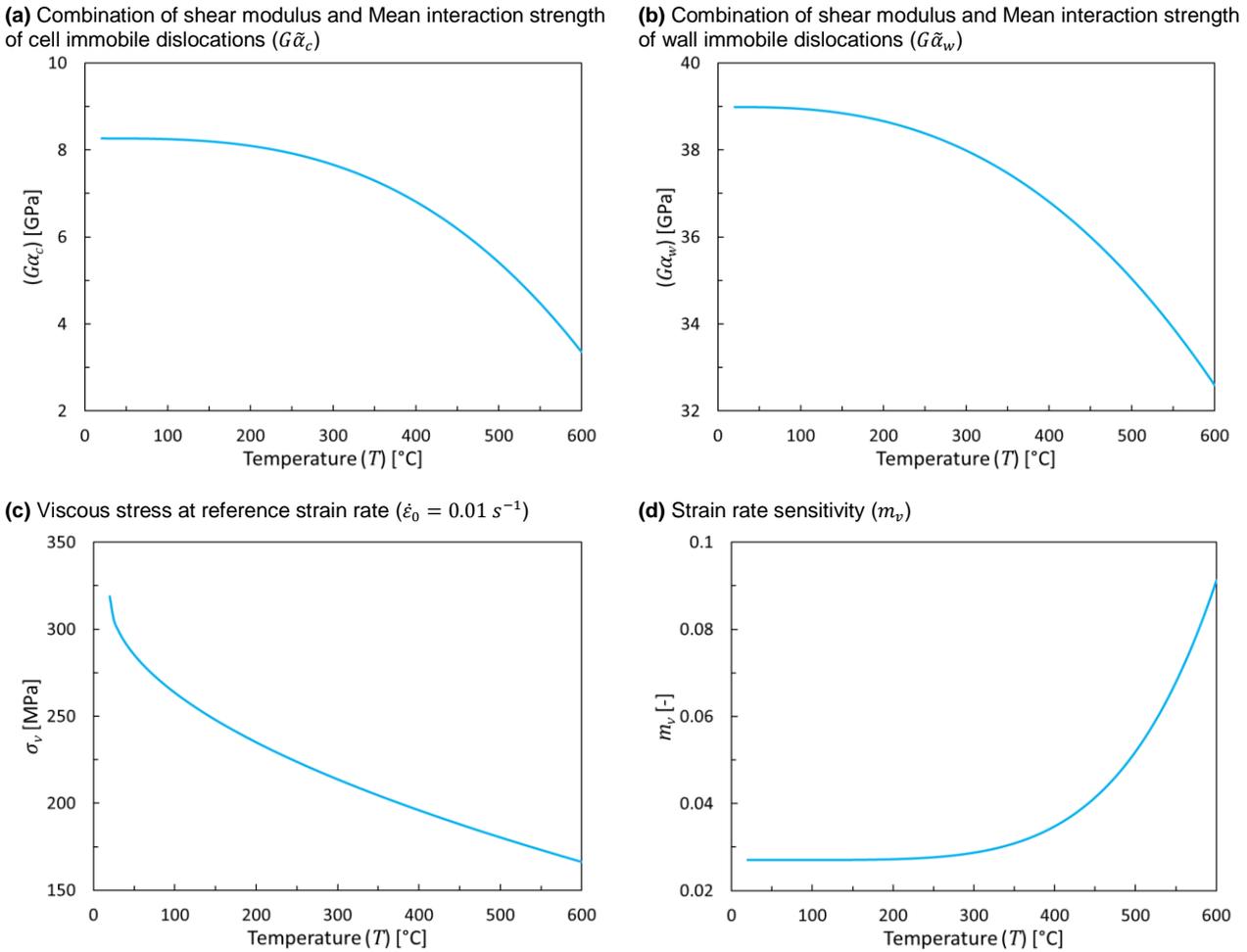

**Fig. 7.** Temperature dependencies of shear modulus, mean interaction strengths, viscous stress, and strain rate sensitivity.

### 5.1. Isothermal uniaxial tests

Experimental and computational flow curves (after optimization by RMVA) at different temperatures (isothermal condition) for compressive deformation at reference strain rate ($\dot{\varepsilon}_0 = 0.01\ s^{-1}$) are shown in Fig. 8.

**(a)** Cold Temperatures ($0\ °C < T < 200\ °C$)

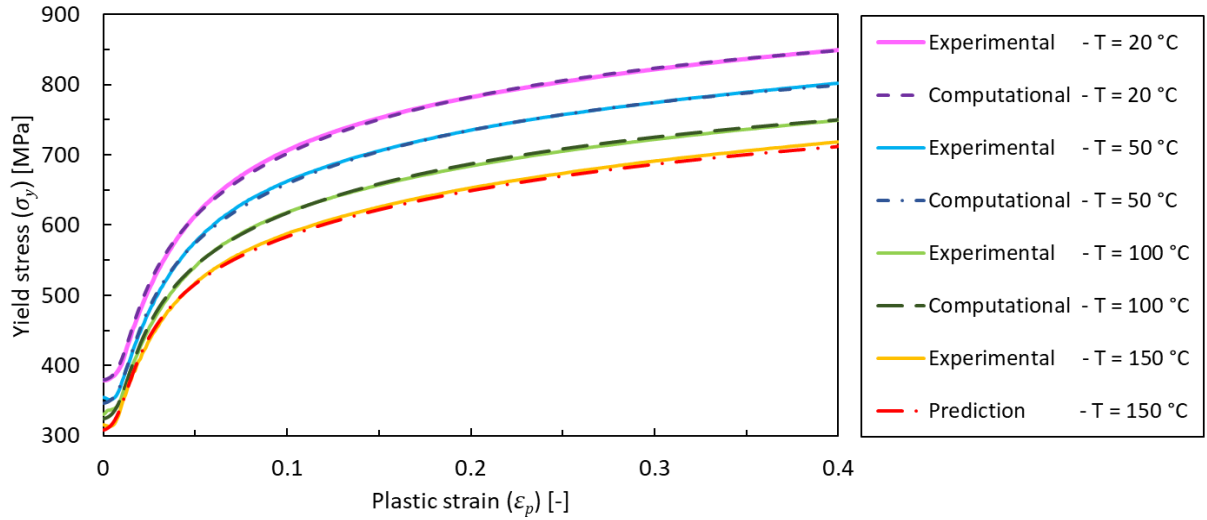

**(b)** Intermediate temperatures ($200\ °C \leq T < 400\ °C$)

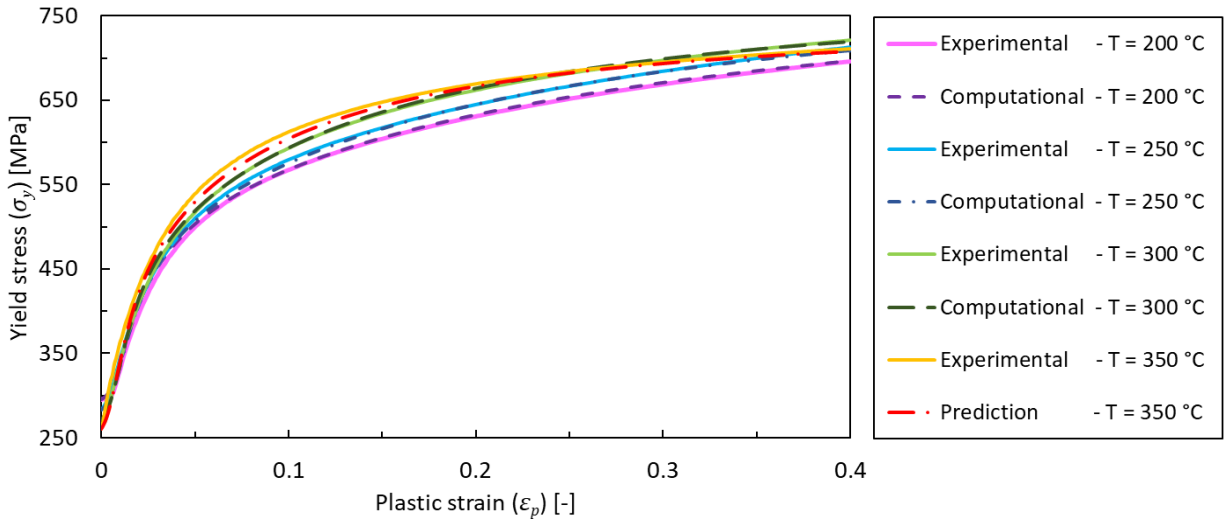

**(c)** Warm temperatures ($400\ °C \leq T \leq 600\ °C$)

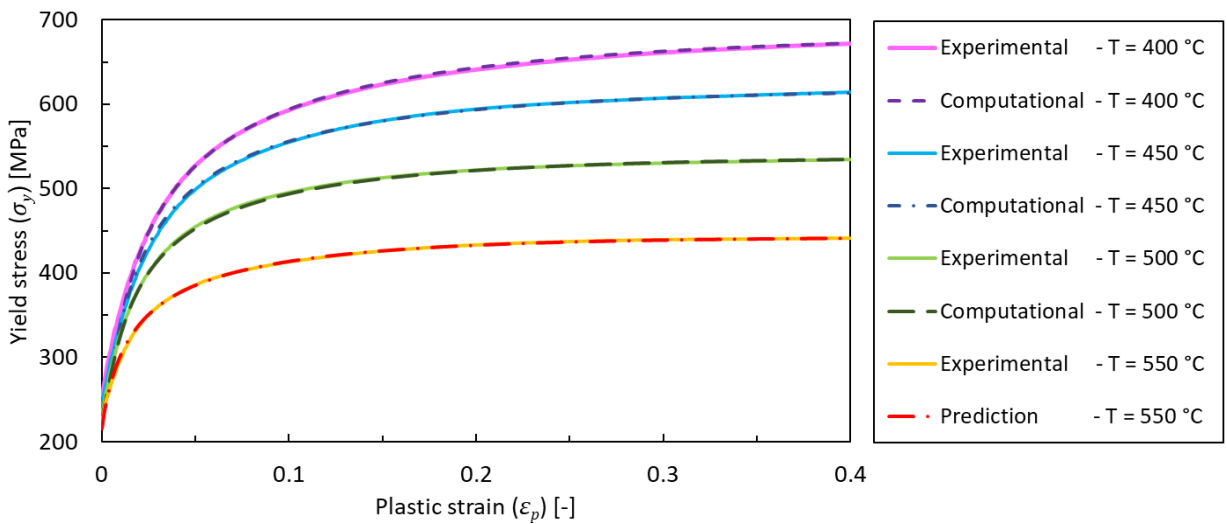

**Fig. 8.** Comparison between experimental and computational flow curves at different temperatures (isothermal condition) for compressive deformation at reference strain rate ($\dot{\varepsilon}_0 = 0.01\ s^{-1}$).

For the RMVA fitting, the experimental data are used only up to the accumulated plastic strain of $\varepsilon_p = 0.4$, because at larger strains the effect of friction between compression tools and specimen becomes dominant. This induces notable inhomogeneous distribution of MSVs and strain rate (in compression specimen) that is in contradiction with the assumptions of homogeneous RMV and uniaxial deformation. Moreover, since experimental measurements carry notable amount of noise, for fitting of constitutive parameters as well as for comparison to corresponding computational curves, they have been smoothed. In each temperature regime (cold, intermediate and warm), one of the experimental flow curves plotted in Fig. 8 has not been used for parameter identification. Those experimental flow curves (isothermal condition at reference strain rate) that are associated with temperatures 150 °C, 350 °C and 550 °C, have been later compared to their computational predictions for validation.

Isothermal flow curves as shown in Fig. 8, are categorized in the aforementioned three temperature groups: cold, intermediate and warm regimes. As shown in Fig. 6 and Fig. 7, in cold temperature regime for the investigated material (20 to 200 °C), temperature dependent constitutive parameters are either almost constant or have approximately linear temperature dependence. However, in intermediate and warm regimes (200 to 600 °C), temperature dependence of most of the thermal material constants abruptly changes. Temperature dependence of flow curves is the result of two competing effects:

- _**Dynamic thermal softening (DTS)**_ due to temperature dependence of viscous stress, shear modulus, interaction strength, dynamic annihilation of dislocations, and dynamic remobilization of immobile dislocations.
- _**Dynamic thermal hardening (DTH)**_ because of DSA (by pinning mechanism), i.e. temperature dependence of dynamic trapping of cell mobile dislocations and dynamic nucleation of wall immobile dislocations.

According to Fig. 6, Fig. 7 and Fig. 8, with respect to relative average difference among flow curves at different temperatures, in cold regime, DTS mostly due to temperature dependence of viscous stress, is clearly stronger than DTH. In cold regime by increasing temperature, DTS gradually decreases while DTH grows. This can be seen from the decreasing trend (by increasing temperature) of relative difference of yield stress in flow curves of cold regime. As shown in Fig. 8 (a), in constant plastic strains by increasing temperature the yield stress decreases due to DTS. However, the amount of reduction by increasing temperature also decreases because the competing DTH becomes stronger as the temperature continues to rise. Thus, the transition between cold and intermediate regimes occurs in the temperature at which DTS and DTH almost neutralize each other.

In intermediate regimes, DTH is slightly more dominant than DTS. As shown in Fig. 8 (b), in constant plastic strains by increasing temperature the yield stress increases as well due to stronger DTH in intermediate regime compared to DTS. Nonetheless, at the upper bound of intermediate regime, the amount of increase in yield stress by increasing temperature diminishes because this time the competing DTS is getting stronger as the temperature continues to grow. In warm regime, the dominant effect again becomes DTS which enhances more than DTH by increasing temperature, mainly due to dynamic annihilation and remobilization processes, mean interaction strengths and shear modulus. As shown in Fig. 8 (c), in constant plastic strains by increasing temperature the yield stress decreases with an increasing rate due to very strong DTS in warm regime that cannot be competed by DTH effect. This implies that, in warm regime by increasing temperature, even though the pinning mechanism is amplified but the DTS impact becomes much greater that macroscopically results in an accelerated reduction of yield stress by increasing temperature. Experimental and computational plastic hardening curves (plastic hardening vs. yield stress or plastic strain) of compression tests in constant reference condition ($T = 20$ °C and $\dot{\varepsilon}_0 = 0.01$ s$^{-1}$) are demonstrated in Fig. 9.

**(a)** Plastic hardening versus yield stress

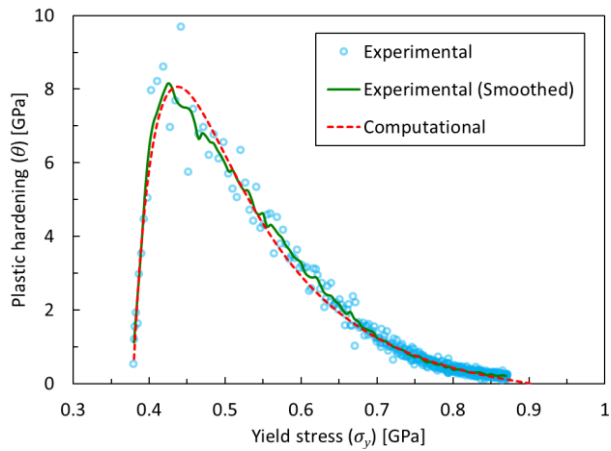

**(b)** Plastic hardening versus plastic strain

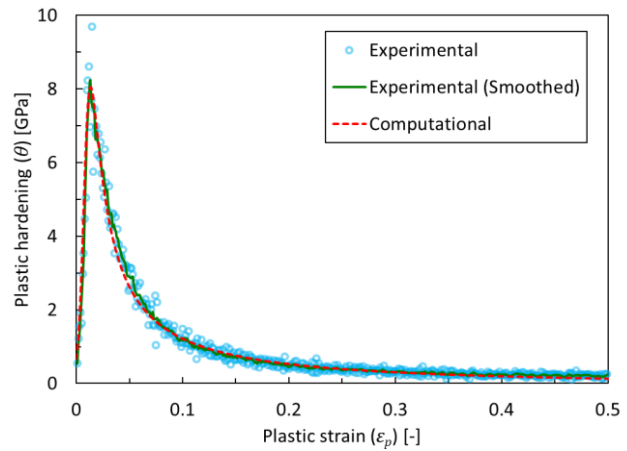

**Fig. 9.** Comparison between experimental and RMV-based computational plastic hardening curves of compression tests at constant reference temperature and strain rate ($T = 20$ °C and $\dot{\varepsilon}_0 = 0.01$ s$^{-1}$).

As plotted in Fig. 9, the plastic hardening increases abruptly at the beginning of deformation of the investigated material at reference temperature and strain rate. As shown in experimental flow curves depicted in Fig. 8, in cold regime and close to lower bound of intermediate regime, yield point elongation occurs which corresponds to this increasing domain in plastic hardening curves (Fig. 9) at small plastic strains. The proposed constitutive model has captured this effect. After reaching a maximum value, plastic hardening begins to decrease with a high rate. The rate of decrease of plastic hardening gradually decreases as the plastic strain continues to accumulate. As shown in Fig. 9 (a), the model predict saturation ($\theta = 0$) at yield stress of $\sigma_y^{sat} \approx 900$ MPa which corresponds to (accumulated) plastic strain of $\varepsilon_p^{sat} \approx 1.1$. The lowest plastic strain at which saturation occurs ($\varepsilon_p^{sat}$) usually has a decreasing trend by increasing temperature. As shown in Fig. 8 (c), for the investigated material in warm regime, the saturation starts at plastic strains of less than 0.4 ($\varepsilon_p^{sat} < 0.4$) which can be measured experimentally. However, for materials similar to the one investigated in the present paper (nearly all steel grades), in cold and intermediate regimes, generally it is not possible to experimentally measure the flow curves until the point of saturation with acceptable accuracy (using uniaxial tests). Therefore, in order to perform FE simulation of metal forming processes in cold and warm regimes with adequate precision, one has to predict/extrapolate the flow curves generally for relatively very large plastic strains (until saturation) for which there are not experimental measurements available. The microstructural constitutive model proposed in this paper can be trusted for such predictions because it can reproduce the experimental flow curves very accurately in the plastic strain range for which the experimental data is available. On the other hand, it has a comprehensive (statistical) physical background that covers all the major dislocation types and processes in cold and warm regimes.

## 5.2. Evolution of dislocation densities at constant strain rate

Numerically calculated evolution of dislocation densities based on RMV versus plastic strain in isothermal condition at various temperatures and reference strain rate ($\dot{\varepsilon}_0 = 0.01$ s$^{-1}$) are plotted in Fig. 10.

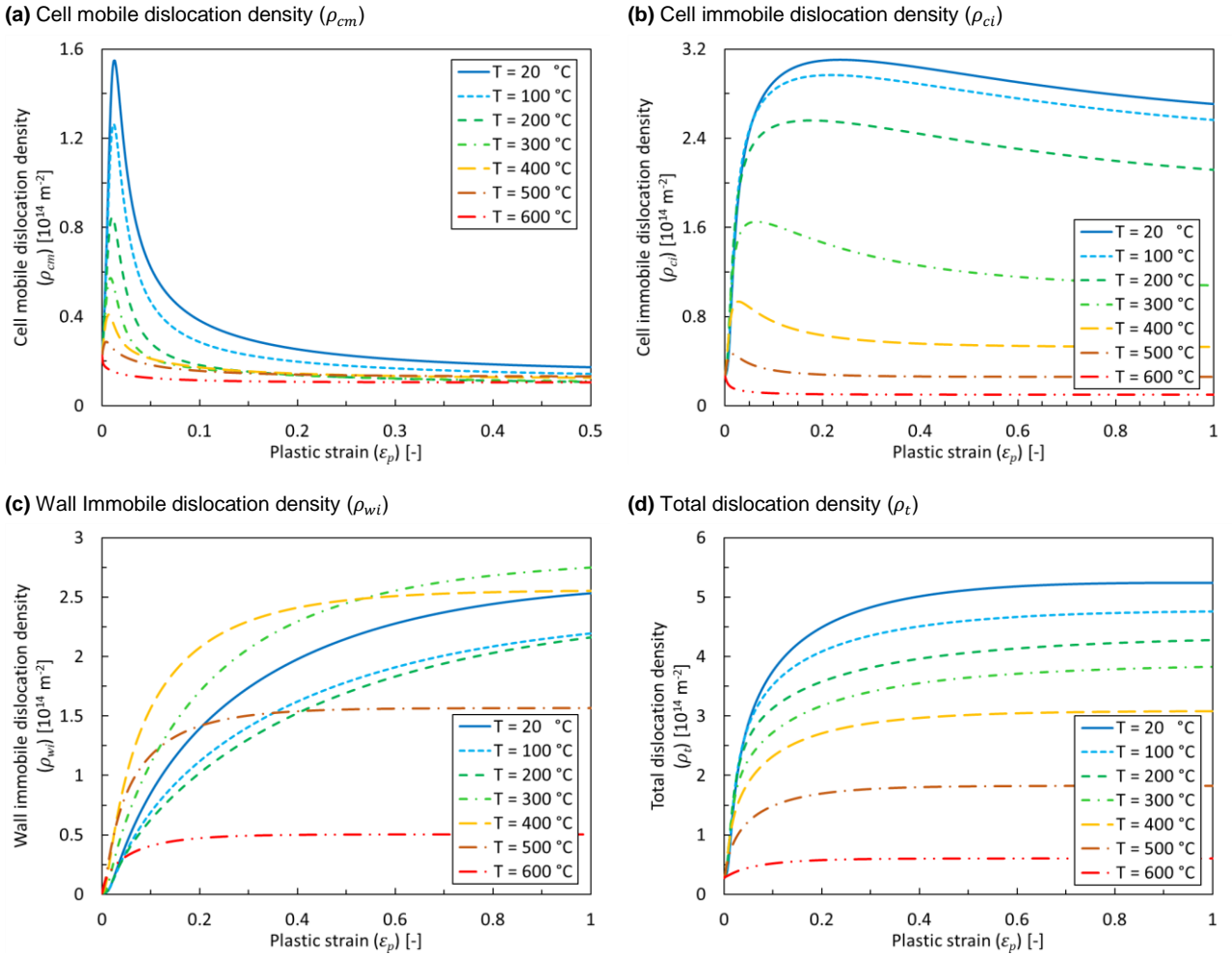

**Fig. 10.** RMV-based computational cell mobile, cell immobile, wall immobile, and total dislocation densities versus plastic strain in isothermal condition at various temperatures and reference strain rate ($\dot{\varepsilon}_0 = 0.01$ s$^{-1}$).

It is emphasized that by increasing temperature in cold and warm regimes, there is trend of reduction in cell mobile, cell immobile and total dislocation densities. However, similar to flow curves of intermediate temperature regime (Fig. 8 (b)), by increasing temperature wall immobile dislocation density increases, but again in warm regime it decreases. This is reasonable because the plastic stress associated with wall (immobile) dislocations ($\sigma_{pw}$) has the largest contribution to the yield stress (postulates (5) and (8)).

As demonstrated in Fig. 10 (b) and (c), in cold regime (0 °C < $T$ < 200 °C), generally (nonlocal) cell immobile dislocation density is higher than wall immobile dislocation density ($\rho_{ci} > \rho_{wi}$). In other words, as mentioned earlier in this section, this material tends to formation of cell immobile dislocation tangles (IDBs) at cold regime (sub-cell forming material). Nevertheless, by increasing temperature from 200 °C to 300 °C (intermediate regime), there is a sudden drop in cell immobile dislocation density and a (increasing) jump in wall immobile dislocation density. In this material, in the intermediate regime, (nonlocal) wall immobile dislocations have higher density than their cell immobile counterparts ($\rho_{ci} < \rho_{wi}$). This implies that, in warm regime, dislocation walls (GNDs) are becoming more frequent in the microstructure (of the investigated material) than cell dislocation pile-ups (IDBs). Temperature-dependent dynamic pinning associated with DSA as an underlying mechanism facilitating wall nucleation process (transforming cell dislocations to wall dislocations), is responsible for this effect. Due to contribution of dynamic pinning, as shown in Fig. 6 (d), at around 200 °C, frequency of wall nucleation process in the investigated material increases exponentially by increasing temperature. This results in transformation of more cell immobile to wall immobile dislocations. As Fig. 10 (d) suggests by increasing temperature, total dislocation density always decreases during plastic deformation at constant temperature and strain rate. At 600 °C which is close to the upper bound of warm regime for the investigated steel and therefore very close to the temperature of the onset of dynamic recrystallization (DRX), total dislocation density during deformation is almost constant. Furthermore, as shown in Fig. 10 (a), at the beginning of viscoplastic deformation, the undeformed/annealed material (low initial dislocation density), generates (cell) mobile dislocations with a remarkably high rate. High generation rate of mobile dislocations subsequently will lead to a maximum cell mobile dislocation density which occurs at relatively low accumulated plastic strains. This maximum is followed by a rapid drop in cell mobile dislocation density, very similar to the plastic hardening behavior (Fig. 9 (b)). As the plastic deformation proceeds, the rate of production of cell mobile dislocations becomes gradually lower until the saturation state at which the rate of production of cell mobile dislocations becomes equal to the rate of their reduction.

### 5.3. GND density measurements

In order to examine and validate the model's prediction of GND density which as mentioned in postulate (3) is equal to wall immobile dislocation density ($\rho_{wi}$), several compression specimens are deformed to different plastic strain at constant reference temperature and strain rate. Along the symmetry axis of a deformed compression specimen, there is a gradient of accumulated (equivalent) plastic strain ranging from a near zero value at the contact surface to a maximum of more than mean/macroscopic/true accumulated plastic strain of the deformed specimen at its center point. Hence, along the symmetry axis there is always a point at which equivalent plastic strain is equal to the mean plastic strain of the deformed specimen. For each of those deformed specimens, coordinates of such points are derived through corresponding FE simulations. It must be noted that there are other points with this property in the cross section of deformed compression specimens; however, the strain state at those points are much more complex (triaxial) compared to the points on the symmetry axis which are almost under uniaxial strain state. Moreover, it can be easily shown that those material points with the aforementioned property ($\bar{\varepsilon}_p = \varepsilon_p$), have also the same average and instantaneous equivalent plastic strain rate as the (constant) prescribed mean/macroscopic/true plastic strain rate of the deforming specimen ($\dot{\bar{\varepsilon}}_p = \dot{\varepsilon}_p$).

For each specimen, around the vicinity of such points (RMV), an EBSD sample is prepared. Sample preparation for EBSD involved standard mechanical polishing to 0.05 μm, followed by electropolishing in a 5% perchloric acid and 95% acetic acid solution (by volume) with an applied voltage of 35 V. Measurements are performed using a field emission gun scanning electron microscope (FEG-SEM), JOEL JSM 7000F, at 20 KeV beam energy, approximately 30 nA probe current, and 100-300 nm step size. A Hikari EBSD camera by Ametek-EDAX, in combination with the OIM software suite (OIM Data Collection and OIM Analysis v7.3) by EDAX-TSL, is used for data acquisition and analysis. Subsequently, at each point, GND density is calculated from kernel average misorientation (KAM) which is the average angular deviation between a point and its neighbors in a distance twice the step size as long as their misorientation does not exceed 5° (Calcagnotto et al., 2010). After mapping KAM values to GND density, over a representative area with the size of 100 × 100 μm, the average GND density is calculated.

Comparison between computationally predicted wall immobile dislocation density and the measured average GND density using high resolution EBSD in different plastic strains at reference condition ($T = 20\,°C$ and $\dot{\varepsilon}_0 = 0.01\ \mathrm{s^{-1}}$) is shown in Fig. 11.

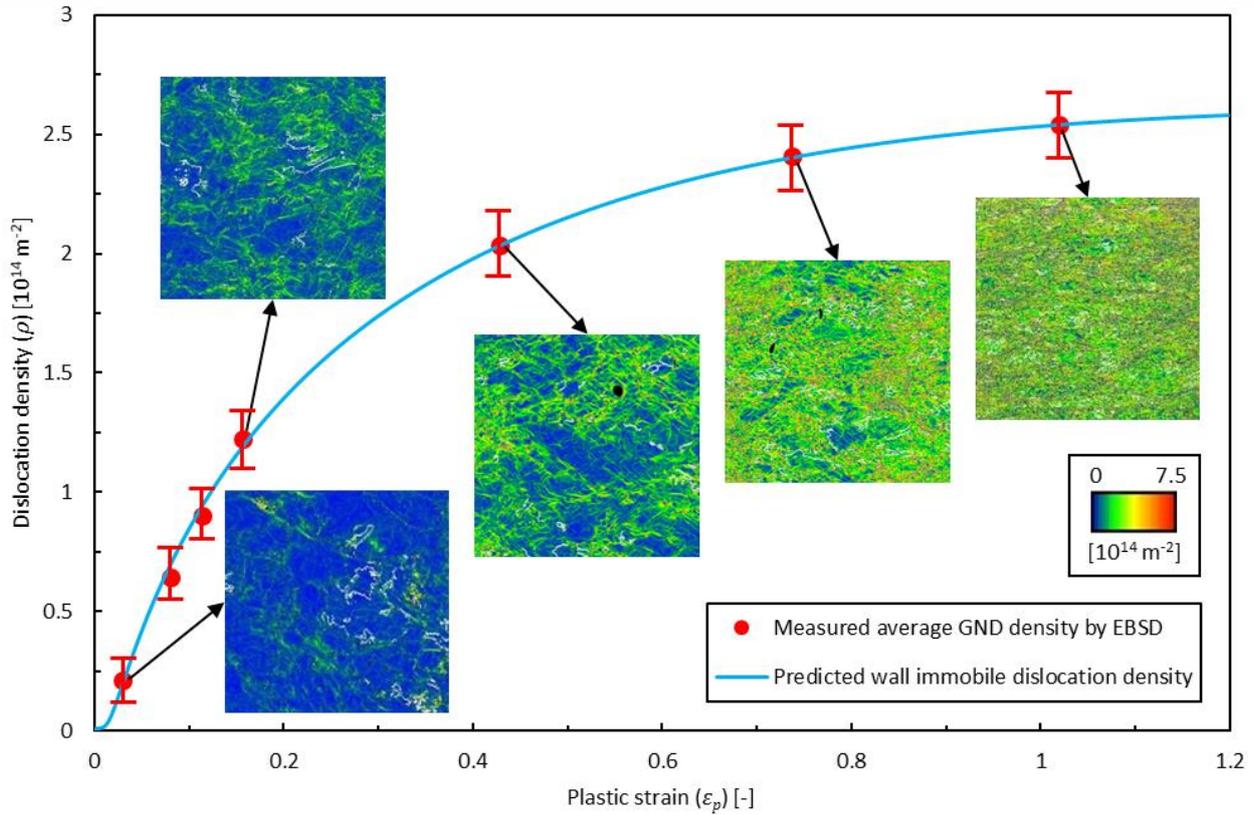

**Fig. 11.** Comparison between computationally predicted wall immobile dislocation density and measured GND density using high resolution EBSD in different compressive plastic strains at reference condition ($T = 20\,°C$ and $\dot{\varepsilon}_0 = 0.01\ \mathrm{s^{-1}}$) along with corresponding EBSD images with GND density distribution (calculated from KAM data) over the representative area.

### 5.4. Multistep uniaxial tests

As mentioned in postulate (1), "*Two identical material samples that are plastically deformed to an equal amount of accumulated plastic strain but with different histories of temperature and strain rate, if again deformed under an equal temperature and strain rate condition, do not necessarily yield the same stress response*". In order to test this statement and further validate the history dependence of the presented microstructural constitutive model, the multistep compression tests are devised. In these tests, first, previously undeformed compression specimens are plastically deformed to a predefined plastic strain in constant temperature $T_i$ with low reference strain rate to make sure temperature remains almost constant during compression (initial compression step). In the next step (final compression step), they are cooled down and again deformed in constant room temperature at the reference strain rate to a certain plastic strain. This cycle is illustrated in the schematic time-temperature diagram shown in Fig. 12 (a). The initial compression step is performed at three constant temperatures, $T_i = 20, 300$ and $400\,°C$ until the accumulated plastic strain $\varepsilon_p = 0.2$. In Fig. 12 (b), flow curves derived from the final compression steps are compared to their corresponding RMV-based computational flow curves predicted by the constitutive model.

All three flow curves shown in Fig. 12 (b) are material responses (yield stress) under plastic deformation at identical constant reference temperature and strain rate with the same initial accumulated plastic strain $\varepsilon_{p0} = 0.2$. However, for each of them, the initial plastic strain was accumulated in different temperatures (different histories of plastic strain accumulation). As shown in Fig. 12 (b), initial yield stress is slightly higher while plastic hardening is considerably much lower for $T_i = 300$ and $400\,°C$ compared to those of $T_i = 20\,°C$. The flow curves related to $T_i = 300$ and $400\,°C$ in the investigated plastic strain domain exhibit yield point elongation (Lüders bands) as they are convex ($\dot{\theta} \equiv \partial^2\sigma_y / \partial\varepsilon_p^2 \geq 0$) instead of typical concave flow curves ($\dot{\theta} \equiv \partial^2\sigma_y / \partial\varepsilon_p^2 \leq 0$), and they have relatively large domain of nonpositive plastic hardening ($\theta \equiv \partial\sigma_y / \partial\varepsilon_p \leq 0$) or low plastic hardening (Pham et al., 2015).

**(a)** Schematic diagram of multistep compression tests

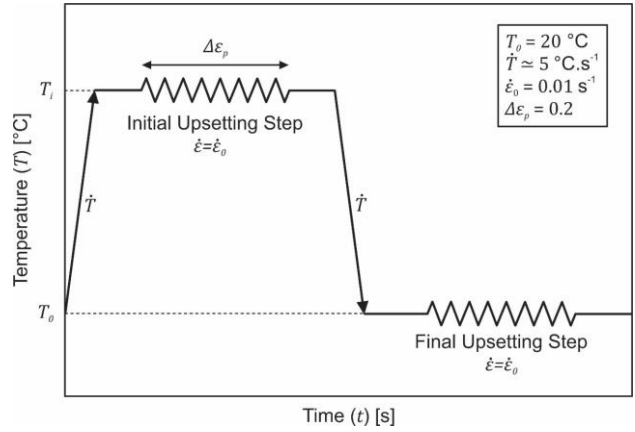

**(b)** Flow curves of final compression step

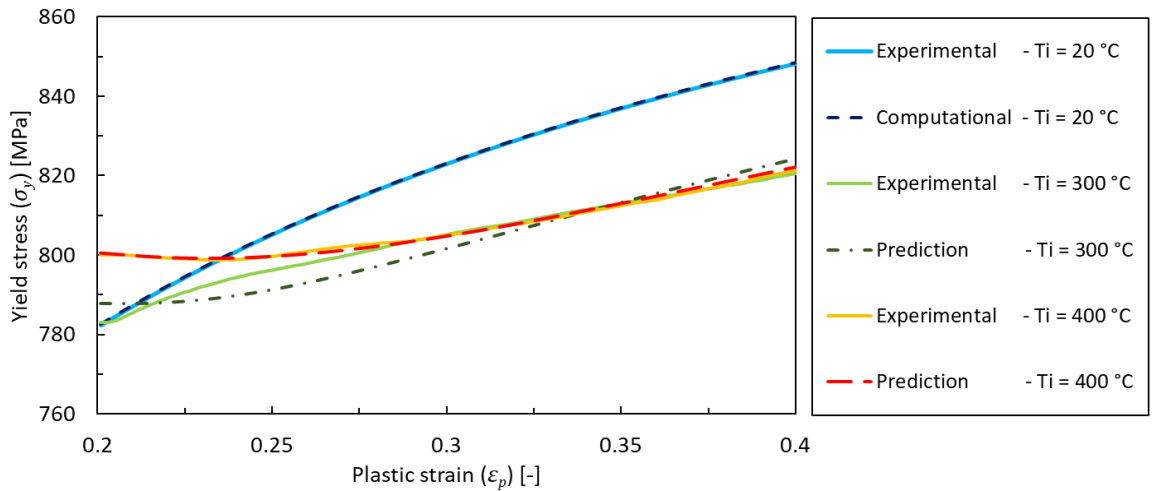

**Fig. 12.** Schematic time-temperature diagram of multistep compression tests and comparison between corresponding experimental and RMV-based computationally predicted flow curves of final compression step.

### 5.5. Nonisothermal uniaxial tests

Nonisothermal compression tests and their respective FE simulations at various strain rates (in intermediate-rate regime) and different initial temperatures (in cold and warm regimes) are conducted for the second step of parameter identification to determine constitutive parameters associated with strain rate sensitivity and dissipation factor. In order to do so, first the presented constitutive model is implemented as a user-defined material subroutine UMAT in ABAQUS/Standard with the material constants identified in the RMV-based fitting step with an initial guess for the parameters yet to be determined in the second step of parameter identification using FEMU technique. Then, thermo-mechanical FE model of compression test with the nominal dimensions of experimental specimens and tools has been created in the implicit ABAQUS/Standard software. Mechanical and physical properties of compression material and tools such as elastic modulus, poisson's ratio, mass density, specific heat capacity, thermal conductivity, and thermal expansion are inputted to the FE model as functions of temperature. Convection and radiation along with thermal contact conductance as a function of pressure of contact interface of compression specimen and tools are also considered. In addition, simple coulomb friction model with friction coefficient of 0.05 corresponding to the experimental condition was used. As mentioned, details of FE implementation of the constitutive model and thermo-mechanical FE simulation of an industrial multistep warm forging of a bevel gear for the same material investigated here are planned to be published later.

Fig. 13 shows FE-simulated distribution of statistical TMM variables, equivalent stress ($\bar{\sigma}$), temperature ($T$), equivalent plastic strain rate ($\dot{\hat{\varepsilon}}_p$), cell mobile dislocation density ($\rho_{cm}$), cell immobile dislocation density ($\rho_{ci}$), and wall immobile dislocation density ($\rho_{wi}$), in radial cross section of compression specimen during compressive plastic deformation at mean plastic strain of $\varepsilon_p = 0.6$ and strain rate of $\dot{\varepsilon} = 1 \text{ s}^{-1}$ with initial temperature of $T_0 = 20 \text{ °C}$. Experimental and FE-simulated yield stress and temperature change ($\Delta T = T - T_0$) due to plastic work for compression tests at different strain rates and initial room temperature are shown in Fig. 14.

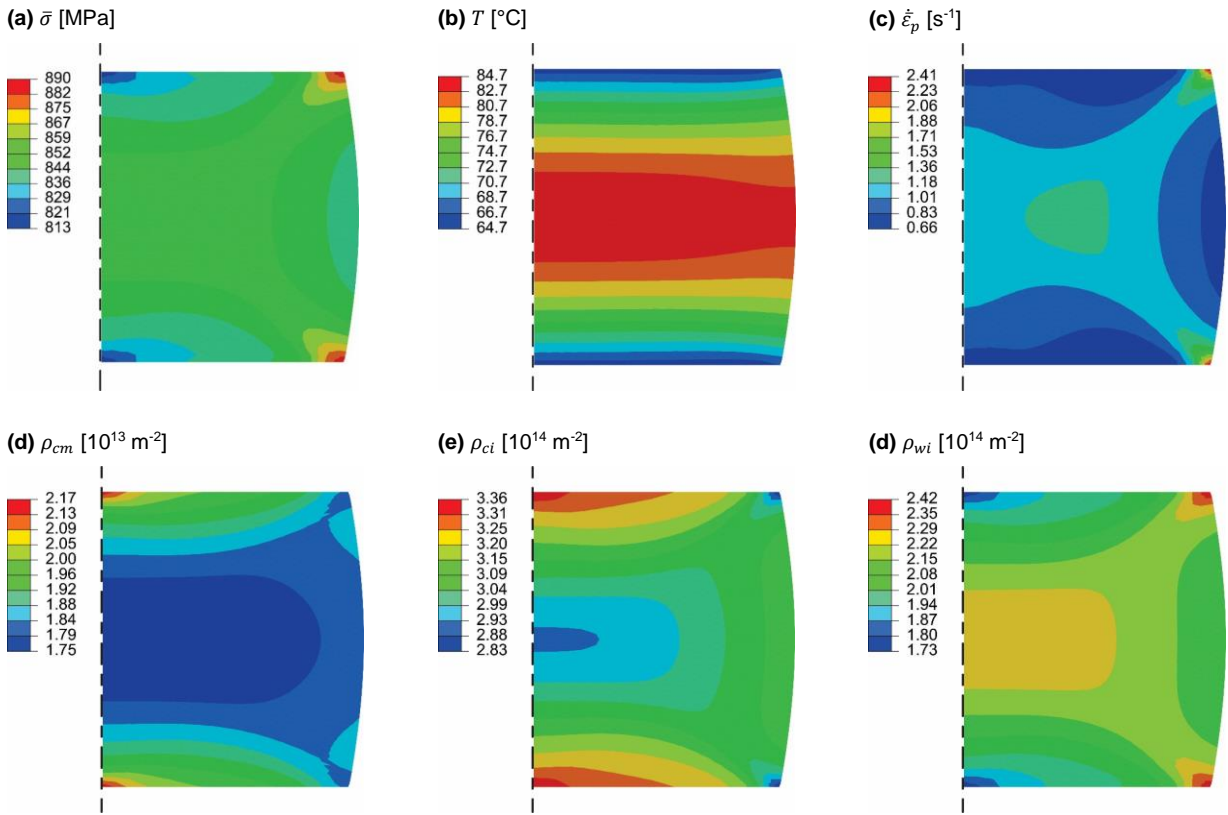

**Fig. 13.** FE-simulated distribution of statistical TMM variables including MSVs in a radial section of compression specimen during deformation at mean plastic strain of $\varepsilon_p = 0.6$ and mean strain rate of $\dot{\varepsilon} = 1$ s$^{-1}$ with initial temperature of $T_0 = 20$ °C.

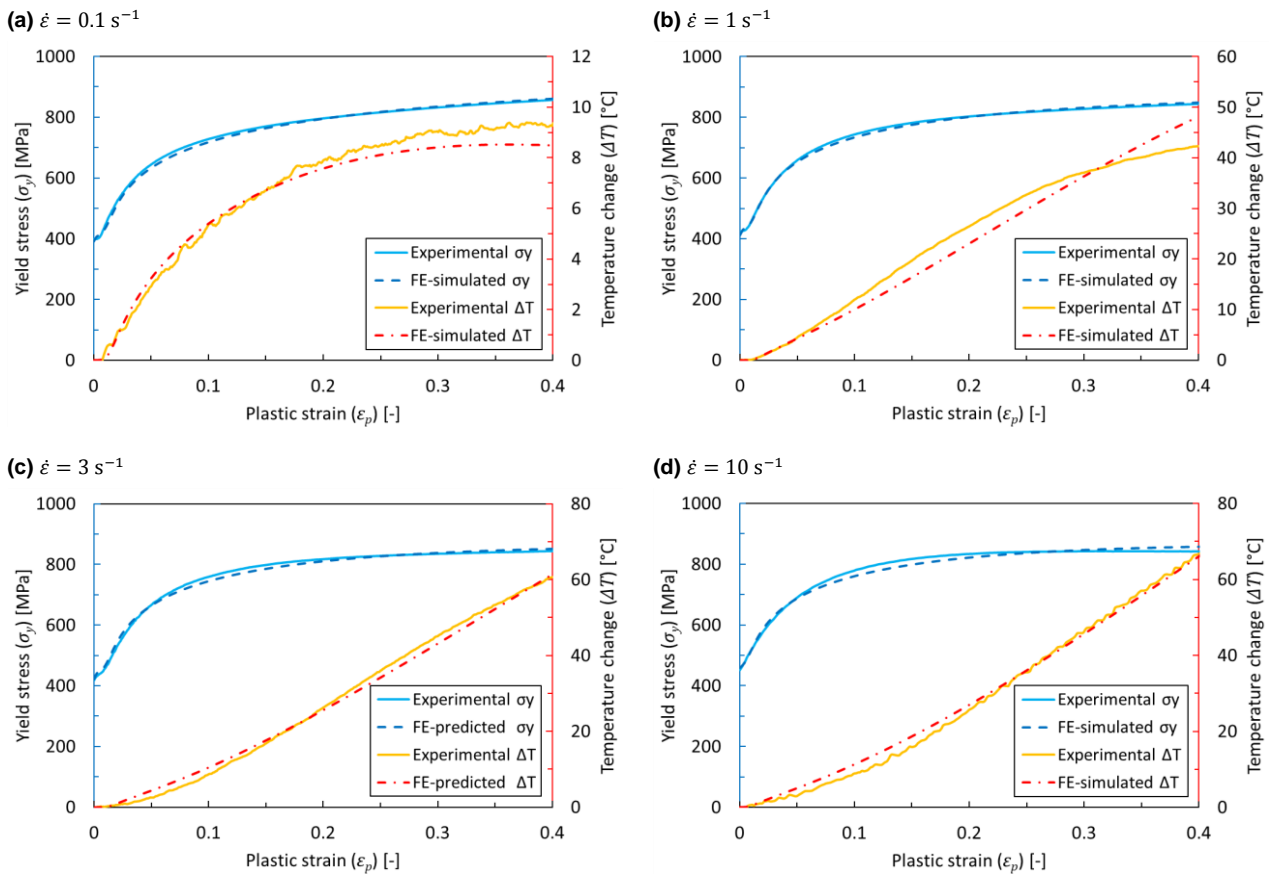

**Fig. 14.** Comparison between experimental and FE-simulated yield stress and temperature change due to plastic work ($\Delta T = T - T_0$) for compression tests at different strain rates ($\dot{\varepsilon} = 0.1, 1, 3, 10$ s$^{-1}$) and initial room temperature ($T_0 = 20$ °C).

In each experimental compression test corresponding to Fig. 14, temperature is measured via a thermocouple welded on the middle point in height direction at the lateral surface of the compression specimen. Likewise, in FE simulation the temperature is read from the respective point. The experimental data (force and temperature versus displacement) obtained from the nonisothermal compression tests at constant strain rates of 0.1, 1.0 and 10.0 $s^{-1}$ are supplied to the second step of parameter identification by FEMU procedure. As shown in Fig. 14, for validation of the model's rate-dependent features, the experimental data corresponding to the nonisothermal compression test at constant strain rate of 3.0 $s^{-1}$ is compared to its counterpart predicted by the FE simulation.

As inferred from Fig. 14, at relatively high strain rates in the investigated rate regimes, since there is not enough time for the generated adiabatic heat to dissipate completely, it remain in the material; and as a result elevates the temperature which in turn leads to spontaneous DTS. Nonisothermal FE simulations of metal forming processes highly depend on the input thermo-physical properties of billet/blank and tools, their pressure-dependent friction and thermal contact conductance, as well as convection and radiation heat transfer with the ambient environment which all of them are temperature-dependent. Hence, large portion of the difference in temperature readings from these FE-simulated nonisothermal compression (virtual) tests compared to their experimental counterparts is attributed to the complexity of the thermomechanical FE simulation of metal forming that is independent from the constitutive model. In addition, the accuracy of experimental temperature measurement with welded thermocouples is highly dependent on the weld quality and position, as well as the thermocouple delay. However, the discrepancy between FE-simulated and experimental temperature increase is much less than expected.

## 6. Outlook

As every continuum plasticity model has limitations, despite its vast scope and outstanding accuracy, the presented constitutive model in this paper is not an exception:

1) It has been tested thus far only for cold and warm regimes. Therefore, application or extension of it to hot temperature regime demands further research.
2) It has been validated up to now, merely for the intermediate-rate regime. Its validity for other strain rate regimes as well as static state must be verified. Moreover, its overall strain rate dependence needs more tests.
3) In its current state, the model does not account for anisotropy and Bauschinger effect. In order to generalize it to account for such effects, some new postulates need to be added while probably some of the existing ones require slight modification. Subsequently, for validation, the model must be implemented in the crystal plasticity framework.
4) It does not account for strain (rate) localization due to pressure-dependent evolution (nucleation and growth) of void volume fraction. Incorporation of strain localization in the constitutive model is crucial for its application to simulation of metal forming processes, as forming operations of especially sheet metal usually need to be designed in such a way to avoid or minimize localized shearing, as this non-uniform deformation is believed to induce notable damage and microstructural heterogeneity in the final product, which is generally considered detrimental to its in-service performance.

Furthermore, the introduced microstructural constitutive model requires further validation using different alloys with various compositions, microstructures and grain size distributions.

## 7. Conclusion

A fully coupled thermo-micro-mechanical constitutive model for isotropic viscoplasticity of polycrystalline metallic materials was proposed based on continuum dislocation dynamics. The constitutive model was designed in order to statistically capture and quantify all statistically-considerable microstructural phenomena influencing dislocations in cold and warm regimes and hence accurately describe macroscopic viscoplastic response of material. The following conclusions are drawn from the present approach:

- The mechanical response of a macroscale material point under viscoplastic deformation is an implicit function of its stochastic microstructural state, plastic strain rate, and temperature. The presented constitutive model is history dependent and it does not depend on accumulated plastic strain. Multistep compression tests were conducted for verification of the accuracy of model's history dependence features.
- Dynamic and static evolutions of different types of dislocations are functions of dynamic and static pinning of dislocations by impurity atoms. Hence, the proposed constitutive model accounts for the effects associated with pinning such as yield point elongation, dynamic strain aging, static aging and bake hardening. However, validation of static evolution of dislocation densities suggested by four postulates is a subject of another paper.

- The constitutive model predicts flow curves at various temperatures and strain rates in cold and warm regimes with an exceptionally high accuracy (mean error of less than 0.4% which is in the same range of experimental scatter). Moreover, the evolution of measured GND density (by EBSD) is very close to that of predicted by the model.
- Comparison of FE simulations of nonisothermal compression tests at various strain rates with their corresponding compression experiments verifies that the postulates suggested for model's rate sensitivity and varying dissipation factor in adiabatic heat generation (due to plastic work) are realistic.
- Implementation of the presented constitutive model as user-defined material subroutines UMAT and VUMAT respectively, in ABAQUS/Standard and ABAQUS/Explicit based on associative isotropic $J_2$ hypoelasto-viscoplasticity, has revealed that the constitutive model is computationally efficient (simulation costs are in the same level as empirical models) in thermo-mechanical FE simulations of cold and warm metal forming processes.

## Acknowledgement

Authors appreciate the support received under the project "IGF-Vorhaben 18531N" in the framework of research program of "Integrierte Umform und Wärmebehandlungssimulation für Massivumformteile (InUWäM)" funded by the German federation of industrial research associations (AiF). The authors also wish to express their gratitude for the support received from the project "Laserunterstütztes Kragenziehen hochfester Bleche" by the research association EFB e.V. funded under the number 18277N by AiF. The help of Mr. Konstantin Schacht with experiments and valuable suggestions of Prof. Wolfgang Bleck and Prof. Wolfgang Blum are gratefully acknowledged.

## References

Abu Al-Rub, R.K., Voyiadjis, G.Z., 2004. Analytical and experimental determination of the material intrinsic length scale of strain gradient plasticity theory from micro- and nano-indentation experiments. International Journal of Plasticity 20, 1139–1182. 10.1016/j.ijplas.2003.10.007.

Adams, B.L., Olson, T., 1998. The mesostructure—properties linkage in polycrystals. Progress in Materials Science 43, 1–87. 10.1016/S0079-6425(98)00002-4.

Allain, S., Chateau, J.-P., Bouaziz, O., 2004. A physical model of the twinning-induced plasticity effect in a high manganese austenitic steel. Materials Science and Engineering: A 387-389, 143–147. 10.1016/j.msea.2004.01.060.

Amodeo, R.J., Ghoniem, N.M., 1990. Dislocation dynamics. I. A proposed methodology for deformation micromechanics. Phys. Rev. B 41, 6958–6967. 10.1103/PhysRevB.41.6958.

Ananthakrishna, G., 2007. Current theoretical approaches to collective behavior of dislocations. Physics Reports 440, 113–259. 10.1016/j.physrep.2006.10.003.

Ananthakrishna, G., Sahoo, D., 1981. A model based on nonlinear oscillations to explain jumps on creep curves. J. Phys. D: Appl. Phys. 14, 2081–2090. 10.1088/0022-3727/14/11/015.

Ardeljan, M., Beyerlein, I.J., Knezevic, M., 2014. A dislocation density based crystal plasticity finite element model: Application to a two-phase polycrystalline HCP/BCC composites. Journal of the Mechanics and Physics of Solids 66, 16–31. 10.1016/j.jmps.2014.01.006.

Argon, A.S., 2012. Strengthening mechanisms in crystal plasticity. Oxford University Press, Oxford, 404 S.

Argon, A.S., Moffatt, W.C., 1981. Climb of extended edge dislocations. Acta Metallurgica 29, 293–299. 10.1016/0001-6160(81)90156-5.

Arsenlis, A., 2004. On the evolution of crystallographic dislocation density in non-homogeneously deforming crystals. Journal of the Mechanics and Physics of Solids 52, 1213–1246. 10.1016/j.jmps.2003.12.007.

Arsenlis, A., Cai, W., Tang, M., Rhee, M., Oppelstrup, T., Hommes, G., Pierce, T.G., Bulatov, V.V., 2007. Enabling strain hardening simulations with dislocation dynamics. Modelling Simul. Mater. Sci. Eng. 15, 553–595. 10.1088/0965-0393/15/6/001.

Arsenlis, A., Parks, D., 1999. Crystallographic aspects of geometrically-necessary and statistically-stored dislocation density. Acta Materialia 47, 1597–1611. 10.1016/S1359-6454(99)00020-8.

Arsenlis, A., Parks, D.M., 2002. Modeling the evolution of crystallographic dislocation density in crystal plasticity. Journal of the Mechanics and Physics of Solids 50, 1979–2009. 10.1016/S0022-5096(01)00134-X.

Ashby, M.F., 1970. The deformation of plastically non-homogeneous materials. Philosophical Magazine 21, 399–424. 10.1080/14786437008238426.

Askari, H., Maughan, M.R., Abdolrahim, N., Sagapuram, D., Bahr, D.F., Zbib, H.M., 2015. A stochastic crystal plasticity framework for deformation of micro-scale polycrystalline materials. International Journal of Plasticity 68, 21–33. 10.1016/j.ijplas.2014.11.001.

Askari, H., Young, J., Field, D., Kridli, G., Li, D., Zbib, H., 2013. A study of the hot and cold deformation of twin-roll cast magnesium alloy AZ31. Philosophical Magazine 94, 381–403. 10.1080/14786435.2013.853884.

Austin, R.A., McDowell, D.L., 2011. A dislocation-based constitutive model for viscoplastic deformation of FCC metals at very high strain rates. International Journal of Plasticity 27, 1–24. 10.1016/j.ijplas.2010.03.002.

Babu, B., Lindgren, L.-E., 2013. Dislocation density based model for plastic deformation and globularization of Ti-6Al-4V. International Journal of Plasticity 50, 94–108. 10.1016/j.ijplas.2013.04.003.

Bailey, J.E., Hirsch, P.B., 1960. The dislocation distribution, flow stress, and stored energy in cold-worked polycrystalline silver. Philosophical Magazine 5, 485–497. 10.1080/14786436008238300.

Bammann, D.J., Aifantis, E.C., 1982. On a proposal for a continuum with microstructure. Acta Mechanica 45, 91–121. 10.1007/BF01295573.

Bardella, L., 2006. A deformation theory of strain gradient crystal plasticity that accounts for geometrically necessary dislocations. Journal of the Mechanics and Physics of Solids 54, 128–160. 10.1016/j.jmps.2005.08.003.

Bay, B., Hansen, N., Hughes, D.A., Kuhlmann-Wilsdorf, D., 1992. Overview no. 96 evolution of f.c.c. deformation structures in polyslip. Acta Metallurgica et Materialia 40, 205–219. 10.1016/0956-7151(92)90296-Q.

Bay, B., Hansen, N., Kuhlmann-Wilsdorf, D., 1989. Deformation structures in lightly rolled pure aluminium. Materials Science and Engineering: A 113, 385–397. 10.1016/0921-5093(89)90325-0.

Bergström, Y., 1970. A dislocation model for the stress-strain behaviour of polycrystalline α-Fe with special emphasis on the variation of the densities of mobile and immobile dislocations. Materials Science and Engineering 5, 193–200. 10.1016/0025-5416(70)90081-9.

Beyerlein, I.J., Tomé, C.N., 2008. A dislocation-based constitutive law for pure Zr including temperature effects. International Journal of Plasticity 24, 867–895. 10.1016/j.ijplas.2007.07.017.

Bhushan, B., Nosonovsky, M., 2003. Scale effects in friction using strain gradient plasticity and dislocation-assisted sliding (microslip). Acta Materialia 51, 4331–4345. 10.1016/S1359-6454(03)00261-1.

Blum, W., Eisenlohr, P., Breutinger, F., 2002. Understanding creep—a review. Metall and Mat Trans A 33, 291–303. 10.1007/s11661-002-0090-9.

Bok, H.-H., Choi, J., Barlat, F., Suh, D.W., Lee, M.-G., 2014. Thermo-mechanical-metallurgical modeling for hot-press forming in consideration of the prior austenite deformation effect. International Journal of Plasticity 58, 154–183. 10.1016/j.ijplas.2013.12.002.

Bouaziz, O., Guelton, N., 2001. Modelling of TWIP effect on work-hardening. Materials Science and Engineering: A 319-321, 246–249. 10.1016/S0921-5093(00)02019-0.

Brinckmann, S., Siegmund, T., Huang, Y., 2006. A dislocation density based strain gradient model. International Journal of Plasticity 22, 1784–1797. 10.1016/j.ijplas.2006.01.005.

Brown, L.M., 2002. A dipole model for the cross-slip of screw dislocations in fcc metals. Philosophical Magazine A 82, 1691–1711. 10.1080/01418610208235684.

Busso, E., 2000. Gradient-dependent deformation of two-phase single crystals. Journal of the Mechanics and Physics of Solids 48, 2333–2361. 10.1016/S0022-5096(00)00006-5.

Busso, E.P., 1998. A continuum theory for dynamic recrystallization with microstructure-related length scales. International Journal of Plasticity 14, 319–353. 10.1016/S0749-6419(98)00008-4.

Byun, T.S., 2003. On the stress dependence of partial dislocation separation and deformation microstructure in austenitic stainless steels. Acta Materialia 51, 3063–3071. 10.1016/S1359-6454(03)00117-4.

Calcagnotto, M., Ponge, D., Demir, E., Raabe, D., 2010. Orientation gradients and geometrically necessary dislocations in ultrafine grained dual-phase steels studied by 2D and 3D EBSD. Materials Science and Engineering: A 527, 2738–2746. 10.1016/j.msea.2010.01.004.

Cereceda, D., Diehl, M., Roters, F., Raabe, D., Perlado, J.M., Marian, J., 2016. Unraveling the temperature dependence of the yield strength in single-crystal tungsten using atomistically-informed crystal plasticity calculations. International Journal of Plasticity 78, 242–265. 10.1016/j.ijplas.2015.09.002.

Chandra, S., Samal, M.K., Chavan, V.M., Raghunathan, S., 2018. Hierarchical multiscale modeling of plasticity in copper: From single crystals to polycrystalline aggregates. International Journal of Plasticity 101, 188–212. 10.1016/j.ijplas.2017.10.014.

Cheong, K.-S., Busso, E.P., 2004. Discrete dislocation density modelling of single phase FCC polycrystal aggregates. Acta Materialia 52, 5665–5675. 10.1016/j.actamat.2004.08.044.

Christian, J.W., Mahajan, S., 1995. Deformation twinning. Progress in Materials Science 39, 1–157. 10.1016/0079-6425(94)00007-7.

Clayton, J.D., McDowell, D.L., Bammann, D.J., 2006. Modeling dislocations and disclinations with finite micropolar elastoplasticity. International Journal of Plasticity 22, 210–256. 10.1016/j.ijplas.2004.12.001.

Columbus, D., Grujicic, M., 2002. A comparative discrete-dislocation/nonlocal crystal-plasticity analysis of plane-strain mode I fracture. Materials Science and Engineering: A 323, 386–402. 10.1016/S0921-5093(01)01397-1.

Cottrell, A.H., 1953. Dislocations and Plastic Flow in Crystals. Clarendon Press, Oxford.

Cottrell, A.H., Bilby, B.A., 1949. Dislocation Theory of Yielding and Strain Ageing of Iron. Proc. Phys. Soc. A 62, 49–62. 10.1088/0370-1298/62/1/308.

Csikor, F.F., Motz, C., Weygand, D., Zaiser, M., Zapperi, S., 2007. Dislocation avalanches, strain bursts, and the problem of plastic forming at the micrometer scale. Science (New York, N.Y.) 318, 251–254. 10.1126/science.1143719.

Cui, Y., Po, G., Ghoniem, N., 2016a. Controlling Strain Bursts and Avalanches at the Nano- to Micrometer Scale. Physical review letters 117, 155502. 10.1103/PhysRevLett.117.155502.

Cui, Y., Po, G., Ghoniem, N., 2016b. Temperature insensitivity of the flow stress in body-centered cubic micropillar crystals. Acta Materialia 108, 128–137. 10.1016/j.actamat.2016.02.008.

Cui, Y., Po, G., Ghoniem, N., 2017. Influence of loading control on strain bursts and dislocation avalanches at the nanometer and micrometer scale. Phys. Rev. B 95. 10.1103/PhysRevB.95.064103.

Derlet, P.M., Maaß, R., 2013. Micro-plasticity and intermittent dislocation activity in a simplified micro-structural model. Modelling Simul. Mater. Sci. Eng. 21, 35007. 10.1088/0965-0393/21/3/035007.

Devincre, B., Kubin, L., Lemarchand, C., Madec, R., 2001. Mesoscopic simulations of plastic deformation. Materials Science and Engineering: A 309-310, 211–219. 10.1016/S0921-5093(00)01725-1.

Doherty, R.D., Hughes, D.A., Humphreys, F.J., Jonas, J.J., Jensen, D., Kassner, M.E., King, W.E., McNelley, T.R., McQueen, H.J., Rollett, A.D., 1997. Current issues in recrystallization: A review. Materials Science and Engineering: A 238, 219–274. 10.1016/S0921-5093(97)00424-3.

Eisenlohr, P., Blum, W., 2005. Bridging steady-state deformation behavior at low and high temperature by considering dislocation dipole annihilation. Materials Science and Engineering: A 400-401, 175–181. 10.1016/j.msea.2005.01.069.

El-Awady, J.A., Wen, M., Ghoniem, N.M., 2009. The role of the weakest-link mechanism in controlling the plasticity of micropillars. Journal of the Mechanics and Physics of Solids 57, 32–50. 10.1016/j.jmps.2008.10.004.

Eringen, A., 1983. Theories of nonlocal plasticity. International Journal of Engineering Science 21, 741–751. 10.1016/0020-7225(83)90058-7.

Essmann, U., Mughrabi, H., 1979. Annihilation of dislocations during tensile and cyclic deformation and limits of dislocation densities. Philosophical Magazine A 40, 731–756. 10.1080/01418617908234871.

Estrin, Y., Braasch, H., Brechet, Y., 1996. A Dislocation Density Based Constitutive Model for Cyclic Deformation. J. Eng. Mater. Technol. 118, 441. 10.1115/1.2805940.

Estrin, Y., Kubin, L.P., 1986. Local strain hardening and nonuniformity of plastic deformation. Acta Metallurgica 34, 2455–2464. 10.1016/0001-6160(86)90148-3.

Estrin, Y., Mecking, H., 1984. A unified phenomenological description of work hardening and creep based on one-parameter models. Acta Metallurgica 32, 57–70. 10.1016/0001-6160(84)90202-5.

Estrin, Y., Mecking, H., 1992. A remark in connection with 'direct versus indirect dispersion hardening'. Scripta Metallurgica et Materialia 27, 647–648. 10.1016/0956-716X(92)90355-I.

Estrin, Y., Tóth, L.S., Molinari, A., Bréchet, Y., 1998. A dislocation-based model for all hardening stages in large strain deformation. Acta Materialia 46, 5509–5522. 10.1016/S1359-6454(98)00196-7.

Evers, L., Brekelmans, W., Geers, M., 2004. Scale dependent crystal plasticity framework with dislocation density and grain boundary effects. International Journal of Solids and Structures 41, 5209–5230. 10.1016/j.ijsolstr.2004.04.021.

Fan, X.G., Yang, H., 2011. Internal-state-variable based self-consistent constitutive modeling for hot working of two-phase titanium alloys coupling microstructure evolution. International Journal of Plasticity 27, 1833–1852. 10.1016/j.ijplas.2011.05.008.

Field, J.E., Walley, S.M., Proud, W.G., Goldrein, H.T., Siviour, C.R., 2004. Review of experimental techniques for high rate deformation and shock studies. International Journal of Impact Engineering 30, 725–775. 10.1016/j.ijimpeng.2004.03.005.

Fleck, N.A., Hutchinson, J.W., 1997. Strain Gradient Plasticity, in: vol. 33. Elsevier, pp. 295–361.

Fleck, N.A., Muller, G.M., Ashby, M.F., Hutchinson, J.W., 1994. Strain gradient plasticity: Theory and experiment. Acta Metallurgica et Materialia 42, 475–487. 10.1016/0956-7151(94)90502-9.

Fleischer, R.L., 1962. Rapid Solution Hardening, Dislocation Mobility, and the Flow Stress of Crystals. Journal of Applied Physics 33, 3504–3508. 10.1063/1.1702437.

Follansbee, P.S., Kocks, U.F., 1988. A constitutive description of the deformation of copper based on the use of the mechanical threshold stress as an internal state variable. Acta Metallurgica 36, 81–93. 10.1016/0001-6160(88)90030-2.

Franciosi, P., 1985. The concepts of latent hardening and strain hardening in metallic single crystals. Acta Metallurgica 33, 1601–1612. 10.1016/0001-6160(85)90154-3.

Frank, F.C., Read, W.T., 1950. Multiplication Processes for Slow Moving Dislocations. Phys. Rev. 79, 722–723. 10.1103/PhysRev.79.722.

Galindo-Nava, E.I., Rae, C., 2016. Microstructure-sensitive modelling of dislocation creep in polycrystalline FCC alloys: Orowan theory revisited. Materials Science and Engineering: A 651, 116–126. 10.1016/j.msea.2015.10.088.

Galindo-Nava, E.I., Sietsma, J., Rivera-Díaz-del-Castillo, P., 2012. Dislocation annihilation in plastic deformation: II. Kocks–Mecking Analysis. Acta Materialia 60, 2615–2624. 10.1016/j.actamat.2012.01.028.

Gao, H., 1999. Mechanism-based strain gradient plasticity: I. Theory. Journal of the Mechanics and Physics of Solids 47, 1239–1263. 10.1016/S0022-5096(98)00103-3.

Gao, H., Huang, Y., 2001. Taylor-based nonlocal theory of plasticity. International Journal of Solids and Structures 38, 2615–2637. 10.1016/S0020-7683(00)00173-6.

Gao, H., Huang, Y., 2003. Geometrically necessary dislocation and size-dependent plasticity. Scripta Materialia 48, 113–118. 10.1016/S1359-6462(02)00329-9.

Gao, H., Huang, Y., Nix, W.D., 1999. Modeling Plasticity at the Micrometer Scale. Naturwissenschaften 86, 507–515. 10.1007/s001140050665.

Gardner, C.J., Adams, B.L., Basinger, J., Fullwood, D.T., 2010. EBSD-based continuum dislocation microscopy. International Journal of Plasticity 26, 1234–1247. 10.1016/j.ijplas.2010.05.008.

Ghosh, G., Olson, G.B., 2002. The isotropic shear modulus of multicomponent Fe-base solid solutions. Acta Materialia 50, 2655–2675. 10.1016/S1359-6454(02)00096-4.

Gilbert, M.R., Queyreau, S., Marian, J., 2011. Stress and temperature dependence of screw dislocation mobility in α-Fe by molecular dynamics. Phys. Rev. B 84, 1052. 10.1103/PhysRevB.84.174103.

Gilman, J.J., 1965. Dislocation Mobility in Crystals. Journal of Applied Physics 36, 3195–3206. 10.1063/1.1702950.

Gottstein, G., Argon, A.S., 1987. Dislocation theory of steady state deformation and its approach in creep and dynamic tests. Acta Metallurgica 35, 1261–1271. 10.1016/0001-6160(87)90007-1.

Greer, J.R., Weinberger, C.R., Cai, W., 2008. Comparing the strength of f.c.c. and b.c.c. sub-micrometer pillars: Compression experiments and dislocation dynamics simulations. Materials Science and Engineering: A 493, 21–25. 10.1016/j.msea.2007.08.093.

Groh, S., Marin, E.B., Horstemeyer, M.F., Zbib, H.M., 2009. Multiscale modeling of the plasticity in an aluminum single crystal. International Journal of Plasticity 25, 1456–1473. 10.1016/j.ijplas.2008.11.003.

Gu, Y., Xiang, Y., Quek, S.S., Srolovitz, D.J., 2015. Three-dimensional formulation of dislocation climb. Journal of the Mechanics and Physics of Solids 83, 319–337. 10.1016/j.jmps.2015.04.002.

Hahn, G., 1962. A model for yielding with special reference to the yield-point phenomena of iron and related bcc metals. Acta Metallurgica 10, 727–738. 10.1016/0001-6160(62)90041-X.

Hähner, P., Zaiser, M., 1999. Dislocation dynamics and work hardening of fractal dislocation cell structures. Materials Science and Engineering: A 272, 443–454. 10.1016/S0921-5093(99)00527-4.

Hall, E.O., 1970. Yield point phenomena in metals and alloys. Plenum Press, New York, 1 online resource (viii, 296.

Hansen, B.L., Beyerlein, I.J., Bronkhorst, C.A., Cerreta, E.K., Dennis-Koller, D., 2013. A dislocation-based multi-rate single crystal plasticity model. International Journal of Plasticity 44, 129–146. 10.1016/j.ijplas.2012.12.006.

Hirth, J.P., 1961. On Dislocation Interactions in the fcc Lattice. Journal of Applied Physics 32, 700–706. 10.1063/1.1736074.

Hirth, J.P., Lothe, J., 1982. Theory of dislocations, 2nd ed. Krieger Pub. Co, Malabar FL, xii, 857.

Hochrainer, T., Sandfeld, S., Zaiser, M., Gumbsch, P., 2014. Continuum dislocation dynamics: Towards a physical theory of crystal plasticity. Journal of the Mechanics and Physics of Solids 63, 167–178. 10.1016/j.jmps.2013.09.012.

Hockett, J.E., Sherby, O.D., 1975. Large strain deformation of polycrystalline metals at low homologous temperatures. Journal of the Mechanics and Physics of Solids 23, 87–98. 10.1016/0022-5096(75)90018-6.

Horstemeyer, M.F., Bammann, D.J., 2010. Historical review of internal state variable theory for inelasticity. International Journal of Plasticity 26, 1310–1334. 10.1016/j.ijplas.2010.06.005.

Huang, M., Zhao, L., Tong, J., 2012. Discrete dislocation dynamics modelling of mechanical deformation of nickel-based single crystal superalloys. International Journal of Plasticity 28, 141–158. 10.1016/j.ijplas.2011.07.003.

Huang, Y., Qu, S., Hwang, K.C., Li, M., Gao, H., 2004. A conventional theory of mechanism-based strain gradient plasticity. International Journal of Plasticity 20, 753–782. 10.1016/j.ijplas.2003.08.002.

Hughes, D., Hansen, N., Bammann, D., 2003. Geometrically necessary boundaries, incidental dislocation boundaries and geometrically necessary dislocations. Scripta Materialia 48, 147–153. 10.1016/S1359-6462(02)00358-5.

Hughes, D.A., 1993. Microstructural evolution in a non-cell forming metal: Al・Mg. Acta Metallurgica et Materialia 41, 1421–1430. 10.1016/0956-7151(93)90251-M.

Hughes, D.A., Hansen, N., 1993. Microstructural evolution in nickel during rolling from intermediate to large strains. MTA 24, 2022–2037. 10.1007/BF02666337.

Hughes, D.A., Kassner, M.E., Stout, M.G., Vetrano, J.S., 1998. Metal forming at the center of excellence for the synthesis and processing of advanced materials. JOM 50, 16–21. 10.1007/s11837-998-0122-z.

Hull, D., Bacon, D.J., 2011. Introduction to dislocations. Butterworth-Heinemann, Amsterdam, 1 online resource (x, 257.

Hunter, A., Preston, D.L., 2015. Analytic model of the remobilization of pinned glide dislocations from quasi-static to high strain rates. International Journal of Plasticity 70, 1–29. 10.1016/j.ijplas.2015.01.008.

Jiang, J., Britton, T.B., Wilkinson, A.J., 2013. Measurement of geometrically necessary dislocation density with high resolution electron backscatter diffraction: effects of detector binning and step size. Ultramicroscopy 125, 1–9. 10.1016/j.ultramic.2012.11.003.

Johnson, G.R., Cook, W.H., 1983. A constitutive model and data for metals subjected to large strains, high strain rates and high temperatures. In Proceedings of the 7th International Symposium on Ballistics 21, 541–547.

Johnston, W.G., Gilman, J.J., 1959. Dislocation Velocities, Dislocation Densities, and Plastic Flow in Lithium Fluoride Crystals. Journal of Applied Physics 30, 129–144. 10.1063/1.1735121.

Kassner, M.E., 2015. Fundamentals of creep in materials. Butterworth-Heinemann, Amsterdam, 1 online resource (1 recurs en línia).

Kassner, M.E., Geantil, P., Levine, L.E., 2013. Long range internal stresses in single-phase crystalline materials. International Journal of Plasticity 45, 44–60. 10.1016/j.ijplas.2012.10.003.

Khan, A.S., Liang, R., 1999. Behaviors of three BCC metal over a wide range of strain rates and temperatures: Experiments and modeling. International Journal of Plasticity 15, 1089–1109. 10.1016/S0749-6419(99)00030-3.

Khan, A.S., Liu, H., 2012. Variable strain rate sensitivity in an aluminum alloy: Response and constitutive modeling. International Journal of Plasticity 36, 1–14. 10.1016/j.ijplas.2012.02.001.

Kibey, S., Liu, J.B., Johnson, D.D., Sehitoglu, H., 2007. Predicting twinning stress in fcc metals: Linking twin-energy pathways to twin nucleation. Acta Materialia 55, 6843–6851. 10.1016/j.actamat.2007.08.042.

Kitayama, K., Tomé, C.N., Rauch, E.F., Gracio, J.J., Barlat, F., 2013. A crystallographic dislocation model for describing hardening of polycrystals during strain path changes. Application to low carbon steels. International Journal of Plasticity 46, 54–69. 10.1016/j.ijplas.2012.09.004.

Knezevic, M., Beyerlein, I.J., Brown, D.W., Sisneros, T.A., Tomé, C.N., 2013. A polycrystal plasticity model for predicting mechanical response and texture evolution during strain-path changes: Application to beryllium. International Journal of Plasticity 49, 185–198. 10.1016/j.ijplas.2013.03.008.

Kocks, U.F., 1966. A statistical theory of flow stress and work-hardening. Philosophical Magazine 13, 541–566. 10.1080/14786436608212647.

Kocks, U.F., 1970. The relation between polycrystal deformation and single-crystal deformation. Metall and Materi Trans 1, 1121–1143. 10.1007/BF02900224.

Kocks, U.F., 1976. Laws for Work-Hardening and Low-Temperature Creep. J. Eng. Mater. Technol. 98, 76. 10.1115/1.3443340.

Kocks, U.F., Argon, A.S., Ashby M. F., 1975. Thermodynamics and kinetics of slip. Progress in Materials Science 19.

Kocks, U.F., Mecking, H., 2003. Physics and phenomenology of strain hardening: The FCC case. Progress in Materials Science 48, 171–273. 10.1016/S0079-6425(02)00003-8.

Koehler, J.S., 1952. The Nature of Work-Hardening. Phys. Rev. 86, 52–59. 10.1103/PhysRev.86.52.

Koyama, M., Sawaguchi, T., Tsuzaki, K., 2015. Deformation Twinning Behavior of Twinning-induced Plasticity Steels with Different Carbon Concentrations – Part 2: Proposal of Dynamic-strain-aging-assisted Deformation Twinning. ISIJ International 55, 1754–1761. 10.2355/isijinternational.ISIJINT-2015-070.

Kubin, L., 2013. Dislocations, Mesoscale Simulations and Plastic Flow. Oxford University Press.

Kubin, L.P., Canova, G., Condat, M., Devincre, B., Pontikis, V., Bréchet, Y., 1992. Dislocation Microstructures and Plastic Flow: A 3D Simulation. SSP 23-24, 455–472. 10.4028/www.scientific.net/SSP.23-24.455.

Kubin, L.P., Chihab, K., Estrin, Y., 1988. The rate dependence of the portevin-Le chatelier effect. Acta Metallurgica 36, 2707–2718. 10.1016/0001-6160(88)90117-4.

Kubin, L.P., Estrin, Y., 1990. Evolution of dislocation densities and the critical conditions for the Portevin-Le Châtelier effect. Acta Metallurgica et Materialia 38, 697–708. 10.1016/0956-7151(90)90021-8.

Kubin, L.P., Fressengeas, C., Ananthakrishna, G., 1979-<2010>. Chapter 57 Collective behaviour of dislocations in plasticity, in: Nabarro, F.R.N., Duesbery, M.S., Hirth, J.P. (Eds.), Dislocations in solids, vol. 11. North-Holland Pub. Co, Amsterdam, New York, pp. 101–192.

Kuhlmann-Wilsdorf, D., Hansen, N., 1991. Geometrically necessary, incidental and subgrain boundaries. Scripta Metallurgica et Materialia 25, 1557–1562. 10.1016/0956-716X(91)90451-6.

Kumar, A., Hauser, F., Dorn, J., 1968. Viscous drag on dislocations in aluminum at high strain rates. Acta Metallurgica 16, 1189–1197. 10.1016/0001-6160(68)90054-0.

Lee, S., Jeong, J., Kim, Y., Han, S.M., Kiener, D., Oh, S.H., 2016. FIB-induced dislocations in Al submicron pillars: Annihilation by thermal annealing and effects on deformation behavior. Acta Materialia 110, 283–294. 10.1016/j.actamat.2016.03.017.

Li, D., Zbib, H., Sun, X., Khaleel, M., 2014. Predicting plastic flow and irradiation hardening of iron single crystal with mechanism-based continuum dislocation dynamics. International Journal of Plasticity 52, 3–17. 10.1016/j.ijplas.2013.01.015.

Li, J.C.M., 1967. Dislocation dynamics in deformation and recovery. Can. J. Phys. 45, 493–509. 10.1139/p67-043.

Lim, H., Lee, M.G., Kim, J.H., Adams, B.L., Wagoner, R.H., 2011. Simulation of polycrystal deformation with grain and grain boundary effects. International Journal of Plasticity 27, 1328–1354. 10.1016/j.ijplas.2011.03.001.

Lin, Y.C., Chen, X.-M., 2011. A critical review of experimental results and constitutive descriptions for metals and alloys in hot working. Materials & Design 32, 1733–1759. 10.1016/j.matdes.2010.11.048.

Lloyd, J.T., Clayton, J.D., Becker, R., McDowell, D.L., 2014. Simulation of shock wave propagation in single crystal and polycrystalline aluminum. International Journal of Plasticity 60, 118–144. 10.1016/j.ijplas.2014.04.012.

Lomer, W.M., 1951. A dislocation reaction in the face-centred cubic lattice. The London, Edinburgh, and Dublin Philosophical Magazine and Journal of Science 42, 1327–1331. 10.1080/14786444108561389.

Luscher, D.J., Addessio, F.L., Cawkwell, M.J., Ramos, K.J., 2017. A dislocation density-based continuum model of the anisotropic shock response of single crystal α -cyclotrimethylene trinitramine. Journal of the Mechanics and Physics of Solids 98, 63–86. 10.1016/j.jmps.2016.09.005.

Lyu, H., Ruimi, A., Zbib, H.M., 2015. A dislocation-based model for deformation and size effect in multi-phase steels. International Journal of Plasticity 72, 44–59. 10.1016/j.ijplas.2015.05.005.

Ma, A., Roters, F., 2004. A constitutive model for fcc single crystals based on dislocation densities and its application to uniaxial compression of aluminium single crystals. Acta Materialia 52, 3603–3612. 10.1016/j.actamat.2004.04.012.

Ma, A., Roters, F., Raabe, D., 2006. A dislocation density based constitutive model for crystal plasticity FEM including geometrically necessary dislocations. Acta Materialia 54, 2169–2179. 10.1016/j.actamat.2006.01.005.

Madec, R., Devincre, B., Kubin, L., Hoc, T., Rodney, D., 2003. The role of collinear interaction in dislocation-induced hardening. Science (New York, N.Y.) 301, 1879–1882. 10.1126/science.1085477.

Mahajan, S., Chin, G.Y., 1973. Formation of deformation twins in f.c.c. crystals. Acta Metallurgica 21, 1353–1363. 10.1016/0001-6160(73)90085-0.

Mecking, H., Kocks, U.F., 1981. Kinetics of flow and strain-hardening. Acta Metallurgica 29, 1865–1875. 10.1016/0001-6160(81)90112-7.

Miller, R.E., Shilkrot, L., Curtin, W.A., 2004. A coupled atomistics and discrete dislocation plasticity simulation of nanoindentation into single crystal thin films. Acta Materialia 52, 271–284. 10.1016/j.actamat.2003.09.011.

Monavari, M., Sandfeld, S., Zaiser, M., 2016. Continuum representation of systems of dislocation lines: A general method for deriving closed-form evolution equations. Journal of the Mechanics and Physics of Solids 95, 575–601. 10.1016/j.jmps.2016.05.009.

Monavari, M., Zaiser, M., 2018. Annihilation and sources in continuum dislocation dynamics. Mater Theory 2, 761. 10.1186/s41313-018-0010-z.

Motz, C., Weygand, D., Senger, J., Gumbsch, P., 2008. Micro-bending tests: A comparison between three-dimensional discrete dislocation dynamics simulations and experiments. Acta Materialia 56, 1942–1955. 10.1016/j.actamat.2007.12.053.

Mughrabi, H., 1983. Dislocation wall and cell structures and long-range internal stresses in deformed metal crystals. Acta Metallurgica 31, 1367–1379. 10.1016/0001-6160(83)90007-X.

Mughrabi, H., 1987. A two-parameter description of heterogeneous dislocation distributions in deformed metal crystals. Materials Science and Engineering 85, 15–31. 10.1016/0025-5416(87)90463-0.

Mughrabi, H., 2006. Deformation-induced long-range internal stresses and lattice plane misorientations and the role of geometrically necessary dislocations. Philosophical Magazine 86, 4037–4054. 10.1080/14786430500509054.

Mukherjee, M., Prahl, U., Bleck, W., 2010. Modelling of Microstructure and Flow Stress Evolution during Hot Forging. steel research int. 81, 1102–1116. 10.1002/srin.201000114.

Mulford, R.A., Kocks, U.F., 1979. New observations on the mechanisms of dynamic strain aging and of jerky flow. Acta Metallurgica 27, 1125–1134. 10.1016/0001-6160(79)90130-5.

Nabarro, F., 1952. Mathematical theory of stationary dislocations. Advances in Physics 1, 269–394. 10.1080/00018735200101211.

Nabarro, F., 1997. Fifty-year study of the Peierls-Nabarro stress. Materials Science and Engineering: A 234-236, 67–76. 10.1016/S0921-5093(97)00184-6.

Nadgornyi, E., 1988. Dislocation dynamics and mechanical properties of crystals. Progress in Materials Science 31, 1–530. 10.1016/0079-6425(88)90005-9.

Nes, E., 1997. Modelling of work hardening and stress saturation in FCC metals. Progress in Materials Science 41, 129–193. 10.1016/S0079-6425(97)00032-7.

Ng, K.S., Ngan, A.H.W., 2008. Stochastic nature of plasticity of aluminum micro-pillars. Acta Materialia 56, 1712–1720. 10.1016/j.actamat.2007.12.016.

Nguyen, T., Luscher, D.J., Wilkerson, J.W., 2017a. A dislocation-based crystal plasticity framework for dynamic ductile failure of single crystals. Journal of the Mechanics and Physics of Solids 108, 1–29. 10.1016/j.jmps.2017.07.020.

Nguyen, T.N., Siegmund, T., Tomar, V., Kruzic, J.J., 2017b. Interaction of rate- and size-effect using a dislocation density based strain gradient viscoplasticity model. Journal of the Mechanics and Physics of Solids 109, 1–21. 10.1016/j.jmps.2017.07.022.

Nix, W.D., Gao, H., 1998. Indentation size effects in crystalline materials: A law for strain gradient plasticity. Journal of the Mechanics and Physics of Solids 46, 411–425. 10.1016/S0022-5096(97)00086-0.

Nix, W.D., Gibeling, J.C., Hughes, D.A., 1985. Time-dependent deformation of metals. MTA 16, 2215–2226. 10.1007/BF02670420.

Nye, J., 1953. Some geometrical relations in dislocated crystals. Acta Metallurgica 1, 153–162. 10.1016/0001-6160(53)90054-6.

Oh, S.H., Legros, M., Kiener, D., Dehm, G., 2009. In situ observation of dislocation nucleation and escape in a submicrometre aluminium single crystal. Nature materials 8, 95–100. 10.1038/nmat2370.

Oren, E., Yahel, E., Makov, G., 2017. Kinetics of dislocation cross-slip: A molecular dynamics study. Computational Materials Science 138, 246–254. 10.1016/j.commatsci.2017.06.039.

Orowan, E., 1934. Zur Kristallplastizitaet. Z. Physik 89, 605–613. 10.1007/BF01341478.

Papanikolaou, S., Cui, Y., Ghoniem, N., 2018. Avalanches and plastic flow in crystal plasticity: An overview. Modelling Simul. Mater. Sci. Eng. 26, 13001. 10.1088/1361-651X/aa97ad.

Parthasarathy, T.A., Rao, S.I., Dimiduk, D.M., Uchic, M.D., Trinkle, D.R., 2007. Contribution to size effect of yield strength from the stochastics of dislocation source lengths in finite samples. Scripta Materialia 56, 313–316. 10.1016/j.scriptamat.2006.09.016.

Patra, A., McDowell, D.L., 2012. Crystal plasticity-based constitutive modelling of irradiated bcc structures. Philosophical Magazine 92, 861–887. 10.1080/14786435.2011.634855.

Pauš, P., Kratochvíl, J., Beneš, M., 2013. A dislocation dynamics analysis of the critical cross-slip annihilation distance and the cyclic saturation stress in fcc single crystals at different temperatures. Acta Materialia 61, 7917–7923. 10.1016/j.actamat.2013.09.032.

Peierls, R., 1940. The size of a dislocation. Proc. Phys. Soc. 52, 34–37. 10.1088/0959-5309/52/1/305.

Perzyna, P., 1966. Fundamental Problems in Viscoplasticity, in: Chernyi, G.G. (Ed.), Advances in applied mechanics, vol. 9. Academic Press, New York, pp. 243–377.

Pham, M.S., Holdsworth, S.R., Janssens, K., Mazza, E., 2013. Cyclic deformation response of AISI 316L at room temperature: Mechanical behaviour, microstructural evolution, physically-based evolutionary constitutive modelling. International Journal of Plasticity 47, 143–164. 10.1016/j.ijplas.2013.01.017.

Pham, M.-S., Iadicola, M., Creuziger, A., Hu, L., Rollett, A.D., 2015. Thermally-activated constitutive model including dislocation interactions, aging and recovery for strain path dependence of solid solution strengthened alloys: Application to AA5754-O. International Journal of Plasticity 75, 226–243. 10.1016/j.ijplas.2014.09.010.

Po, G., Mohamed, M.S., Crosby, T., Erel, C., El-Azab, A., Ghoniem, N., 2014. Recent Progress in Discrete Dislocation Dynamics and Its Applications to Micro Plasticity. JOM 66, 2108–2120. 10.1007/s11837-014-1153-2.

Püschl, W., 2002. Models for dislocation cross-slip in close-packed crystal structures: A critical review. Progress in Materials Science 47, 415–461. 10.1016/S0079-6425(01)00003-2.

Qiu, X., Huang, Y., Nix, W.D., Hwang, K.C., Gao, H., 2001. Effect of intrinsic lattice resistance in strain gradient plasticity. Acta Materialia 49, 3949–3958. 10.1016/S1359-6454(01)00299-3.

Rivera-Díaz-del-Castillo, P., Huang, M., 2012. Dislocation annihilation in plastic deformation: I. Multiscale irreversible thermodynamics. Acta Materialia 60, 2606–2614. 10.1016/j.actamat.2012.01.027.

Rosakis, P., Rosakis, A.J., Ravichandran, G., Hodowany, J., 2000. A thermodynamic internal variable model for the partition of plastic work into heat and stored energy in metals. Journal of the Mechanics and Physics of Solids 48, 581–607. 10.1016/S0022-5096(99)00048-4.

Roters, F., 2011. Advanced Material Models for the Crystal Plasticity Finite Element Method: Development of a general CPFEM framework. Habilitation thesis.

Roters, F., Raabe, D., Gottstein, G., 2000. Work hardening in heterogeneous alloys—a microstructural approach based on three internal state variables. Acta Materialia 48, 4181–4189. 10.1016/S1359-6454(00)00289-5.

Rusinek, A., Klepaczko, J.R., 2001. Shear testing of a sheet steel at wide range of strain rates and a constitutive relation with strain-rate and temperature dependence of the flow stress. International Journal of Plasticity 17, 87–115. 10.1016/S0749-6419(00)00020-6.

Rusinek, A., Rodríguez-Martínez, J.A., 2009. Thermo-viscoplastic constitutive relation for aluminium alloys, modeling of negative strain rate sensitivity and viscous drag effects. Materials & Design 30, 4377–4390. 10.1016/j.matdes.2009.04.011.

Rusinek, A., Rodríguez-Martínez, J.A., Arias, A., 2010. A thermo-viscoplastic constitutive model for FCC metals with application to OFHC copper. International Journal of Mechanical Sciences 52, 120–135. 10.1016/j.ijmecsci.2009.07.001.

Sandfeld, S., Thawinan, E., Wieners, C., 2015. A link between microstructure evolution and macroscopic response in elasto-plasticity: Formulation and numerical approximation of the higher-dimensional continuum dislocation dynamics theory. International Journal of Plasticity 72, 1–20. 10.1016/j.ijplas.2015.05.001.

Sandström, R., Lagneborg, R., 1975. A model for hot working occurring by recrystallization. Acta Metallurgica 23, 387–398. 10.1016/0001-6160(75)90132-7.

Seeger, A., 1955. The generation of lattice defects by moving dislocations, and its application to the temperature dependence of the flow-stress of F.C.C. crystals. The London, Edinburgh, and Dublin Philosophical Magazine and Journal of Science 46, 1194–1217. 10.1080/14786441108520632.

Seeger, A., Diehl, J., Mader, S., Rebstock, H., 1957. Work-hardening and work-softening of face-centred cubic metal crystals. Philosophical Magazine 2, 323–350. 10.1080/14786435708243823.

Sherby, O.D., Burke, P.M., 1968. Mechanical behavior of crystalline solids at elevated temperature. Progress in Materials Science 13, 323–390. 10.1016/0079-6425(68)90024-8.

Shiari, B., Miller, R.E., Curtin, W.A., 2005. Coupled Atomistic/Discrete Dislocation Simulations of Nanoindentation at Finite Temperature. J. Eng. Mater. Technol. 127, 358. 10.1115/1.1924561.

Steenackers, G., Devriendt, C., Guillaume, P., 2007. On the use of transmissibility measurements for finite element model updating. Journal of Sound and Vibration 303, 707–722. 10.1016/j.jsv.2007.01.030.

Steif, P.S., Clifton, R.J., 1979. On the kinetics of a Frank-Read source. Materials Science and Engineering 41, 251–258. 10.1016/0025-5416(79)90145-9.

Steinmetz, D.R., Jäpel, T., Wietbrock, B., Eisenlohr, P., Gutierrez-Urrutia, I., Saeed-Akbari, A., Hickel, T., Roters, F., Raabe, D., 2013. Revealing the strain-hardening behavior of twinning-induced plasticity steels: Theory, simulations, experiments. Acta Materialia 61, 494–510. 10.1016/j.actamat.2012.09.064.

Stricker, M., Sudmanns, M., Schulz, K., Hochrainer, T., Weygand, D., 2018. Dislocation multiplication in stage II deformation of fcc multi-slip single crystals. Journal of the Mechanics and Physics of Solids 119, 319–333. 10.1016/j.jmps.2018.07.003.

Stricker, M., Weygand, D., 2015. Dislocation multiplication mechanisms – Glissile junctions and their role on the plastic deformation at the microscale. Acta Materialia 99, 130–139. 10.1016/j.actamat.2015.07.073.

Sung, J.H., Kim, J.H., Wagoner, R.H., 2010. A plastic constitutive equation incorporating strain, strain-rate, and temperature. International Journal of Plasticity 26, 1746–1771. 10.1016/j.ijplas.2010.02.005.

Tang, M., Marian, J., 2014. Temperature and high strain rate dependence of tensile deformation behavior in single-crystal iron from dislocation dynamics simulations. Acta Materialia 70, 123–129. 10.1016/j.actamat.2014.02.013.

Taylor, G.I., 1934. The Mechanism of Plastic Deformation of Crystals. Part I. Theoretical. Proceedings of the Royal Society A: Mathematical, Physical and Engineering Sciences 145, 362–387. 10.1098/rspa.1934.0106.

Taylor, G.I., 1938. Plastic strain in metals. J. Inst. Metals, 307–325.

Taylor, G.I., Quinney, H., 1934. The Latent Energy Remaining in a Metal after Cold Working. Proceedings of the Royal Society A: Mathematical, Physical and Engineering Sciences 143, 307–326. 10.1098/rspa.1934.0004.

Thompson, N., 1953. Dislocation Nodes in Face-Centred Cubic Lattices. Proc. Phys. Soc. B 66, 481–492. 10.1088/0370-1301/66/6/304.

van den Beukel, A., Kocks, U.F., 1982. The strain dependence of static and dynamic strain-aging. Acta Metallurgica 30, 1027–1034. 10.1016/0001-6160(82)90211-5.

van der Giessen, E., Needleman, A., 1995. Discrete dislocation plasticity: A simple planar model. Modelling Simul. Mater. Sci. Eng. 3, 689–735. 10.1088/0965-0393/3/5/008.

Vegge, T., Jacobsen, K.W., 2002. Atomistic simulations of dislocation processes in copper. J. Phys.: Condens. Matter 14, 2929–2956. 10.1088/0953-8984/14/11/309.

Venables, J., 1964. The nucleation and propagation of deformation twins. Journal of Physics and Chemistry of Solids 25, 693–700. 10.1016/0022-3697(64)90178-7.

Viatkina, E.M., Brekelmans, W., Geers, M., 2007. Modelling the evolution of dislocation structures upon stress reversal. International Journal of Solids and Structures 44, 6030–6054. 10.1016/j.ijsolstr.2007.02.010.

Voyiadjis, G.Z., Abed, F.H., 2005. Microstructural based models for bcc and fcc metals with temperature and strain rate dependency. Mechanics of Materials 37, 355–378. 10.1016/j.mechmat.2004.02.003.

Voyiadjis, G.Z., Al-Rub, R.K.A., 2005. Gradient plasticity theory with a variable length scale parameter. International Journal of Solids and Structures 42, 3998–4029. 10.1016/j.ijsolstr.2004.12.010.

Webster, G.A., 1966. A widely applicable dislocation model of creep. Philosophical Magazine 14, 775–783. 10.1080/14786436608211971.

Weygand, D., 2014. Mechanics and Dislocation Structures at the Micro-Scale: Insights on Dislocation Multiplication Mechanisms from Discrete Dislocation Dynamics Simulations. MRS Proc. 1651, 188. 10.1557/opl.2014.362.

Wong, S.L., Madivala, M., Prahl, U., Roters, F., Raabe, D., 2016. A crystal plasticity model for twinning- and transformation-induced plasticity. Acta Materialia 118, 140–151. 10.1016/j.actamat.2016.07.032.

Yamakov, V., Wolf, D., Phillpot, S.R., Mukherjee, A.K., Gleiter, H., 2002. Dislocation processes in the deformation of nanocrystalline aluminium by molecular-dynamics simulation. Nature materials 1, 45–48. 10.1038/nmat700.

Yang, H.K., Zhang, Z.J., Tian, Y.Z., Zhang, Z.F., 2017. Negative to positive transition of strain rate sensitivity in Fe-22Mn-0.6C-x(Al) twinning-induced plasticity steels. Materials Science and Engineering: A 690, 146–157. 10.1016/j.msea.2017.02.014.

Yuan, S., Huang, M., Zhu, Y., Li, Z., 2018. A dislocation climb/glide coupled crystal plasticity constitutive model and its finite element implementation. Mechanics of Materials 118, 44–61. 10.1016/j.mechmat.2017.12.009.

Zaiser, M., 2013. Statistical aspects of microplasticity: Experiments, discrete dislocation simulations and stochastic continuum models. Journal of the Mechanical Behavior of Materials 22. 10.1515/jmbm-2012-0006.

Zbib, H.M., La Diaz de Rubia, T., 2002. A multiscale model of plasticity. International Journal of Plasticity 18, 1133–1163. 10.1016/S0749-6419(01)00044-4.

Zbib, H.M., Rhee, M., Hirth, J.P., 1998. On plastic deformation and the dynamics of 3D dislocations. International Journal of Mechanical Sciences 40, 113–127. 10.1016/S0020-7403(97)00043-X.

Zecevic, M., Knezevic, M., 2015. A dislocation density based elasto-plastic self-consistent model for the prediction of cyclic deformation: Application to AA6022-T4. International Journal of Plasticity 72, 200–217. 10.1016/j.ijplas.2015.05.018.

Zehnder, A.T., 1991. A model for the heating due to plastic work. Mechanics Research Communications 18, 23–28. 10.1016/0093-6413(91)90023-P.

Zhang, J.-l., Zaefferer, S., Raabe, D., 2015. A study on the geometry of dislocation patterns in the surrounding of nanoindents in a TWIP steel using electron channeling contrast imaging and discrete dislocation dynamics simulations. Materials Science and Engineering: A 636, 231–242. 10.1016/j.msea.2015.03.078.

Zhou, C., Biner, S.B., LeSar, R., 2010. Discrete dislocation dynamics simulations of plasticity at small scales. Acta Materialia 58, 1565–1577. 10.1016/j.actamat.2009.11.001.

Zhu, T., 2004. Predictive modeling of nanoindentation-induced homogeneous dislocation nucleation in copper. Journal of the Mechanics and Physics of Solids 52, 691–724. 10.1016/j.jmps.2003.07.006.